\documentclass[draft,a4paper,twocolumn,prd,superscriptaddress,showpacs,%
floatfix,nobalancelastpage,preprintnumbers,nofootinbib]{revtex4-1}

\usepackage[final]{graphics,graphicx}
\usepackage{amsmath,amssymb} 
\usepackage{verbatim}
\usepackage{epsfig} 
\usepackage{bm} 
\usepackage{color}
\usepackage{fixme}
\usepackage[normalem]{ulem}

\newcommand{\Mei}{{8.8~$M_\odot$}}
\newcommand{\Mten}{{10.8~$M_\odot$}}
\newcommand{\Met}{{18.0~$M_\odot$}}
\newcommand{\bone}{{$\bar1$}} \newcommand{\btwo}{{$\bar2$}}
\newcommand{\bthree}{{$\bar3$}}

\begin{document}

\title{Combining collective, MSW, and turbulence effects in 
supernova neutrino flavor evolution}

\author{Tina Lund}
\affiliation{Department of Physics, North
Carolina State University, 2401 Stinson Drive, Raleigh, USA}

\author{James P. Kneller}
\affiliation{Department of Physics, North 
Carolina State University, 2401 Stinson Drive, Raleigh, USA}

\date{\today}


\begin{abstract}
In order to decode the neutrino burst signal from a Galactic
core-collapse supernova (ccSN) and reveal the complicated inner
workings of the explosion we need a thorough understanding of
the neutrino flavor evolution from the proto-neutron star
outwards. The flavor content of the signal evolves due to both
neutrino collective effects and matter effects which can lead
to a highly interesting interplay and distinctive spectral
features. In this paper we investigate the supernova neutrino 
flavor evolution in three different progenitors and include
collective flavor effects, the evolution of the Mikheyev, 
Smirnov \& Wolfenstein (MSW) conversion due
to the shock wave passage through the star, and the impact of 
turbulence. We consider both normal and inverted neutrino
mass hierarchies and a value of $\theta_{13}$ close to
the current experimental measurements.  
In the Oxygen-Neon-Magnesium (ONeMg) supernova we find that the 
impact of turbulence is both brief and slight during a window 
of 1--2 seconds post bounce. This is because the shock races 
through the star extremely quickly and the turbulence amplitude 
is expected to be small, less than 10\%, since these stars do 
not require multi-dimensional physics to explode. Thus the 
spectral features of collective and shock effects in the 
neutrino signals from Oxygen-Neon-Magnesium supernovae may be 
almost turbulence free making them the easiest to interpret.
For the more massive progenitors we again find that small amplitude
turbulence, up to 10\%, leads to a minimal modification of the
signal, and the emerging neutrino spectra retain both collective
and MSW features. However, when larger amounts of turbulence is
added, 30\% and 50\%, which is justified by the requirement of
multi-dimensional physics in order to make these stars explode,
the features of collective and shock wave effects in the 
high (H) density resonance channel are almost completely obscured
at late times. 
Yet at the same time we find the other mixing 
channels - the low (L) density resonance channel and the non-resonant
channels - begin to develop turbulence signatures. 
Large amplitude turbulent motions in the outer layers of more massive,
iron core-collapse supernovae may obscure the most obvious 
fingerprints of collective and shock wave effects in the
neutrino signal but cannot remove them completely, and 
additionally bring about new features in the signal.
\end{abstract}


\maketitle

\section{Introduction}	       		 \label{sec:Intro}
Our understanding of how massive stars explode continues to
evolve at a frenetic pace due to advances in the hydrodynamical
modelling  
\cite{Janka:2012wk,Burrows:2012ew,2011ApJ...742...74M,2012arXiv1210.5241D,
2012arXiv1210.6674O,2012ApJ...755..138H,2012ApJ...746..106P,
2012ApJ...761...72M,2012ApJ...749...98T,2013arXiv1301.1326L,
Ugliano:2012kq,Ott:2008jb,O'Connor:2012am,Kuroda:2012nc,
Bruenn:2012mj,Lentz:2012xc,Lentz:2011aa,
Fischer:2009,Muller:2011yi,Nakazato:2012qf,Huedepohl:2009wh}.
The long-sought goal of simulating successful explosions based 
upon first-principle physics appears to be imminent,
and emerging from that is a basic paradigm with the collapse 
of the core followed by the
formation of a shock which propagates out to $r\sim 200\;{\rm
km}$ before stalling. Due to a mixture of neutrino heating
and/or the Standing Accretion Shock Instability (SASI)
\cite{2011ApJ...742...74M,2012arXiv1210.5241D,2012arXiv1210.6674O,
2012ApJ...755..138H,2012ApJ...746..106P,2012ApJ...761...72M,
2012ApJ...749...98T,2013arXiv1301.1326L} the outward motion of
the shock is revived after a delay of up to $t \sim 500\;{\rm ms}$.
The shock then makes its way through the mantle of the star to
eventually reach the surface creating the spectacular fireworks
we observe. 

As impressive as the optical emission from core-collapse
supernovae (ccSNe) may be, the neutrino burst from the next ccSN
in our Galaxy will outshine the rest of the Universe in
neutrinos and represents an unparalleled opportunity to learn
about the dynamics at the core of the explosion. The potential
of the signal to answer outstanding questions in physics and
astrophysics was recently reviewed by Scholberg
\cite{2012ARNPS..62...81S}. For example, the neutrino emission
during the accretion phase leaves tell-tale oscillatory features
which may be observed in large water/ice Cerenkov detectors such
as IceCube as shown by Lund \emph{et al.}
\cite{2010PhRvD..82f3007L,2012PhRvD..86j5031L}.  Decoding the 
neutrino burst signal will not be easy, however, because our 
detectors are sensitive to the neutrino flavor, 
and the flavor content of
the signal is a function of both time, energy and emission point
from the proto-neutron star. As the neutrinos propagate through
the supernova mantle two time dependent flavor transformation
processes can occur: collective (self-interaction) effects
during the first $\sim 1000\; {\rm km}$ or so, and the Mikheyev,
Smirnov \& Wolfenstein (MSW) \cite{M&S1986,Wolfenstein1977}
effect which occurs when the matter density is in the range of
$1 \; {\rm g/cm^3} \leq \rho \leq 10^{4} \; {\rm g/cm^3}$ for
neutrino energies of order $1-100\; {\rm MeV}$. 

Neutrino collective effects are a very active field of study
with significant and ongoing progress.  Initially investigations
were primarily in terms of effective two flavor calculations
(see e.g.~\cite{Pantaleone:Gamma1292eq,Samuel:1993uw,Pastor:2001iu,
Duan:2005cp,Hannestad:2006nj,Duan:2006an,Raffelt:2007cb,
2008PhRvD..78l5015R,2011PhRvD..84h5023R,Choubey:2010,Fogli:2009tern})
but lately more and more investigations consider three flavors
(see e.g.~\cite{Dasgupta:2010,Friedland:2010,Duan:2011,
2011PhRvL.107o1101C,2012JPhG...39c5201G,2012PhRvL.108z1104C,
2012PhRvD..85k3007S,2012PhRvL.108w1102M,Mirizzi:2012wp}).
The phenomenology of collective effects with three flavors has
been found to be much richer than for two with new effects
appearing such as multiple splits for both neutrinos and
anti-neutrinos in either hierarchy 
(e.g.\ Dasgupta {\it et al.}~\cite{Dasgupta:2010}).
Over time it has also been realized that the standard set of
luminosity and energy values employed (equipartition of
luminosities and a strong hierarchy of energies) is a special 
case rather than being a generalizable choice of values. 
In calculations of collective effects typical values used 
for the mean energies are
$E_{\nu_e}= 10-12$~MeV, $E_{\bar\nu_e}= 15-16$~MeV and
$E_{\nu_x}= 16-27$~MeV \cite{Duan:2006an,Schirato:2002tg,
Choubey:2010,Dasgupta:2010} 
along with equipartition of the luminosities. 
Smaller energy differences are found in recent long term simulations, 
e.g.\ H\"udepohl {\it et al.} \cite{Huedepohl:2009wh} and 
Nakazato {\it et al.} \cite{Nakazato:2012qf}, where energies are 
on the order of $E_{\nu_e}= 9-10.1$~MeV, 
$E_{\bar\nu_e}= 11.5-12.9$~MeV and $E_{\nu_x} \sim 13$~MeV. 
Among others, Roberts {\it et al.} \cite{Roberts:2012um}
and Horowitz {\it et al} \cite{Horowitz:2012us} have pointed out
that the neutrino opacities and interaction cross sections used 
by the modeling community are not as correct as they could be. 
Previous statements by Reddy {\it et al.} \cite{Reddy:1997yr} 
along the same lines lead to recalculations
by Mart\'inez-Pinedo {\it et al.}~\cite{MartinezPinedo:2012rb}
of some of the simulations from \cite{Fischer:2009},
although not to post bounce times as late as those available from
Fischer {\it et al.} \cite{Fischer:2009}. The findings of
Mart\'inez-Pinedo {\it et al.}~\cite{MartinezPinedo:2012rb} 
indicate that the inclusion of
the correct opacities will lead to an overall lowering
of the luminosities, and a reduction, respectively increase, of
the $\nu_e$ and $\bar\nu_e$ energies. A similar conclusion
was reached by Horowitz {\it et al.} \cite{Horowitz:2012us}
in their short duration 1D simulations. 
As a result of the ongoing improvements to the simulations
the energy and luminosity values are continously adjusted.
Investigations have scanned parts of the parameter space 
of luminosities and energies
(e.g.\ Fogli {\it et al.}~\cite{Fogli:2009tern}) but they 
have not been exhaustive.  Even
more recently it has been found there are new flavor
instabilities due to flavor dependent angular distributions at
the neutrinosphere \cite{2012PhRvL.108w1102M}, and possibly an
additional effect, known as the neutrino halo, due to the
scattering of the neutrinos
\cite{2012PhRvL.108z1104C,2012PhRvD..85k3007S,Cherry:2013}.  

At the present time the understanding of the neutrino
self-interaction is not sufficient to be able to predict results
except in the broadest sense. The linear stability analysis of
Banerjee {\it et al.} \cite{Banerjee:2011fj} and Sarikas {\it et al.} 
\cite{Sarikas:2011am,Sarikas:2012ad}, extended in 
Mirizzi \& Serpico \cite{2012PhRvL.108w1102M,Mirizzi:2012wp}
and Saviano {\it et al.} \cite{Saviano:2012yh}
allows one to analyze the system and demonstrate the existence
of the conditions that lead to the collective phenomena but not
the details.  The non-linear nature of the neutrino
self-interaction and the strong dependence that has been
discovered on even small differences in L and E means we do not
yet possess the analytical predictive power over the resulting
features and their behavior. We therefore have to primarily rely
on numerical calculations to expand our knowledge.

The flavor transformation due to the MSW effect is also
non-trivial because the passage of the shock wave through the
star leaves an impression in the signal
\cite{Schirato:2002tg,Takahashi:2002yj,Fogli:2003dw,Tomas:2004gr,
Choubey:2006aq,Kneller:2007kg,Gava:2009pj}.
A further complication is the possible presence of density
fluctuations/turbulence
\cite{Loreti:1995ae,Fogli:2006xy,Friedland:2006ta,Kneller:2010,
Kneller:2010ky,2011PhRvD..84h5023R,2011PhRvD..84j5034C} which
one would expect to be created by the large scale
inhomogeneities generated during the accretion phase.
A turbulent density profile would also imply neutrinos of the
same energy arriving at the same time in a detector but emitted
from different locations at the proto-neutron star will not have
experienced the same flavor density profile history
\cite{Kneller&Mauney,2013arXiv1302.3825K}.  
Finally, once the star is left behind
and the neutrinos have reached Earth recent studies indicate
that fortunately/unfortunately (depending upon one's point of
view) Earth matter effects on the neutrino signal may be minimal
\cite{2012PhRvD..86h3004B}. 

For the most part the various neutrino flavor transformation
processes have been studied in isolation, see Gava \emph{et al.}
\cite{Gava:2009pj} for an exception, but of course the neutrino
signal is the denouement involving all these protagonists. In
order to understand the explosion narrative we must determine
which process left which features in the signal and how they
interacted.  The aim of this paper is to study the interplay of
the various neutrino flavor transformation processes that can
occur in core-collapse supernovae by exploring the features each
engenders separately and in combination.  For our calculations
we use the density profiles and neutrino spectra from the
hydrodynamical simulations by Fischer \emph{et al.}
\cite{Fischer:2009}, 
consider both
normal and inverted hierarchies and use a value of $\theta_{13}$
close to the recent experimental results from T2K
\cite{2011PhRvL.107d1801A}, Double Chooz
\cite{2012PhRvL.108m1801A}, RENO \cite{2012PhRvL.108s1802A} and
Daya Bay \cite{2012PhRvL.108q1803A}. 

The organization of this paper is as follows.  In order to
assist the reader we first describe in
Sec.~\ref{sec:calc_descript} the similarities and differences
between the density profiles, and the neutrino spectra from the
three simulations pointing out important features relevant for
the neutrino flavor evolution. We then describe the evolution
calculations and how we modified the profiles to steepen the
shocks and insert turbulence. In Sec.~\ref{sec:results} we walk
the reader through the different signatures inserted into the
signal by each transformation process at a given snapshot during
the explosion. Then we run through how these features evolve
with time in Sec.~\ref{sec:time_evolv}.  We present our
conclusions in section Sec.~\ref{sec:conclusions}.

\begin{table*} 
\centering 
\caption{\label{tab:datachart} Characteristics of our 3 numerical 
models. All times are post
bounce. Further details can be found in \cite{Fischer:2009}.}
\begin{tabular*}{\linewidth}{@{\extracolsep{\fill}}l r r r }
\toprule 
Model   & 8.8~$M_\odot$	& 10.8~$M_\odot$
& $18.0~M_\odot$   \\ \colrule Simulation end time [s] & 4.541
& 10.545  & 21.804 \\ shock revival time [ms]     & $\sim$ 30 &
$\simeq$ 300  & $\simeq$ 300     \\
EoS                     & Shen      & Shen  & Shen \\ Progenitor
from         & Nomoto (83, 84 \& 87)      & Woosley {\it et al.}
(02)       &   Woosley {\it et al.} (02)   \\ 
\botrule
\end{tabular*} 
\end{table*}

\section{Description of the calculations}
\label{sec:calc_descript}
\subsection{Theory and background}
In order to construct the neutrino burst signals here at Earth
there are many calculated components one needs to put together.
Our first task is to calculate the probability that a neutrino
in a particular initial state emerges from the supernova  in a
given final state. These probabilities depend upon the basis and
for supernova neutrinos the most useful transition probabilities
are those linking the initial flavor state $\nu_\alpha$ to the mass
eigenstate $\nu_i$. The mass eigenstates are the states which
diagonalize the vacuum Hamiltonian and the reason we need the
probability of emerging in these states - and not the
probability that the neutrino emerges in a flavor state - is
because the neutrino wave packet will decohere on its passage to
Earth and will arrive as separate mass states.  This probability,
$P_{i\alpha} = P(\nu_{\alpha} \rightarrow \nu_i)$, can be found
quite easily from the $S$-matrix linking the initial neutrino
flavor states to the final mass states i.e.\ $P_{i\alpha} =
|S_{i\alpha}|^{2}$.  In this paper we shall use $\bar
P_{i\alpha}$ for the anti-neutrino probabilities and the symbol
$\bar{S}$ for the anti-neutrino $S$-matrix.
There is a third basis one often sees in the literature known as
the matter basis \cite{Kneller:2009km,2012JPhG...39c5201G}.  
This basis is the most useful one for actually
doing the calculations plus the matter basis states closely
align with the flavor states at the neutrinosphere and the mass
states in the vacuum. Throughout this paper we will show results
in the matter basis transition probabilities and we refer the
reader to \cite{Kneller:2009km,2012JPhG...39c5201G} for the
definition of this basis and its detailed connection with the other two
bases. 
Briefly put the matter basis is related to the flavor basis by
the usual mixing matrix, but with the the mixing angles
modified. 

The $S$-matrix is found by solving the Schr\"odinger equation 
\begin{equation} \imath \frac{dS}{dx} = H S \end{equation}     
where $H$ is the Hamiltonian. For neutrino propagation in
supernova the Hamiltonian is composed of three parts: the vacuum
$H_V$, the MSW contribution $H_{MSW}$ and the neutrino
self-interaction $H_{\nu\nu}$. 

The vacuum Hamiltonian $H_V$ in the mass basis is diagonal and
parameterized by two mass squared differences $\delta m_{ij}^2 =
m_i^2 - m_j^2$ and the neutrino energy $E$. The vacuum
Hamiltonian in the flavor basis is related to the mass basis by
the Maki-Nakagawa-Sakata-Pontecorvo
\cite{Maki:1962mu,Nakamura:2010zzi} unitary matrix parameterized
by the three mixing angles, $\theta_{12}$, $\theta_{13}$ and
$\theta_{23}$, a CP phase and two Majorana phases. The Majorana
phases have no effect upon the evolution
\cite{1987NuPhB.282..589L,2012JPhG...39c5201G} and will be 
ignored.  In our calculations we will employ the following 
values for the mass splittings and mixing angles: 
$\theta_{12} = 34.4^\circ $,
$\theta_{13} = 9^\circ$, $\theta_{23} = 45^\circ$, $\delta
m_{21}^2 = 7.59\times 10^{-5}$ eV$^2$ and $\delta m_{32}^2 =
2.43\times 10^{-3}$ eV$^2$, and we take $\delta$, the 
CP-violating phase, to be zero.

\subsection{The MSW potential and the density profiles}
The MSW potential describes the effect of the background matter
upon the neutrino and this contribution is diagonal in the
flavor basis.  In this paper we shall only consider the effect
of matter upon the electron neutrino and anti-neutrino and
ignore the small $\mu\tau$ potential which, in the standard
model, is a factor of $\sim 10^{-5}$ smaller.
Consequently only the $e,e$ component of $H_{MSW}$ is non-zero
and is equal to $\sqrt{2} G_F n_e({\bf r})$ where $G_F$ is the
Fermi constant and $n_e({\bf r})$ the electron density.  The
effect of matter upon the anti-neutrinos is opposite that of the
neutrinos i.e. $\bar{H}_{MSW} = -H_{MSW}$.  The electron density
is calculated from matter density profiles generated  by the
Basel simulation group \cite{Fischer:2009}.  A thorough
explanation of the simulations can be found in
\cite{Fischer:2009}. We therefore refrain from going into
details about them here, and merely give the most basic
information on each progenitor in Table~\ref{tab:datachart}.

The density profiles come from three one-dimensional numerical
simulations of progenitors with masses of \Mei, \Mten{} and
\Met. The simulations ran to a post bounce (pb) time of 4.5~s,
10.5~s and 21.8~s respectively.  Not every snapshot from the
simulations are used. We sample the profiles at 1~second
intervals after bounce for a total of 5, 11 and 11 profiles for
the \Mei, \Mten{} and the \Met{} model respectively (see
Table~\ref{tab:times}).

\begin{figure}
\includegraphics[width=\linewidth]{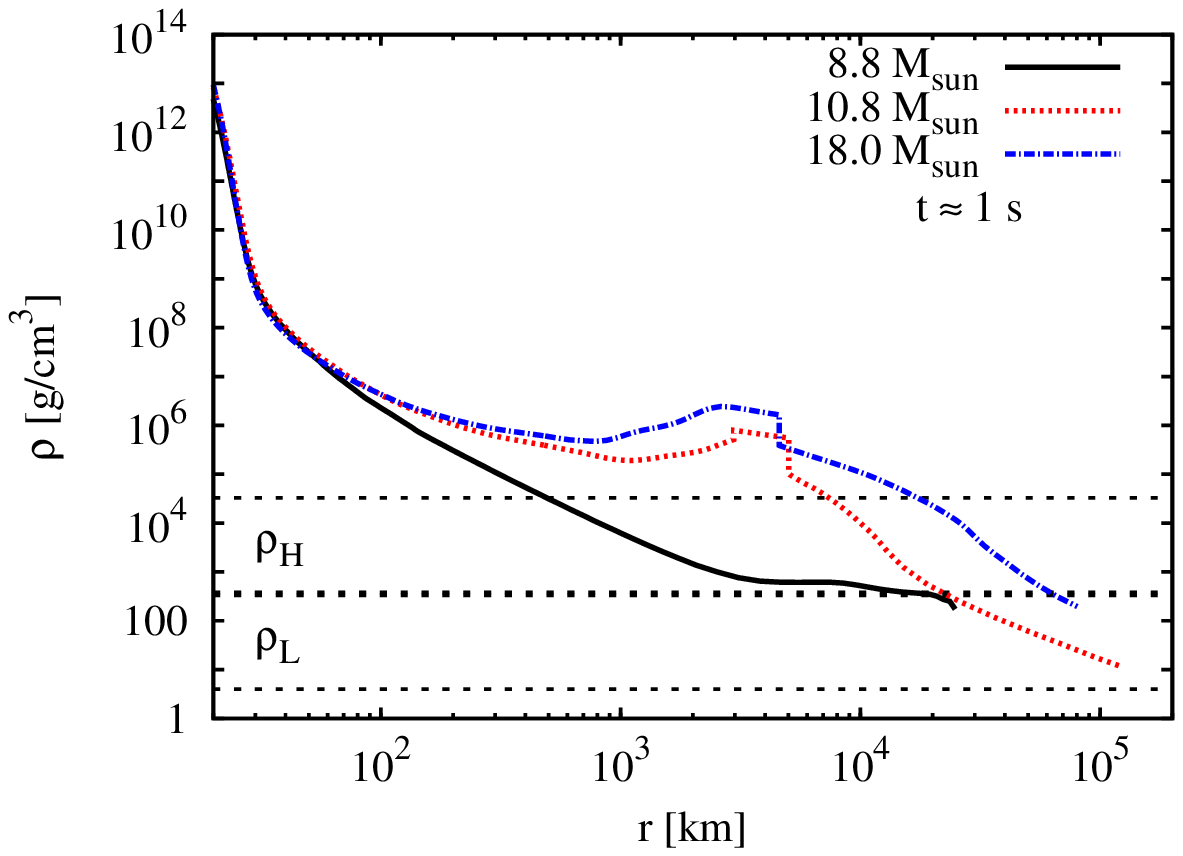}
\includegraphics[width=\linewidth]{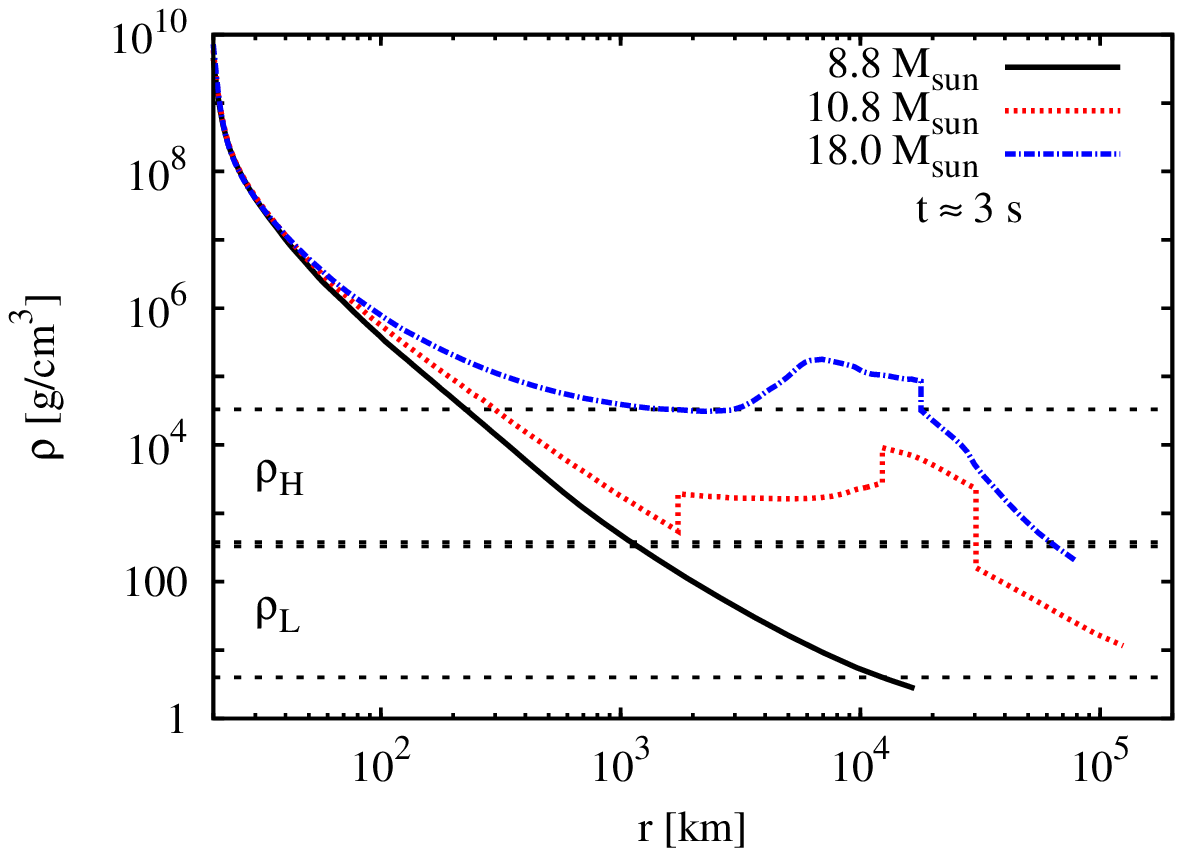}
\caption{\label{fig:rhos} (color online).  Density profiles at
$\sim$1~s ({\it top}) and $\sim$3~s ({\it bottom}) for our three
progenitor models: 8.8~$M_\odot$ (black solid line),
10.8~$M_\odot$ (red dashed line) and 18.0~$M_\odot$ (blue
dot-dashed line).  The horizontal gray dashed lines encompass
the MSW resonant densities for neutrinos with energies in the
range 1--100~MeV.  The upper band corresponds to the H resonance
and the lower band to the L resonance.} 
\end{figure}

\begin{figure}
\includegraphics[width=\linewidth]{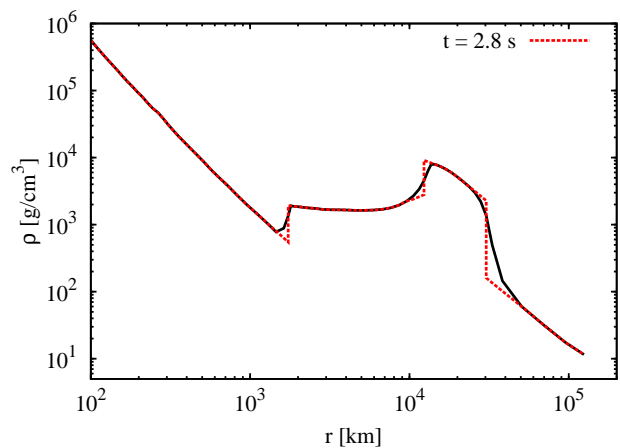}
\caption{\label{fig:steepenedRho} (color online).  Initial
density profile (black solid line) and steepened version (red
dashed). The profile is from the \Mten{} model at 2.8~s.}
\end{figure}

\begin{figure*}
\includegraphics[width=.49\linewidth]{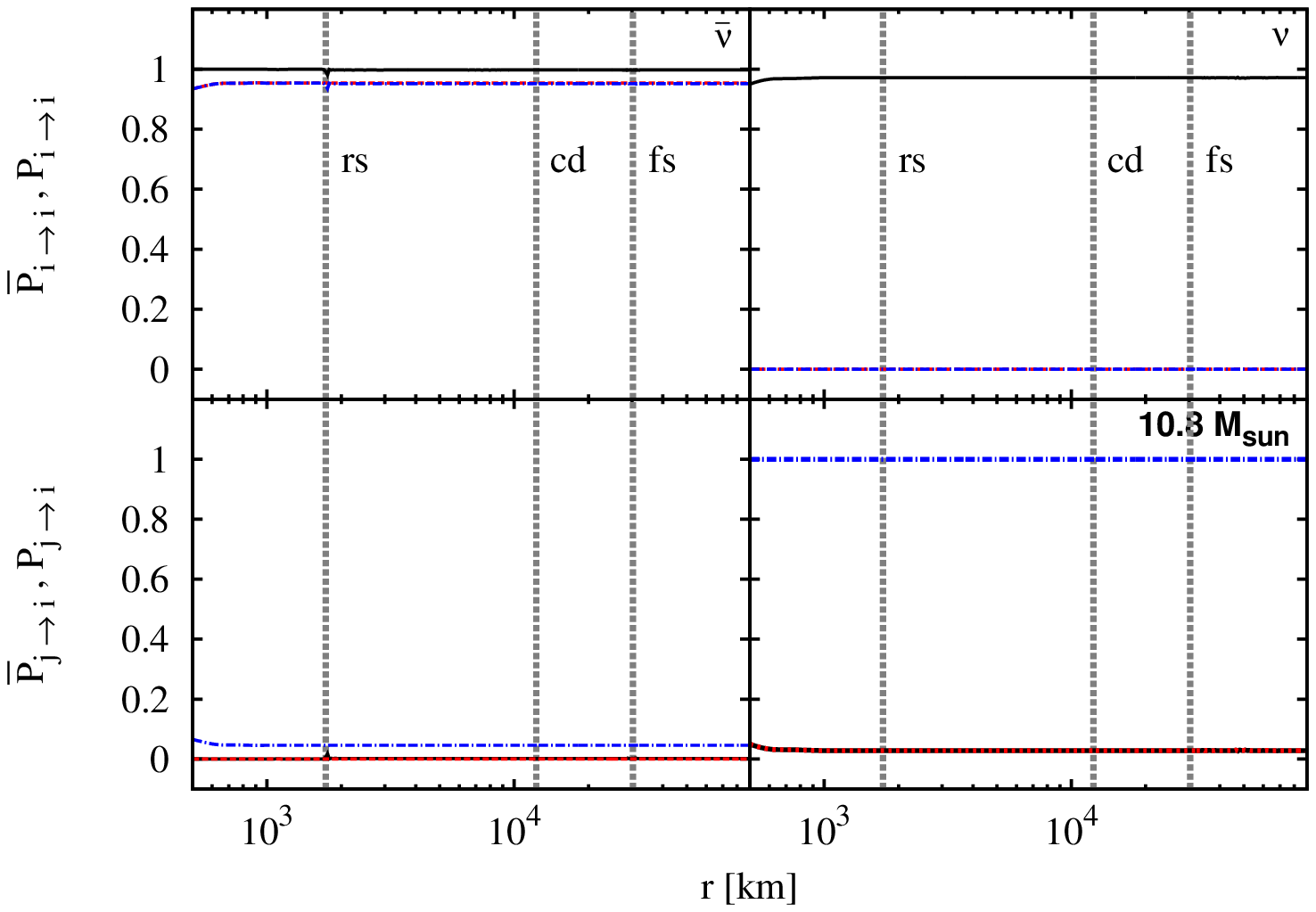}
\includegraphics[width=.49\linewidth]{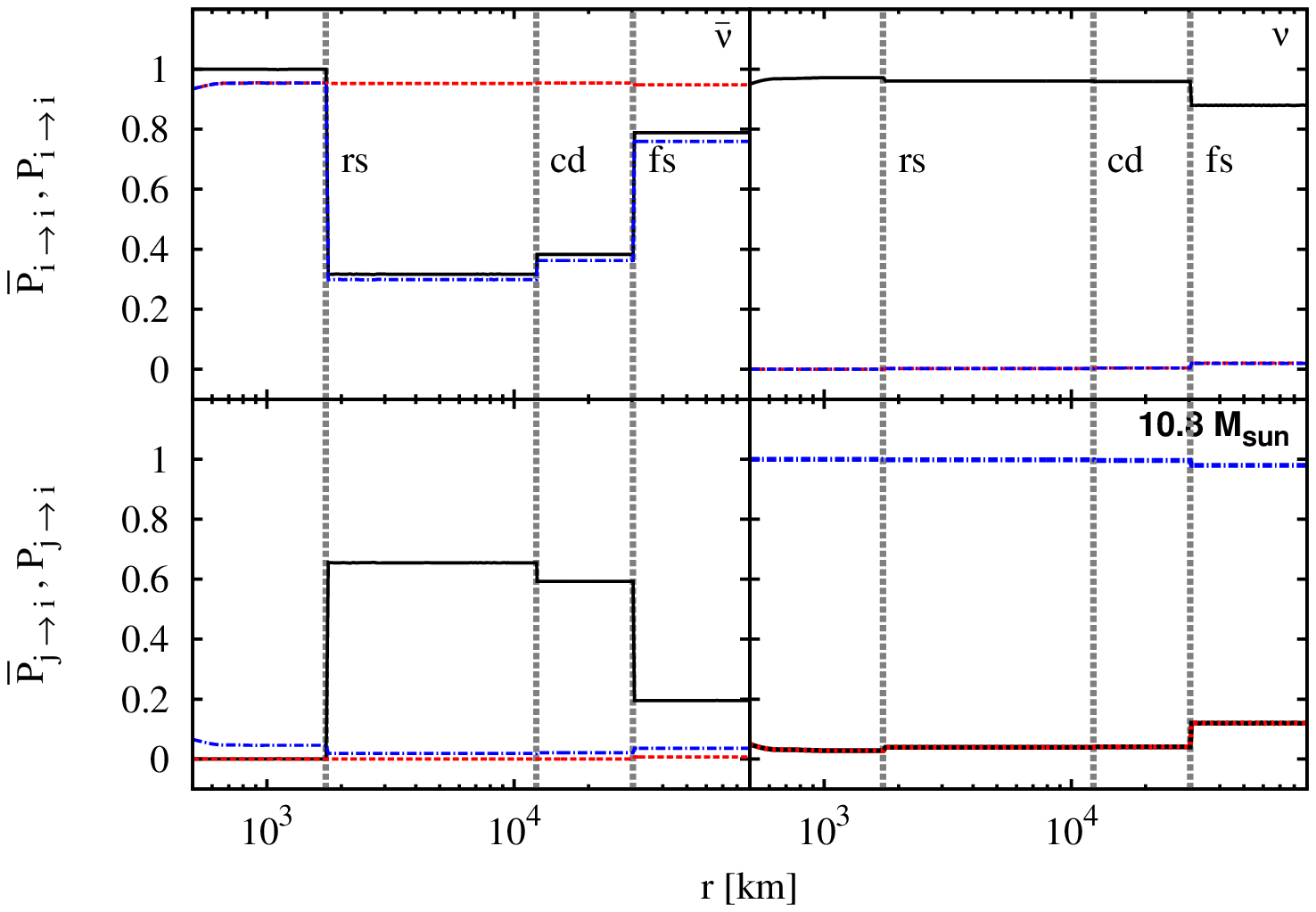}
\caption{\label{fig:probs_stpVSnon} (color online).  
Matter state survival and transition probabilities 
for $\bar\nu$ ({\it left} panels of each quartet, $\bar P$) 
and $\nu$ ({\it right} panels of each quartet, $P$) for a calculation 
from the PNS surface to the end of the density profile at 2.8~s pb
from the \Mten{} model.
Matter survival probabilities $P_{\nu_i \rightarrow \nu_i}$ are
shown in the upper panels of each quartet, and transition
probabilities $P_{\nu_i \rightarrow \nu_j}$ are shown in the
bottom ones. The solid black lines indicate the probabilities
for ending in matter state 1 (or anti-matter state \bone{} in
the case of anti-neutrinos), the red dashed lines give the
probabilities of ending in matter state 2 (\btwo) and finally
the blue dot-dashed lines give the probabilities for ending in
matter state 3 (\bthree) at the end of the calculation domain.
See the text for more explanation.
The left quartets correspond to the original density profile,
the right quartets to the steepened density profile, both in the IH.
The vertical gray dashed lines mark the positions of the reverse
shock (rs), the contact discontinuity (cd) and the forward shock
(fs).}
\end{figure*}

\begin{table}[b!] 
\caption{\label{tab:times} Post bounce times
for the investigated density profiles of our three progenitor
models.} 
\begin{tabular*}{\linewidth}{@{\extracolsep{\fill}}l r r r } 
\toprule 
& \Mei{} & \Mten{} & \Met{} \\ Post bounce time [s] & & \\ 
\colrule 
Bounce  & 0.000 & 0.000 & 0.000 \\ 1       &
1.006 & 0.815 & 0.980 \\ 2       & 2.006 & 1.814 & 1.982 \\ 3
& 3.005 & 2.816 & 3.007 \\ 4       & 3.992 & 3.811 & 4.005 \\
Last    & 4.491 &       &       \\ 5       &       & 4.829 &
5.000 \\ 6       &       & 5.809 & 5.979 \\ 7       &       &
6.807 & 7.024 \\ 8       &       & 7.813 & 7.996 \\ 9       &
& 8.814 & 8.996 \\ 10      &       & 9.815 & 9.985 \\ Last    &
& 10.545& 10.985 \\ 
\botrule 
\end{tabular*} 
\end{table}

In Fig.~\ref{fig:rhos} we show the 6 density profiles that will
be the main focus of this paper. The top panel shows the 1
second profiles for all three progenitors; a solid black line
for the \Mei{} progenitor, a dashed red line for the \Mten{}
model and in dot-dashed blue the \Met{} model. The bottom panel
similarly shows the density profiles at 3 s pb, which is well
into the cooling phase.
In the profiles for the \Met{} model (blue dot-dashed line) a
forward shock is present at both 1 and 3 seconds. The same is
the case with the \Mten{} model (red dashed line). Here
additionally though, the contact discontinuity has developed,
and in the 3 second profile also the reverse shock has
materialized. None of these features are present in the \Mei{}
model profiles at the times we use.

We have had to modify slightly the profiles plotted in
Fig.~\ref{fig:rhos}. The original density profiles did not have
sufficiently steep shocks, which is a known complication due to
insufficient radial resolution in numerical simulations. 
We therefore steepened by hand both shocks and the contact
discontinuity into actual discontinuous jumps. 
In Fig.~\ref{fig:steepenedRho} we show an original density profile
(solid black line) and the steepened version of the same profile
(red dashed line).

To illustrate precisely how crucial the steepness of the 
density profile is with the currently favored large value of
$\theta_{13}$, 
we show in Fig.~\ref{fig:probs_stpVSnon} the matter state
probabilities for the original density profile at 2.8~s for
the \Mten{} progenitor and for the steepened version. The
probabilities are for a calculation with a 20~MeV 
(anti-)neutrino propagating from the PNS surface to the
end of the density profile without turbulence. We clearly see
with the profile that was not steepened (left quartet) that
nothing happens to the probabilities as the neutrino pass the 
shocks or the contact discontinuity because all of them are too 
adiabatic. On the other hand as the neutrino traverses the
steepened profile (right quartet) we see how the diabatic 
passage at the shocks and the contact discontinuity leads to
mixing of the neutrino states.
In the left quartet we see that the divergence from unit
survival probability has begun before we reach the radius of the
reverse shock, and it can therefore be attributed to collective
effects. As the neutrino passes through features in the density
profile nothing new happens in the probabilities. 
In the right quartet we see the same small divergence from unit 
survival probability at low r values, and once again this is 
caused by the collective neutrino interaction (this will be
demonstrated further in Sect.~\ref{sec:3s_results}).
As the radius of the reverse shock is reached we see as expected
in the IH the enhanced mixing of anti-neutrino states \bone{} and 
\bthree{} caused by the MSW H resonance. Consequently the survival
probability of states \bone{} (black solid line) and \bthree{}
(dot-dashed blue line) drops from unity as
they are converted into each other. When the forward shock
is traversed the resonant enhancement reverts the previous
mixing and the survival probability returns to almost unity.
At the forward shock we also observe a decrease in the survival 
probability of the neutrino state 1, which is caused by the MSW 
L resonance (This will be discussed further in
Sect.~\ref{sec:time_evolv}.)

With the plethora of differences between the survival
probabilities of the steepened and the un-steepend density
profiles correct calculations of the neutrino flavor evolution
obviously requires steep density profiles to ensure the
diabatic resonance crossing we know should take place.

\subsection{Turbulence}		\label{sec:theo_turb} 
The turbulence we include in the calculations will enter through 
the MSW potential. Our approach follows that 
of \cite{Kneller:2010,Kneller:2007kg}
and many others whereby we multiply the smooth, turbulence free,
density profiles from the one-dimensional hydrodynamical
supernova simulations, which we call $\langle n_e \rangle$, with
a Gaussian random field $1+F(r)$. The turbulence is placed into
the three models slightly differently.  For the \Mten{} model we
place one turbulence field between the forward shock and contact
discontinuity and then another between the contact discontinuity
and the reverse shock if the reverse shock is present.  For the
\Met{} model we let one turbulence field cover the entire region
behind the forward shock since neither a contact discontinuity
nor a reverse shock develops in this model. Finally, for the
\Mei{} model we again insert the turbulence behind the shock but
the shock in this model quickly runs out of the simulation
domain and thereafter we allow the turbulence to cover all of
the profile.  At no point is turbulence placed into the profile
ahead of the forward shock. The amplitudes of the turbulence
seen in the simulations by Meakin \& Arnett \cite{Meakin:2006uh}
are very small, typically between $\sim 10^{-5}$ to $\sim
10^{-3}$,  and we have verified that this is too small to affect
the neutrinos.  

The Gaussian field is modelled as a Fourier series with a
normalized power spectrum $E(k)$ multiplied by two damping
factors to suppress fluctuations close to the shocks and contact
discontinuity and prevent discontinuities. Concretely, we use 
\begin{equation}\label{eq:F} 
\begin{split} F(r) &
=C_{\star}\,\tanh\left(\frac{r-r_r}{\lambda}\right)\,
\tanh\left(\frac{r_s-r}{\lambda}\right) \\ &
\times\sum_{n=1}^{N_k}\,\sqrt{V_{n}}\left\{ A_{n}
\cos\left(k_{n}\,r\right) + B_{n} \sin\left(k_{n}r\right)
\right\}.  
\end{split}
\end{equation}
for radii between $r_r \leq r \leq r_s$ and zero outside this
range.  The damping scale $\lambda$ is set to $\lambda =
100\;{\rm km}$ and the parameter $C_{\star}$ sets the amplitude.
Each of the $N_{k}$ co-efficients in the sets $\{A\}$ and
$\{B\}$ are independent standard Gaussian random variates with
zero mean thus ensuring a vanishing expectation value of $F$. 
\begin{figure}[t!]
\includegraphics[width=\linewidth]{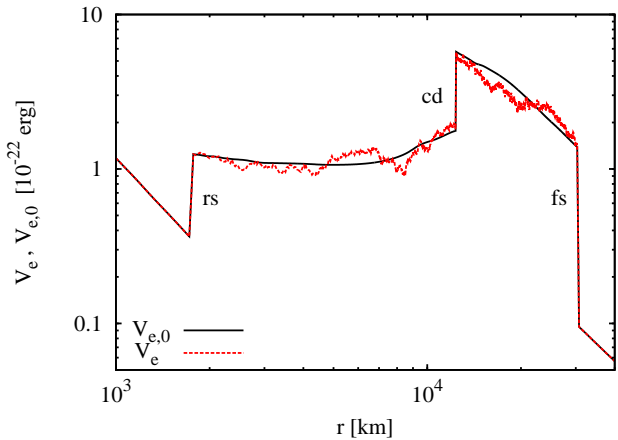}
\caption{\label{fig:turbRho} (color online).  Close up of the
unperturbed potential $V_{e,0}$ (black solid line) at 2.8~s for
our 10.8~$M_\odot$ model and the potential with turbulence
added, $V_e$ (red dashed line).  The two areas with added
turbulence are: between the forward shock (fs) and the contact
discontinuity (cd), and between the contact discontinuity and
the reverse shock (rs).} 
\end{figure}
Finally, the parameters $V_{n}$ are k-space volume
co-efficients.  To generate the $N_k$ $k$'s, $V$'s, $A$'s and
$B$'s for a realization of $F$ we use `Variant C' of the
Randomization Method the reader can find in Kramer,
Kurbanmuradov, \& Sabelfeld \cite{2007JCoPh.226..897K}.  The
algorithm behind this randomization method is to partition
k-space into $N_{k}$ regions and from each select a random
wavenumber using the power-spectrum, $E(k)$, as a probability
distribution.  The volume parameters $V_{n}$ are the integrals
of the power spectrum for each partition.  Variant C of the
Randomization Method divides the k-space so that the number of
partitions per decade is uniform over $N_d$ decades starting
from a cutoff scale $k_{\star}$. We shall use a wavenumber
cutoff $k_{\star}$ set to $k_{\star}=\pi/\Delta r$. 
Where $\Delta r$ is the distance between the discontinuities 
under consideration. This
logarithmic distribution of the modes is designed so that the
quality of the realizations is uniform over the range of
length scales considered, i.e.~it is scale invariant. This
feature is important because the oscillation wavelength of the
neutrinos is constantly changing as the density evolves. All the
results in this paper shall adopt $N_d=4$ and $N_k=40$ but to
reassure the reader we have checked several of our results with
the combination $N_d=5$ and $N_k=50$. 

In Fig.~\ref{fig:turbRho} we show an example from the \Mten{}
model where 10\% turbulence has been added to the potential at
2.8~s pb. The unperturbed potential $V_{e,0}$ is shown in
solid black and the turbulent potential $V_e$ is shown in red
dashed. We have furthermore marked the reverse shock (rs), the 
forward shock (fs) and the contact discontinuity (cd), which are 
all present in this particular profile.

\begin{figure*}
\includegraphics[width=\textwidth]{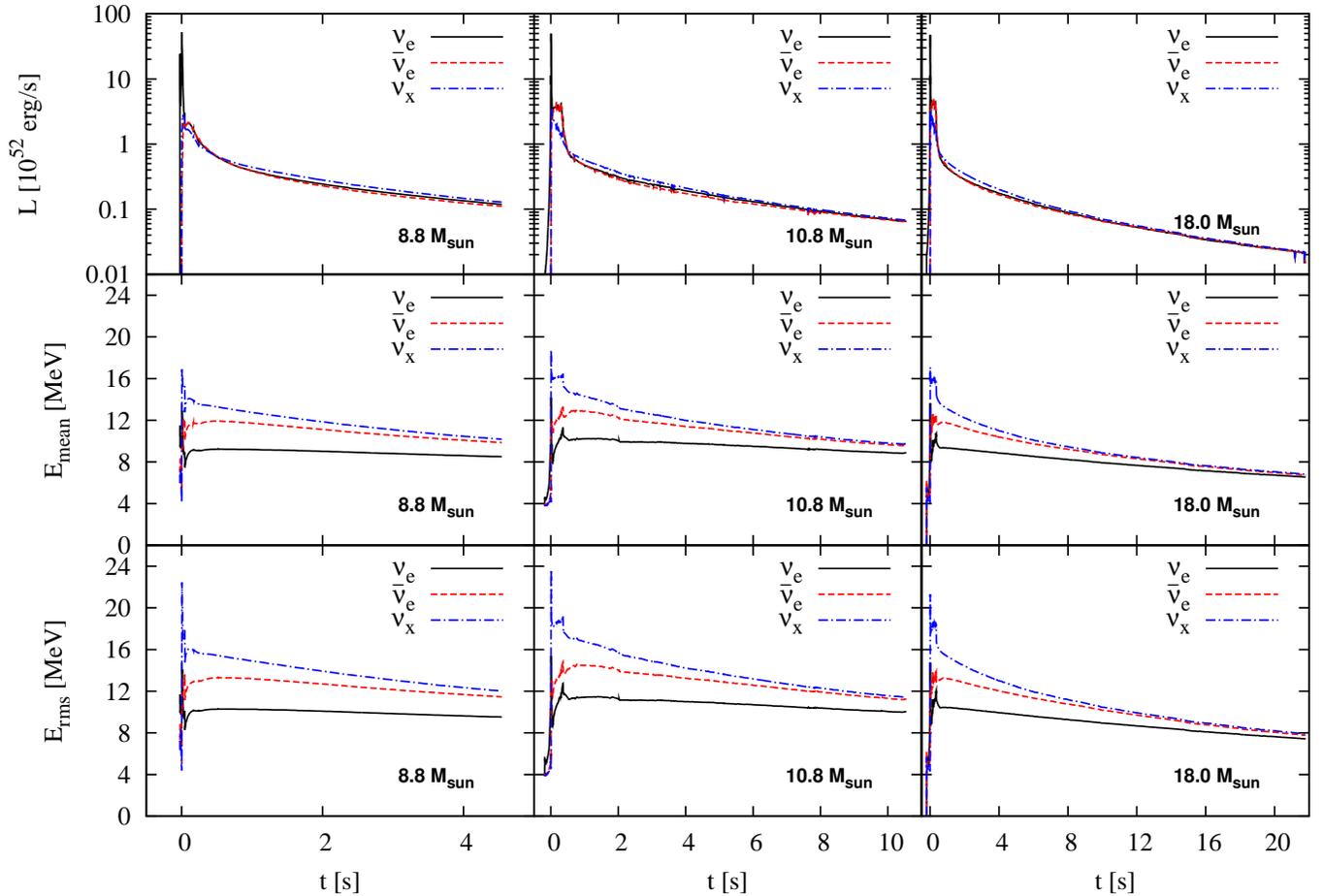}
\caption{\label{fig:LandE} (color online). Luminosity ({\it
top}), mean energy ({\it middle}) and rms energy ({\it bottom})
for the three models; 8.8~$M_\odot$ ({\it left}), 10.8~$M_\odot$
({\it middle}) and 18.0~$M_\odot$ ({\it right}).  Black solid
lines designate electron neutrinos $\nu_e$.  Red dashed lines
show electron anti-neutrinos $\bar\nu_e$.  Blue dot-dashed lines
show non-electron flavor neutrinos $\nu_x$.  The non-electron
anti-neutrino, $\bar\nu_x$, quantities are almost identical to
the ones of their neutrino counterparts.} 
\end{figure*}

\subsection{Neutrino self-interactions}
Finally, the neutrino density in the innermost regions of the
supernova is so high that neutrinos become a background to
themselves leading to a non-linear self-coupling. The extended
source of the neutrinos means that at some given radial position
$r$ above the neutrinosphere one will find neutrinos propagating
in a wide swath of directions relative to the radial direction.
In principle the evolution of each has to be calculated
simultaneously but often the reader will observe the use of the
single angle approximation which treats all outward
directions as being equivalent. At the present time the validity
of the single-angle approximation is unclear.  Duan \& Friedland
\cite{Duan:2011} compared single angle and ``multi-angle''
calculations and observed a rapid onset of neutrino
transformation due to collective effects in their single angle
calculations that did not occur in multi-angle calculations. At
the same time it was shown in \cite{Sarikas:2012ad}
that single angle
calculations can match multi-angle calculations.  It is not
our intention to wade into this debate here. We shall adopt the
single angle approximation because it makes the numerous
calculations we must undertake feasible and its results are
``realistic'' in the sense that they give the same (or similar)
features as the multi-angle calculations. 

The single-angle self-interaction Hamiltonian in the flavor
basis is of the form 
\begin{widetext} \begin{equation} H_{\nu\nu} =
\frac{\sqrt{2}\,G_{F}}{2\,\pi\,R_{\nu}^{2}}\,C(r/R_{\nu})\,\left\{
\int dE_{\nu}
S(r,E_{\nu})\rho(r_{0},E_{\nu})S^{\dagger}(r,E_{\nu}) - \int
dE_{\bar{\nu}} \left(
\bar{S}(r,E_{\bar{\nu}})\bar{\rho}(r_{0},E_{\bar{\nu}})
\bar{S}^{\dagger}(r,E_{\bar{\nu}})\right)^{\star} \right\}
\end{equation} \end{widetext}
where $\rho(r_{0},E_{\nu})$ is the energy dependent density
matrix for the neutrinos at the initial point $r_{0}$ and
similarly $\bar{\rho}(r_{0},E_{\bar{\nu}})$ is for the
anti-neutrinos.  We shall adopt $r_0 = 70\;{\rm km}$ and we have
verified that our results do not depend upon the initial
starting radius. The radius of the neutrinosphere is $R_{\nu}$
and the function $C(r)$ is commonly known as the geometric
factor. For this paper we shall adopt the form given in
\cite{Dasgupta:2008} \begin{displaymath} C(r) = 4 \;
\frac{R_\nu^2}{r^2} \; \left[ \frac{1-\sqrt{1 - R_\nu^2 /
r^2}}{R_\nu^2 / r^2} \right]^2 
- \frac{R_\nu^2}{r^2} \quad.  \end{displaymath}

The neutrino spectral information is buried inside the two
density matrices $\rho(r_{0},E_{\nu})$ and
$\bar{\rho}(r_{0},E_{\bar{\nu}})$.  To model the neutrino
spectra we use the `pinched' spectra found in Keil \emph{et al.}
\cite{Keil:2003}, and the luminosities and mean energies
supplied from the Basel group \cite{Fischer:2012PM}.  The pinch
parameters are computed from the ratio of rms to mean energy.
Luminosities and energies are shown in Fig.~\ref{fig:LandE}.
During the accretion phase (up to $\sim$ 0.5 s) the electron
neutrino luminosities, $L_{\nu_e}$, dominates, but during the
cooling phase the non-electron neutrino luminosities,
$L_{\nu_x}$, are largest. At very late times all luminosities
become very similar.  The mean energies show a distinct
hierarchy at early times, $E_{\nu_x,\bar\nu_x} > E_{\bar\nu_e} >
E_{\nu_e}$, while at later times the $E_{\bar\nu_e}$ and
$E_{\nu_x,\bar\nu_x}$ become much more similar and their
dominance over $E_{\nu_e}$ becomes smaller.

As we mentioned in the introduction, the accuracy of simulations
is continously improving, particularly with respect to 
the neutrino opacities.
We acknowledge these improvements in more recent simulations but
we find that the disagreement on the exact size and impact is
still large, and that the difference due to including the 
updated cross sections or not is on the order of changing the 
equation of state of the progenitor, as can be seen from  
O'Connor \& Ott \cite{O'Connor:2012am}. 
Furthermore, using luminosities, energies and density profiles
from the same simulation allows us to carry out our calculations 
selfconsistently.
Therefore we will use the older simulations \cite{Fischer:2009} 
covering the longer post bounce times.

\section{Results}	\label{sec:results}
In order to pick apart a neutrino burst signal and be confident
we have identified features caused by the neutrino
self-interactions from those generated by the MSW effect or the
turbulence we must first understand the features each engenders
separately before seeing how they combine.  But with so many
simulation snapshots, three different progenitors and two
hierarchies we find a virtual zoo of phenomena.  In order for
the reader to understand our later results we begin by walking
through our results for a single case.     
To keep things simple we chose the profiles of each simulation at 
3~seconds post bounce when the shock has propagated far enough
into the star that it begins to affect the MSW resonance. We
then undertake four different calculations for this simulation
snapshot of each model: 
\begin{itemize} 
\item an `inner region' calculation from the proto-neutron star 
up to 1000~km where the collective effects are expected to be 
dominant.
\item an `outer region' calculation that covers from 1000~km to 
the end of the profiles where the MSW effect dominate the flavor 
evolution.  
\item a `turbulence free' calculation which covers the entire 
profile but with no added turbulence,
\item a `turbulent' calculation which again covers the entire
profile but now various amounts of turbulence is added according
to the prescription given previously.  
\end{itemize}

\begin{figure}[t!]
\includegraphics[width=\linewidth]{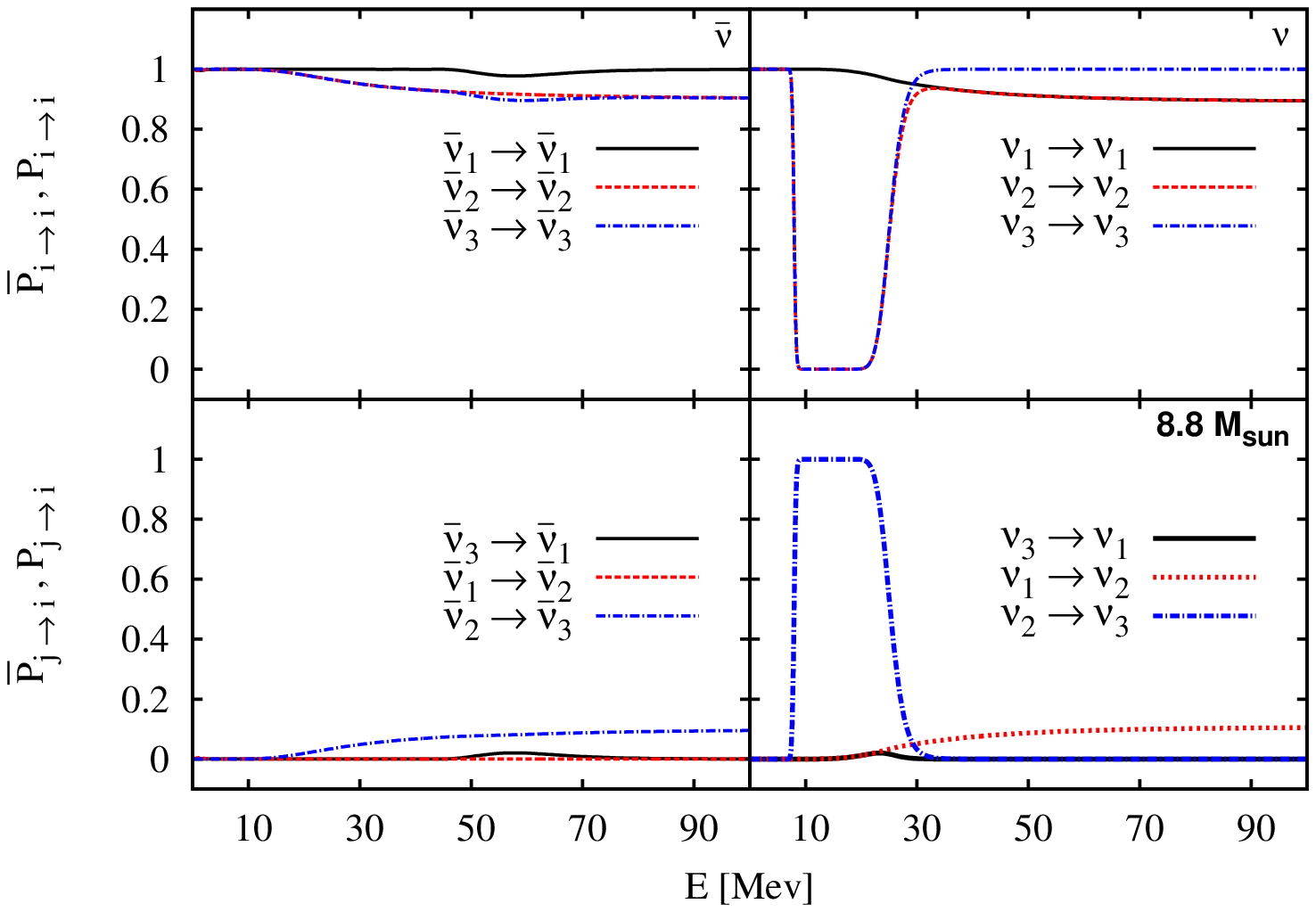}
\includegraphics[width=\linewidth]{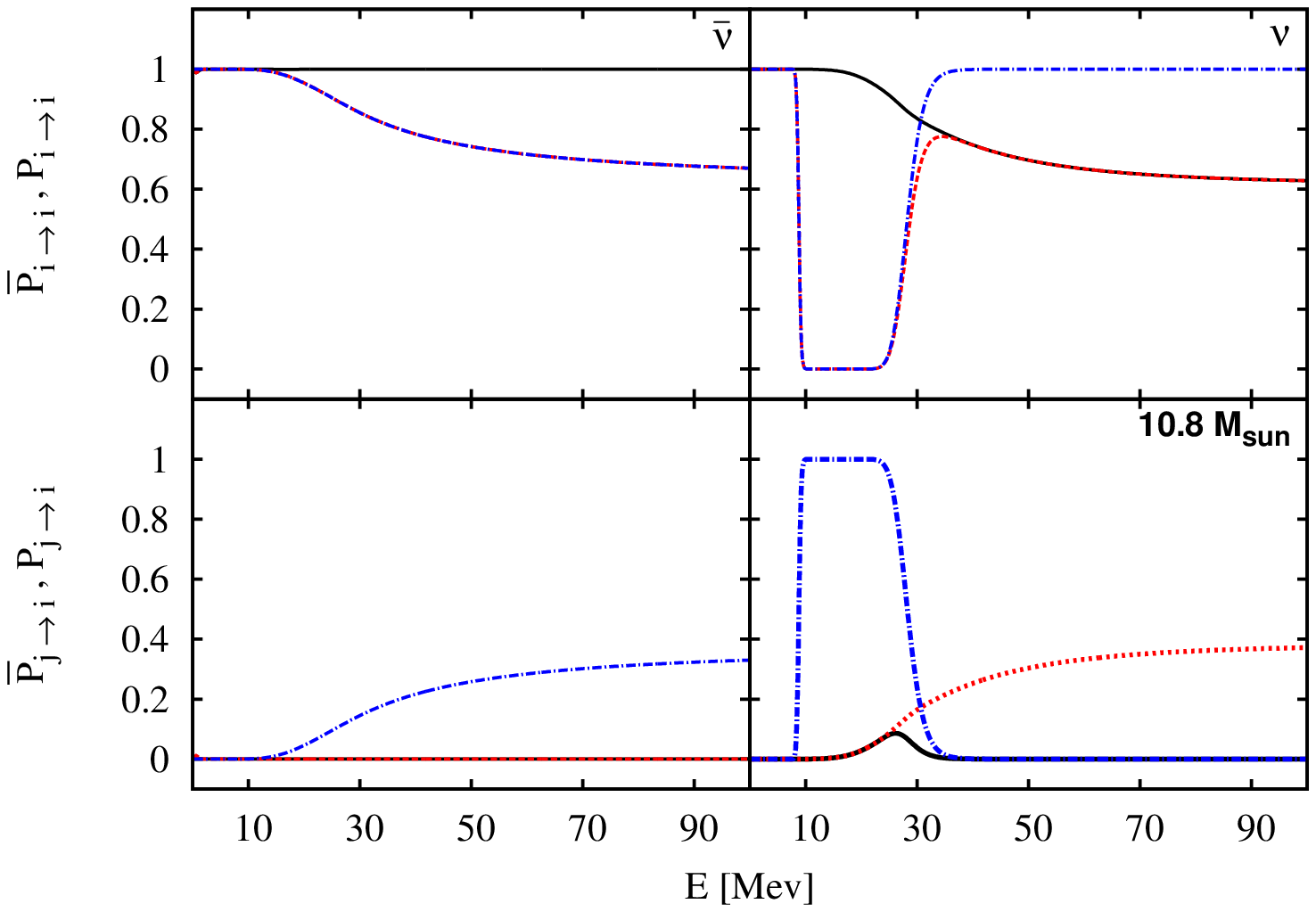}
\includegraphics[width=\linewidth]{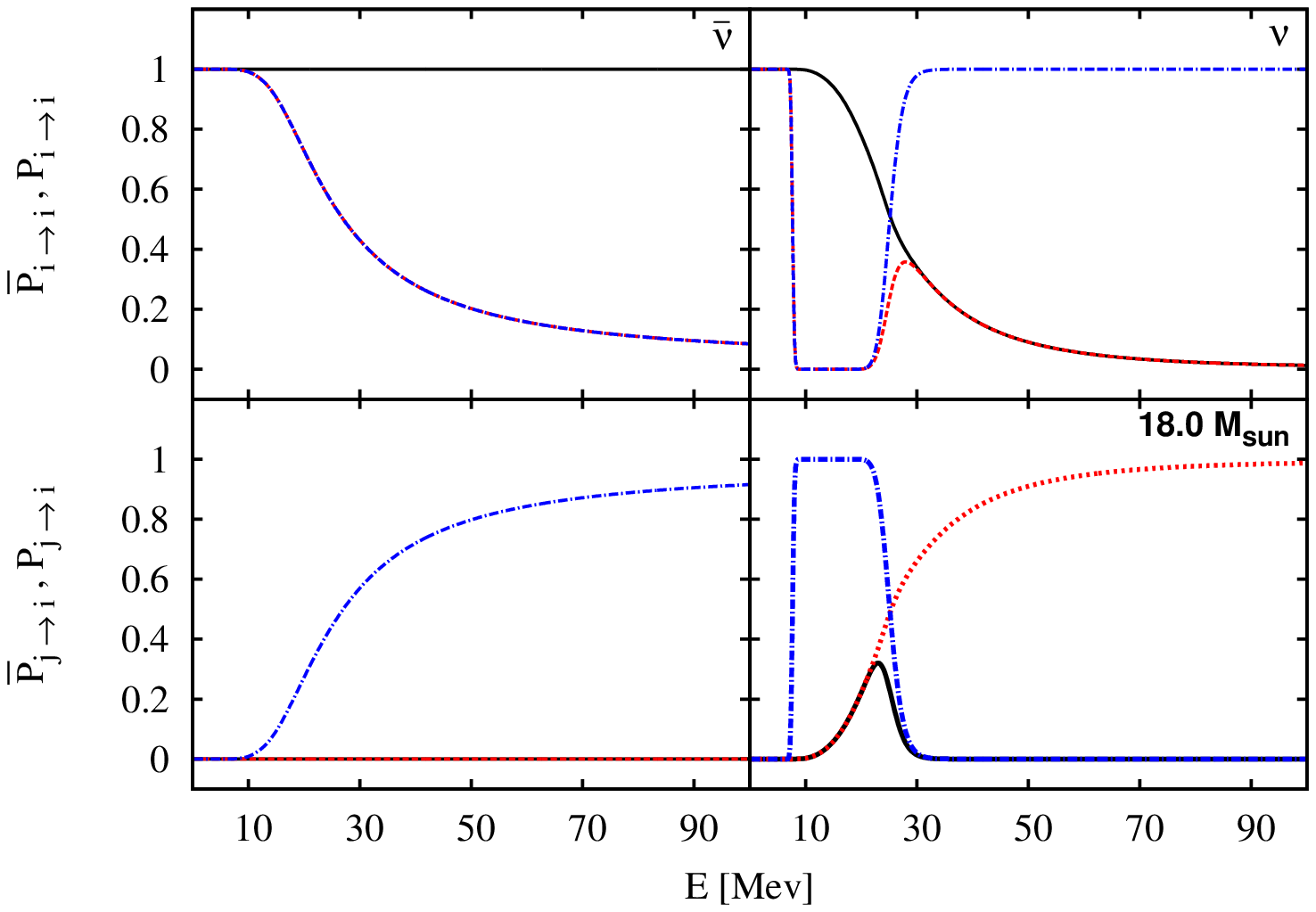}
\caption{\label{fig:coIH_3sec} (color online).  Matter state
survival and transition probabilities at 3 s pb for $\bar\nu$
({\it left panels}, $\bar P$) and $\nu$ ({\it right panels},
$P$) in the Inverted Hierarchy for the inner region. The top
four panels belong to the \Mei{} model, the middle four to the
\Mten{} progenitor and the bottom four to the \Met{} progenitor.
Matter survival probabilities $P_{\nu_i \rightarrow \nu_i}$ are
shown in the upper panels of each quartet, and transition
probabilities $P_{\nu_i \rightarrow \nu_j}$ are shown in the
bottom ones. The solid black lines indicate the probabilities
for ending in matter state 1 (or anti-matter state \bone{} in
the case of anti-neutrinos), the red dashed lines give the
probabilities of ending in matter state 2 (\btwo) and finally
the blue dot-dashed lines give the probabilities for ending in
matter state 3 (\bthree) at the end of the calculation domain.
See the text for more explanation.} 
\end{figure}
\begin{figure}[t!]
\includegraphics[width=\linewidth]{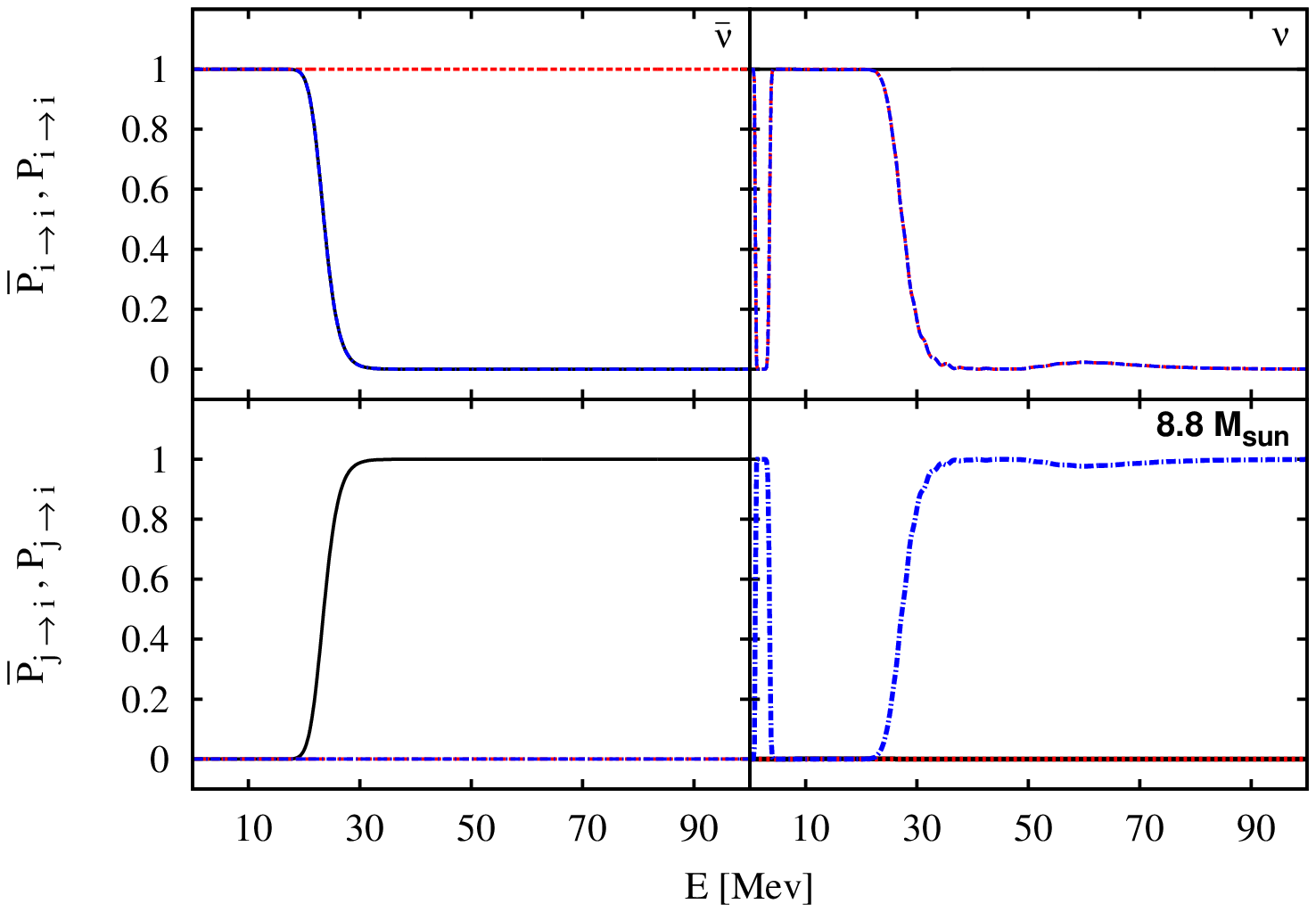}
\includegraphics[width=\linewidth]{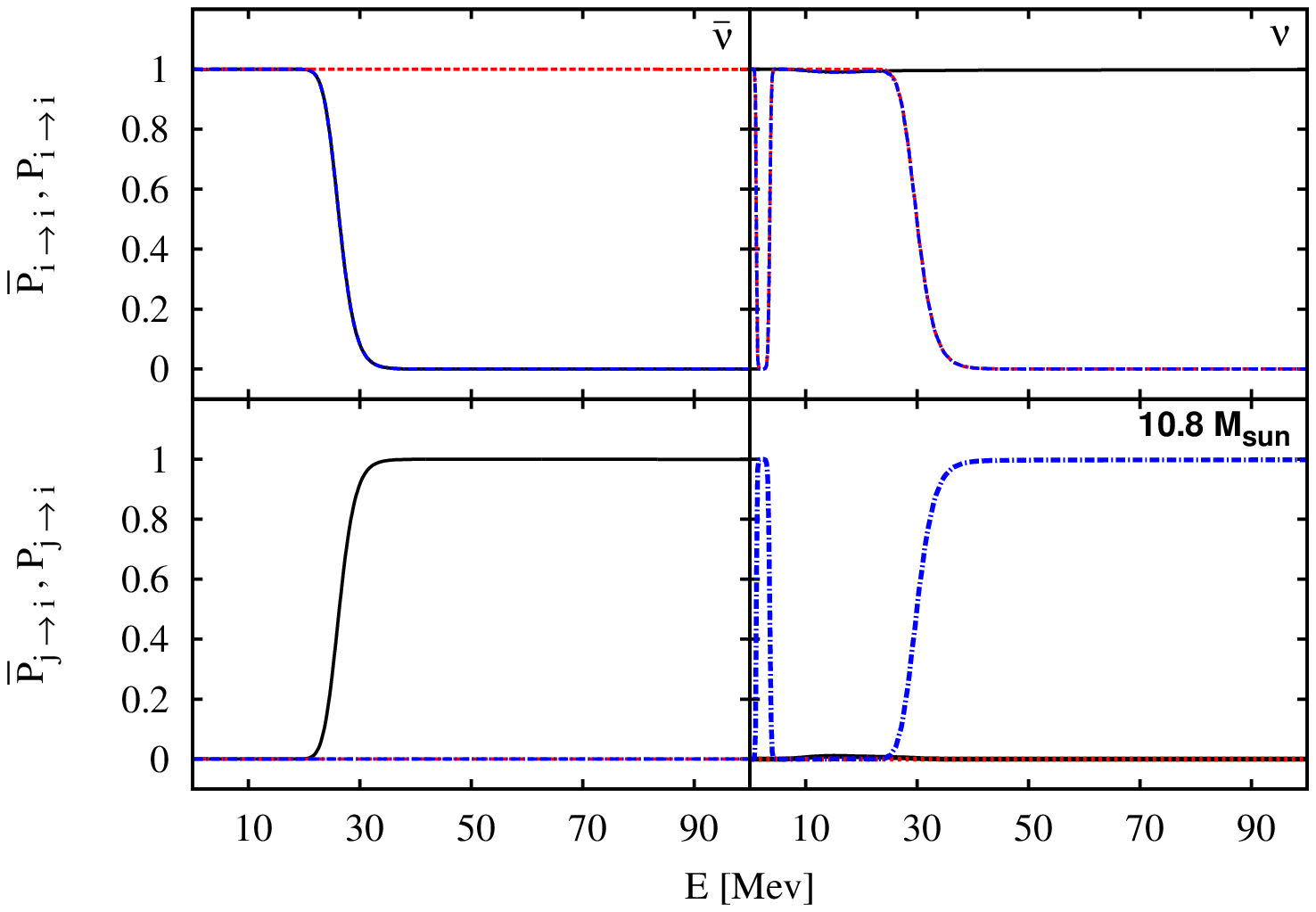}
\includegraphics[width=\linewidth]{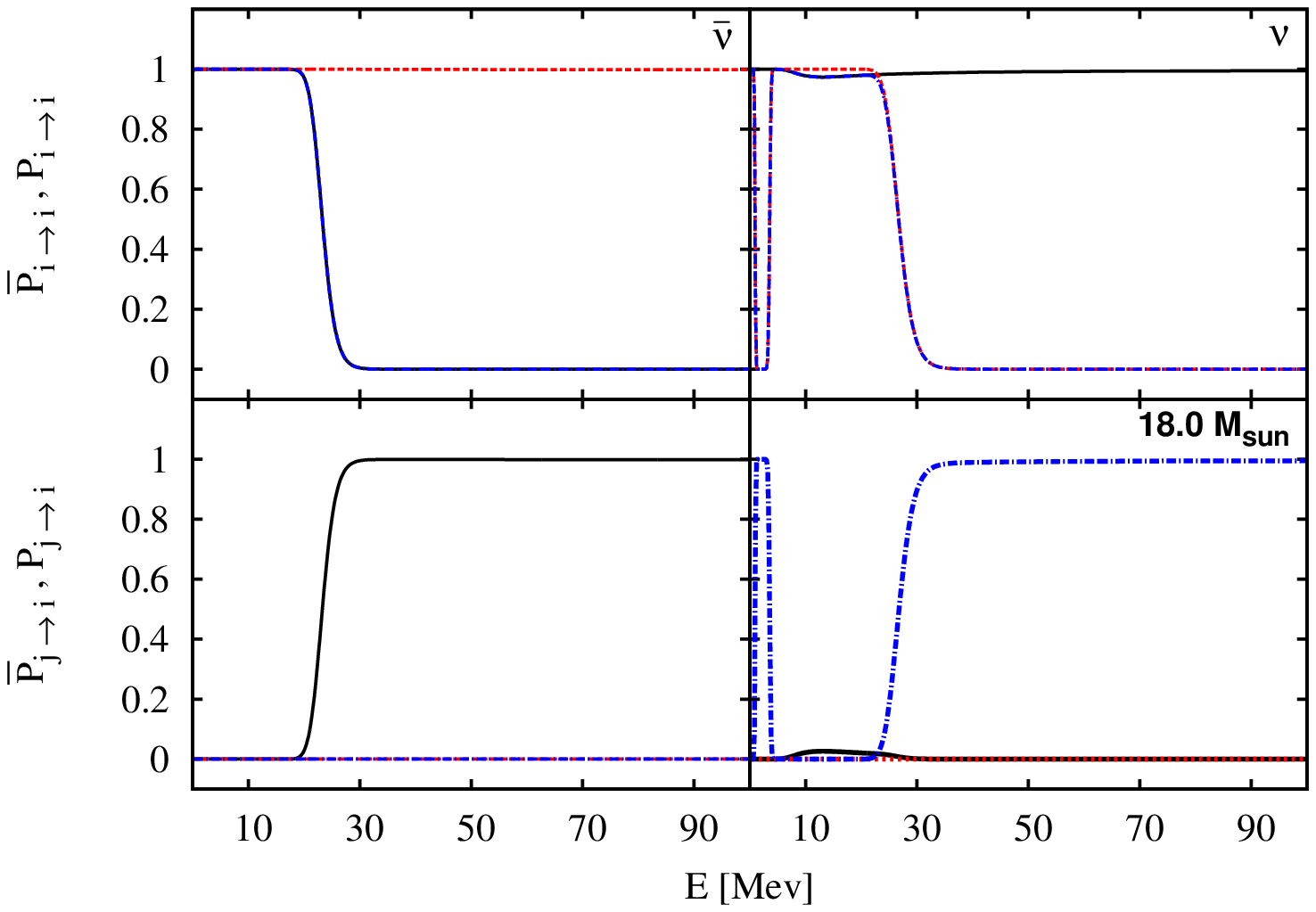}
\caption{\label{fig:coNH_3sec} (color online).   Matter state
survival and transition probabilities at 3 s pb for $\bar\nu$
({\it left panels}, $\bar P$) and $\nu$ ({\it right panels},
$P$) in the Normal Hierarchy for the inner region. The top four
panels belong to the \Mei{} model, the middle four to the
\Mten{} progenitor and the bottom four to the \Met{} progenitor.
Matter survival probabilities $P_{\nu_i \rightarrow \nu_i}$ are
shown in the upper panels of each quartet, and transition
probabilities $P_{\nu_i \rightarrow \nu_j}$ are shown in the
bottom ones. The solid black lines indicate the probabilities
for ending in matter state 1 (or anti-matter state \bone{} in
the case of anti-neutrinos), the red dashed lines give the
probabilities of ending in matter state 2 (\btwo) and finally
the blue dot-dashed lines give the probabilities for ending in
matter state 3 (\bthree) at the end of the calculation domain.}
\end{figure}

\subsection{Profiles at 3~seconds}     \label{sec:3s_results}
\begin{table*}[t!] 
\caption{\label{tab:LandE} Luminosities and
energies for our three progenitors at 1 and 3 s post bounce.}
\begin{tabular*}{\linewidth}{@{\extracolsep{\fill}} l r r r | l r r r } 
\toprule 
At 1 s:& \Mei{} & \Mten{} (0.81 s) & \Met{} &
At 3 s: &\Mei{} & \Mten{} (2.82 s) & \Met{} \\ 
\colrule
$L_{\nu_e}$ [$10^{51}$ erg/s]   & 3.858 & 5.344 & 4.321 & 
& 1.751 & 2.503 & 2.158 \\ 
$L_{\bar\nu_e}$ [$10^{51}$ erg/s] & 3.826 & 5.410 & 4.537 & 
& 1.617 & 2.276 & 2.056 \\ 
$L_{\nu_x}$ [$10^{51}$ erg/s]   & 4.382 & 6.271 & 5.292 &
& 1.969 & 2.863 & 2.539 \\ 
$L_{\bar\nu_x}$ [$10^{51}$ erg/s] & 4.416 & 6.319 & 5.333 &        
& 1.977 & 2.874 & 2.545 \\ 
\hline 
$E_{mean,\nu_e}$ [MeV] & 9.189 & 10.19 & 9.322 &
& 8.804 & 9.890 & 9.007 \\ 
$E_{mean,\bar\nu_e}$ [MeV] & 11.72 & 12.90 & 11.73 &         
& 10.55 & 11.82 & 10.77 \\ 
$E_{mean,\nu_x}$ [MeV] & 12.75 & 14.46 & 13.10 &       
& 11.09 & 12.65 & 11.53 \\ 
$E_{mean,\bar\nu_x}$ [MeV] & 12.84 & 14.57 & 13.20 &        
& 11.13 & 12.69 & 11.57 \\ 
\hline
$E_{rms,\nu_e}$ [MeV] & 10.26 & 11.37 & 10.42 &        
& 9.848 & 11.12 & 10.10 \\ 
$E_{rms,\bar\nu_e}$ [MeV] & 13.20 & 14.48 & 13.20 &        
& 12.15 & 13.64 & 12.38 \\ 
$E_{rms,\nu_x}$ [MeV] & 14.90 & 16.92 & 15.35 &        
& 13.07 & 14.98 & 13.62 \\ 
$E_{rms,\bar\nu_x}$ [MeV] & 15.04 & 17.08 & 15.49 &
& 13.13 & 15.06 & 13.69 \\ 
\botrule 
\end{tabular*}
\end{table*}
\subsubsection{Inner region, 70 -- 1000 km: Collective
dominated}{\label{sec:3s_inner}} The results of our calculations
for inner region, where collective effects dominate, for the
simulation snapshot at 3~s pb are shown in
Fig.~\ref{fig:coIH_3sec} for the Inverted Hierarchy (IH) and in
Fig.~\ref{fig:coNH_3sec} for the Normal Hierarchy (NH).  In each
figure the top four panels are for the \Mei{} progenitor, the
middle four panels are for the \Mten{} progenitor and the bottom
four panels are for the \Met{} progenitor. Then, within each 
quartet of panels, the left two panels show the probabilities 
for the
anti-neutrino matter states and the right two panels show the
neutrino matter state probabilities. The top two panels of a
quartet show the survival probabilities ($P_{\nu_i \rightarrow
\nu_i}$), and the bottom two show selected transition
probabilities ($P_{\nu_i \rightarrow \nu_j}$). The remaining
transition probabilities can be found from probability
conservation.
In the top right (left) panels of each quartet the solid black
line gives the probability that neutrino (anti-neutrino) matter
state 1 ($\bar 1$) goes to matter state 1 ($\bar 1$). In the
lower right (left) panels of each quartet the black line
signifies the probability that neutrino (anti-neutrino) matter
state 3 ($\bar 3$) goes to matter state 1 ($\bar 1$).  Similarly
the red dashed line in the top panels are matter state 2 ($\bar
2$) to 2 ($\bar 2$), and the blue dot-dashed line is matter
state 3 ($\bar 3$) going to 3 ($\bar 3$). In the lower panels
the red dashed line indicates matter state 1 ($\bar 1$) going to
matter state 2 ($\bar 2$), and the blue dot-dashed line
represents matter state 2 ($\bar 2$) going to matter state 3
($\bar 3$). The legend is the same for each quartet in
subsequent figures in this paper which is why it only appears in
the top left quartet of each pair of figures.\\

From Figures~\ref{fig:coIH_3sec} and \ref{fig:coNH_3sec} we see
that in both hierarchies all three models have at least two
spectral splits (also known as `swaps') in the neutrino sector.
Spectral splits are sudden changes in the transition
probabilities as a function of neutrino or anti-neutrino energy
and have been observed in self-interaction calculations starting
from Duan {\it et al.} \cite{Duan:2006an}.  Sets of complete and incomplete
swaps\footnote{We follow the terminology of A.~Friedland (2010) 
\cite{Friedland:2010} and define an incomplete swap to be when 
the survival probability is neither zero nor one, but instead 
decreases gradually with neutrino energy, i.e.\ the swap 
probability increases gradually with neutrino energy.} between 
the neutrino states are very typical of all
self-interaction calculations and the results depend upon
the hierarchy.  In the IH (Fig.~\ref{fig:coIH_3sec}) the splits
are between neutrino matter states 2 and 3 occurring at 7--8~MeV
and 24--28~MeV depending on the progenitor. From the top right
panels of each quartet we see that between these two energies
the survival probabilities of the two states drop to zero.
Looking in the bottom right panels of each quartet we learn that
in this energy range the matter state 2 goes to matter state 3,
and that 3 does not go (significantly) to state 1, thus it must
transform into matter state 2. Therefore, we conclude that a
full swap of matter states 2 and 3 takes place between the two
swap energies in the spectrum.  In the NH
(Fig.~\ref{fig:coNH_3sec}) we observe a third split in the 
neutrinos between matter states 2 and 3, and the energies where 
we find the three splits are 2, 4 and 27--30~MeV .
Finally, for just the \Met{} model, there is an additional soft
split between neutrino matter states 1 and 2 at 26~MeV in the
IH.

The anti-neutrinos display a sharp spectral split between states
\bone{} and \bthree{} for all three progenitors in the NH at
energies of 24--26~MeV depending on the progenitor. In the IH
all three models have an incomplete swap between anti-neutrino 
states \btwo{} and \bthree{} above $\sim$10--15~MeV.

What is remarkable about these figures is how similar they are
even though the progenitors are very different.  The results
shown in Figures~\ref{fig:coIH_3sec} and \ref{fig:coNH_3sec} are
with luminosities and energies that appear to be far from the
``standard'' set of values \cite{Duan:2006an,Schirato:2002tg,
Choubey:2010,Dasgupta:2010}. We find multiple 
\clearpage
splits in both
hierarchies for both neutrinos and anti-neutrinos (depending on
what time we are investigating). The placement of the splits are
not the same as was found in previous investigations
(e.g.~\cite{Dasgupta:2010}) but based on the general
understanding that any specific choice of values will always be
a special case, we believe the 
discrepancies are due to the
slight differences to our luminosity and energy values.  We
give the luminosities, mean and rms energies for our models at
1 and 3~s in table~\ref{tab:LandE} and we emphasize that the strong
hierarchy in the luminosities seen in other simulations
is not found here. Furthermore the energies are
overall lower than in other simulations and the differences
between the electron, anti-electron and non-electron flavor
values are smaller. From the luminosities and mean energies
given in table~\ref{tab:LandE} we can also compute the number 
fluxes of neutrinos 
($\Phi_f = L_{\nu_f} / E_{mean,\nu_f}$)
which turns out to be very similar for all 3 progenitors. This
explains why our results for the inner regions
(Figures~\ref{fig:coIH_3sec} and \ref{fig:coNH_3sec})
are so
relatively similar across our otherwise quite different
progenitors.  Likewise the density profiles of each model, shown
in Fig.~\ref{fig:rhos}, are also very similar especially in
the region $r \lesssim 200\;{\rm km}$. Since these `inner
region' calculations are collective dominated and the
luminosities, mean energies and profiles are so similar for each
model, the results end up being very similar. If these
collective features make their way through to the observed 
signal then
from the viewpoint of decoding that signal the similarity is
both good and bad: good in the sense that the features are
robust and we can make definite statements about the neutrino
hierarchy, but bad in the sense that there is no information
about the star because the source seems to be standard.

\subsubsection{Outer region, 1000 km -- profile end: MSW
dominated}{\label{sec:3s_outer}}
Next we consider the evolution of the neutrinos through the
outer layers of the star where the MSW effect takes over.  Now
the neutrino luminosities and mean energies are irrelevant and
only the density profile matters.  As shown in
Fig.~\ref{fig:rhos}, the density profiles in the outer layers of
the stars at 3~s pb are very different: The forward shock in
the \Mei{} model has raced off the simulation grid leaving a
very steep density profile; The forward shock of the \Mten{} 
model has
propagated out to $r \sim 3\times 10^{4}\;{\rm km}$ and the
presence of the forward shock, reverse shock and contact
discontinuity are affecting the H resonances of 
the intermediate and higher
energies and the L resonances of the lower, and in the \Met{} 
model the forward shock has propagated to a similar $r \sim
2\times 10^{4}\;{\rm km}$ but is only just beginning to affect
the H resonances of the lower energies. 

\begin{figure}[t!]
\includegraphics[width=\linewidth]{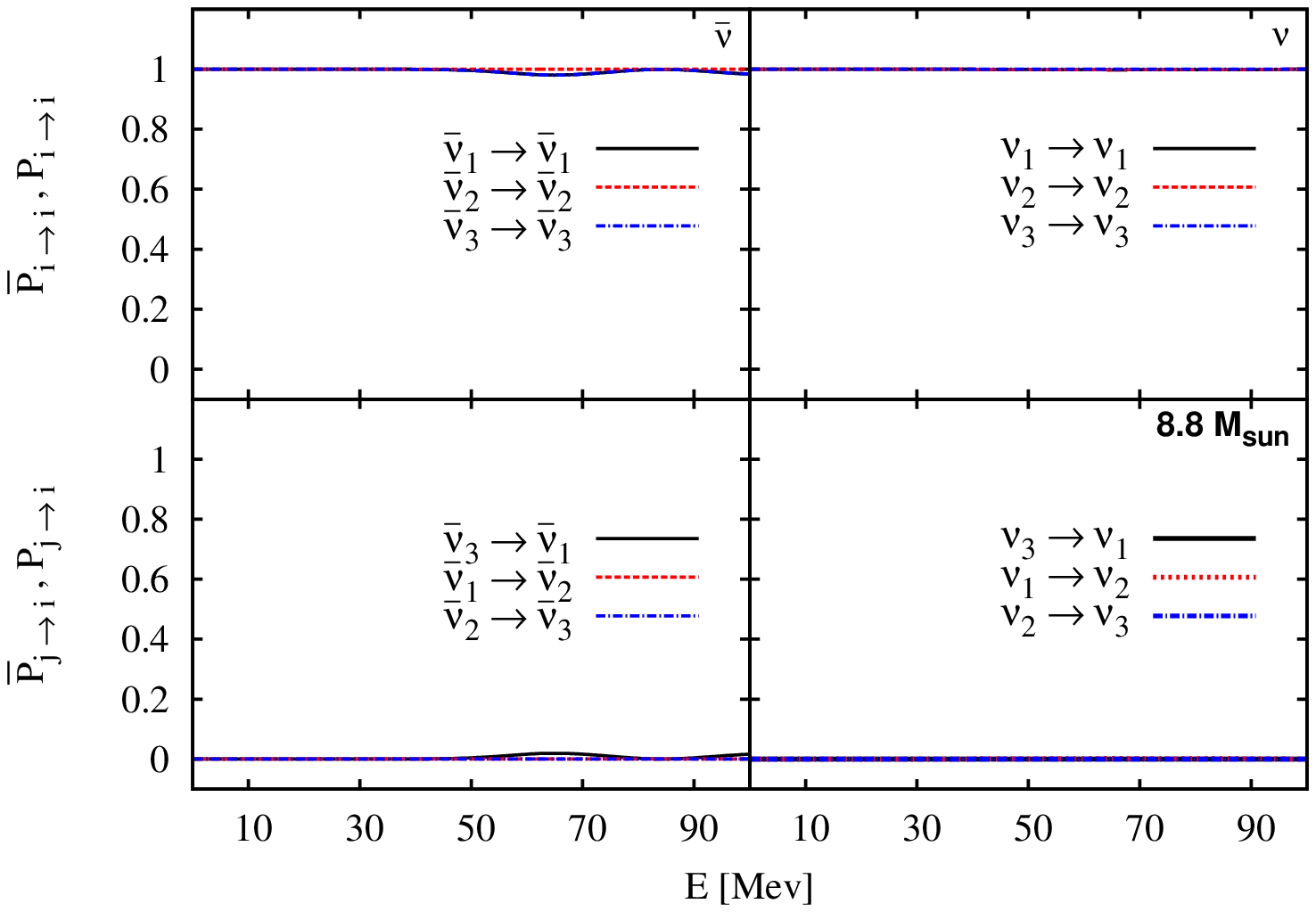}
\includegraphics[width=\linewidth]{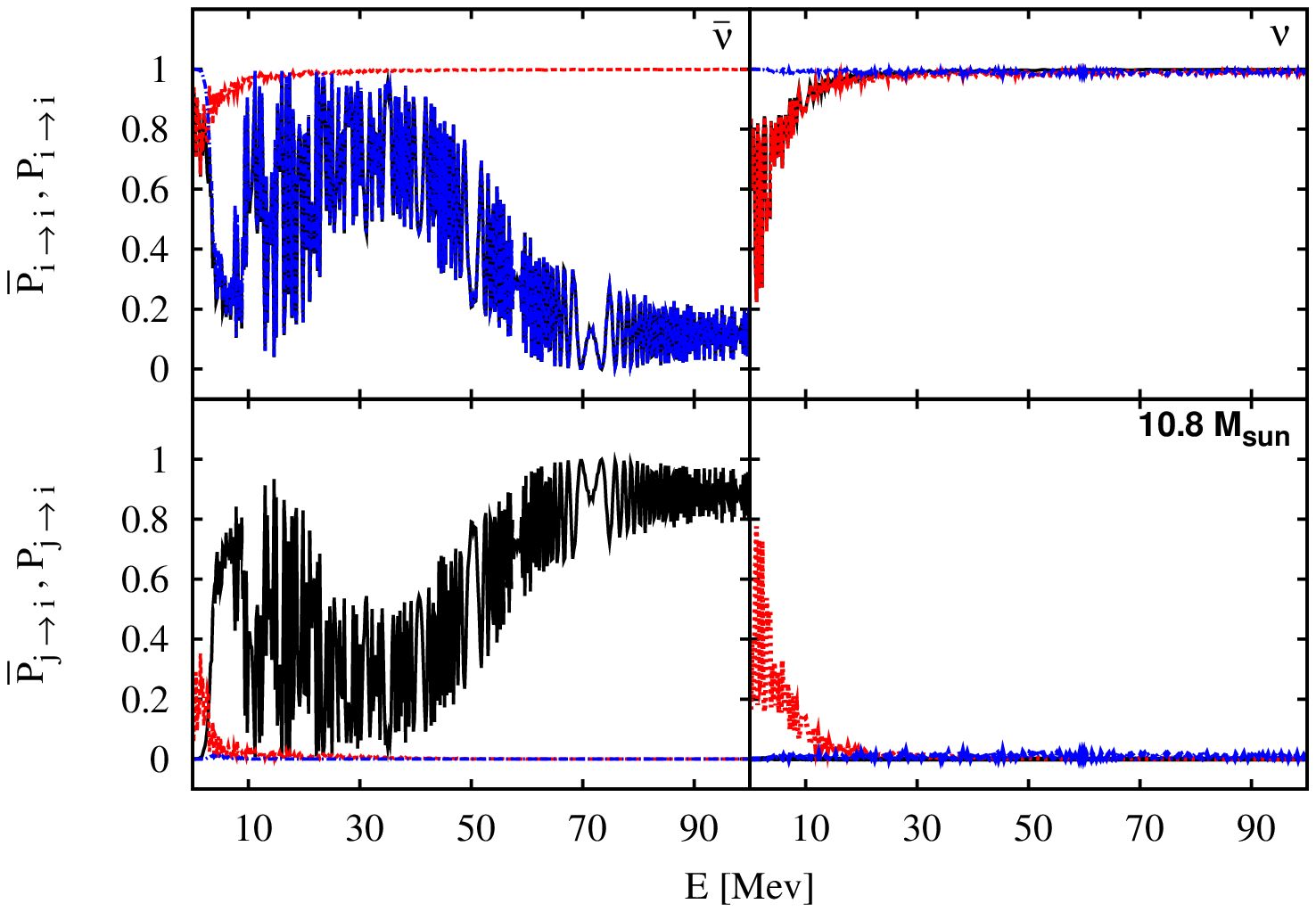}
\includegraphics[width=\linewidth]{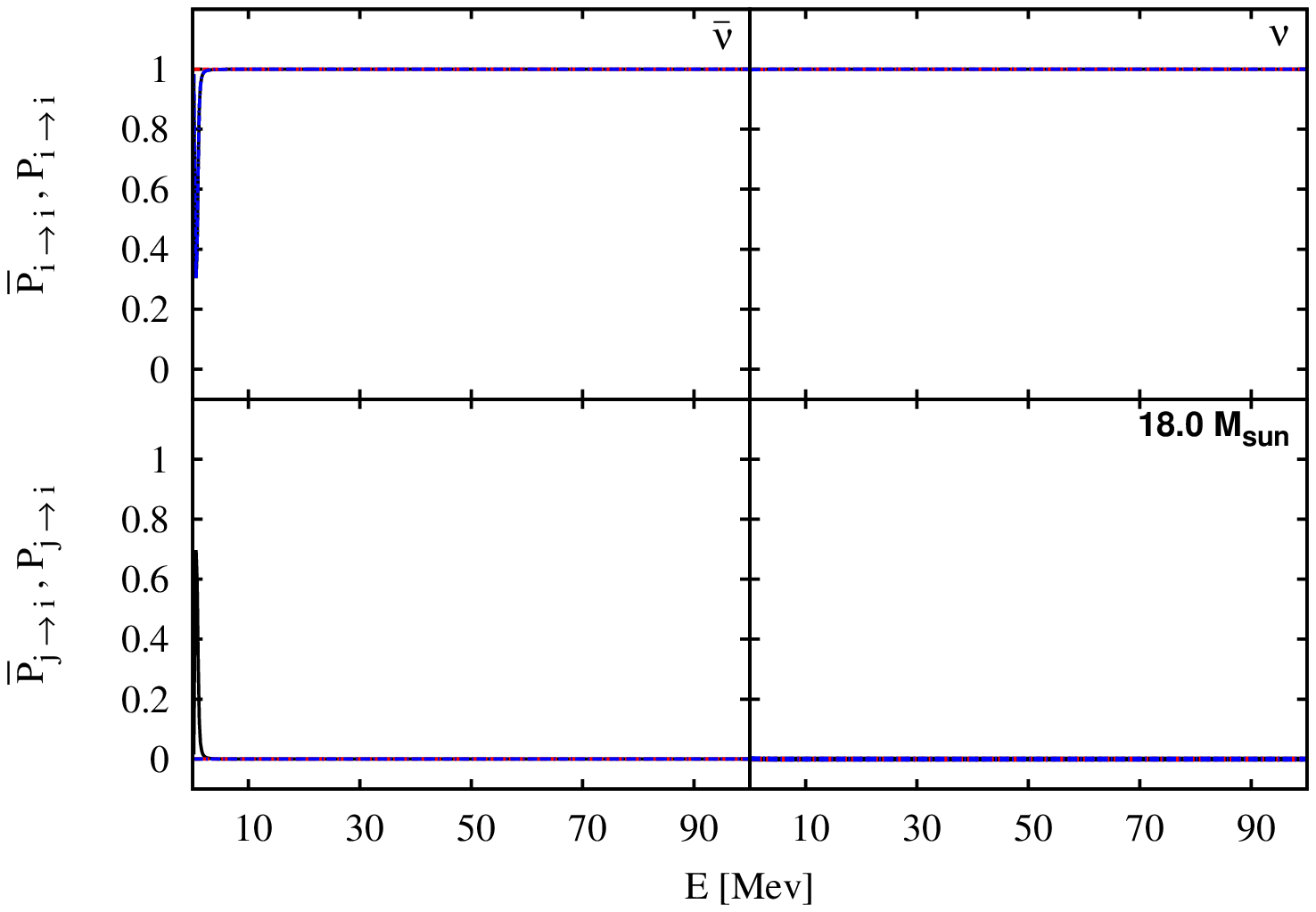}
\caption{\label{fig:moIH_3sec} (color online).  As
Fig.~\ref{fig:coIH_3sec} but for the outer region.  Inverted
Hierarchy.} 
\end{figure}

\begin{figure}[t!]
\includegraphics[width=\linewidth]{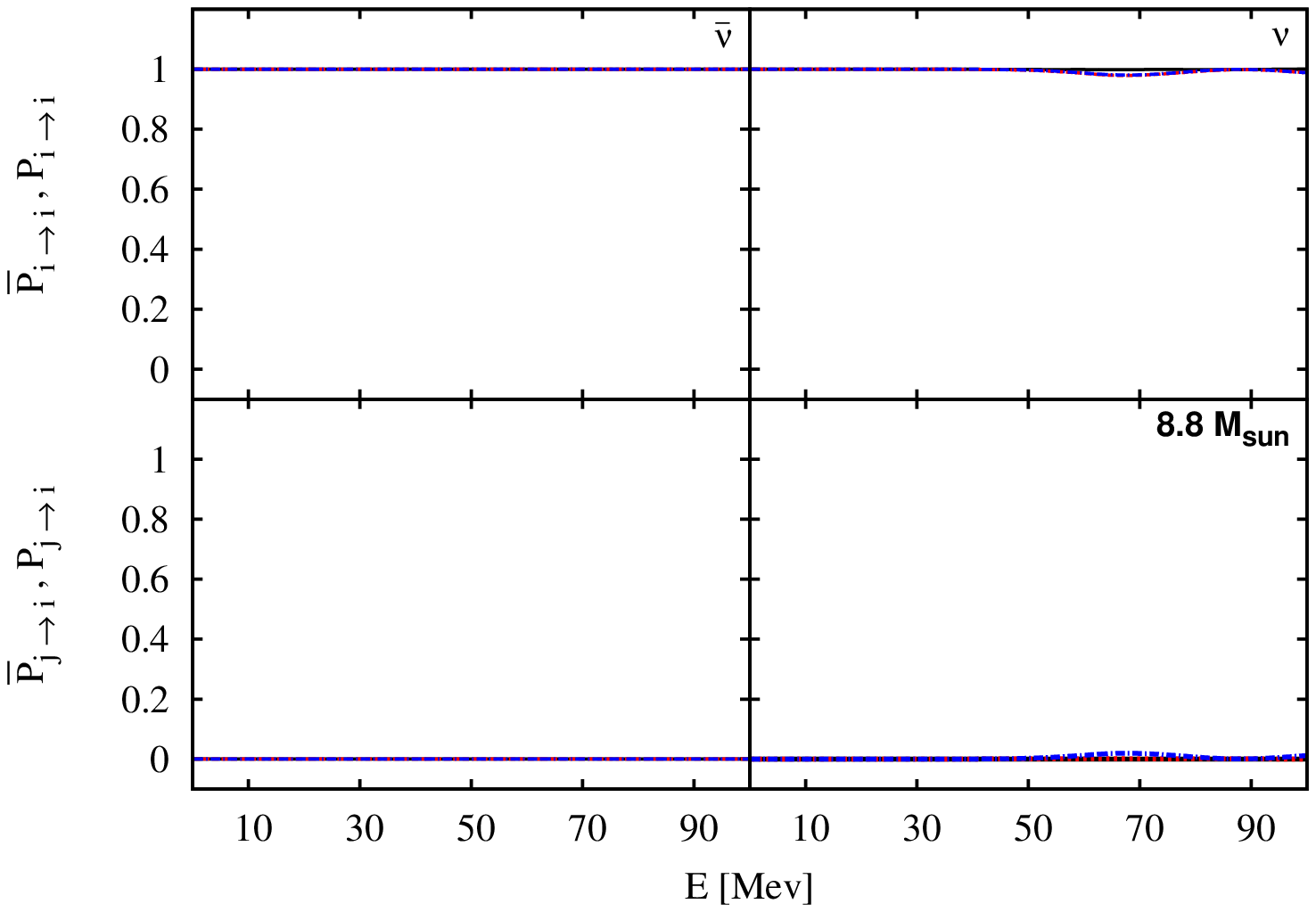}
\includegraphics[width=\linewidth]{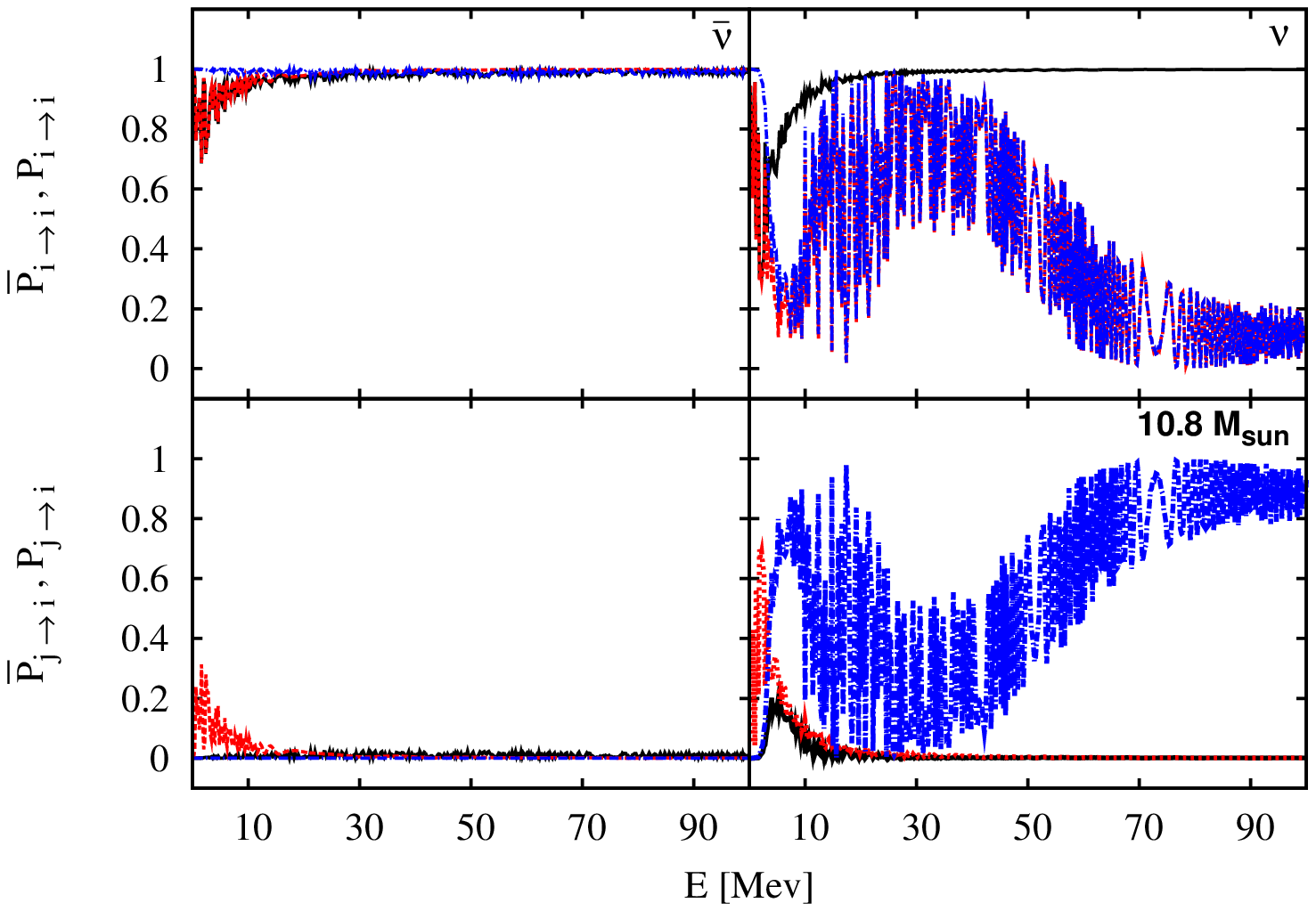}
\includegraphics[width=\linewidth]{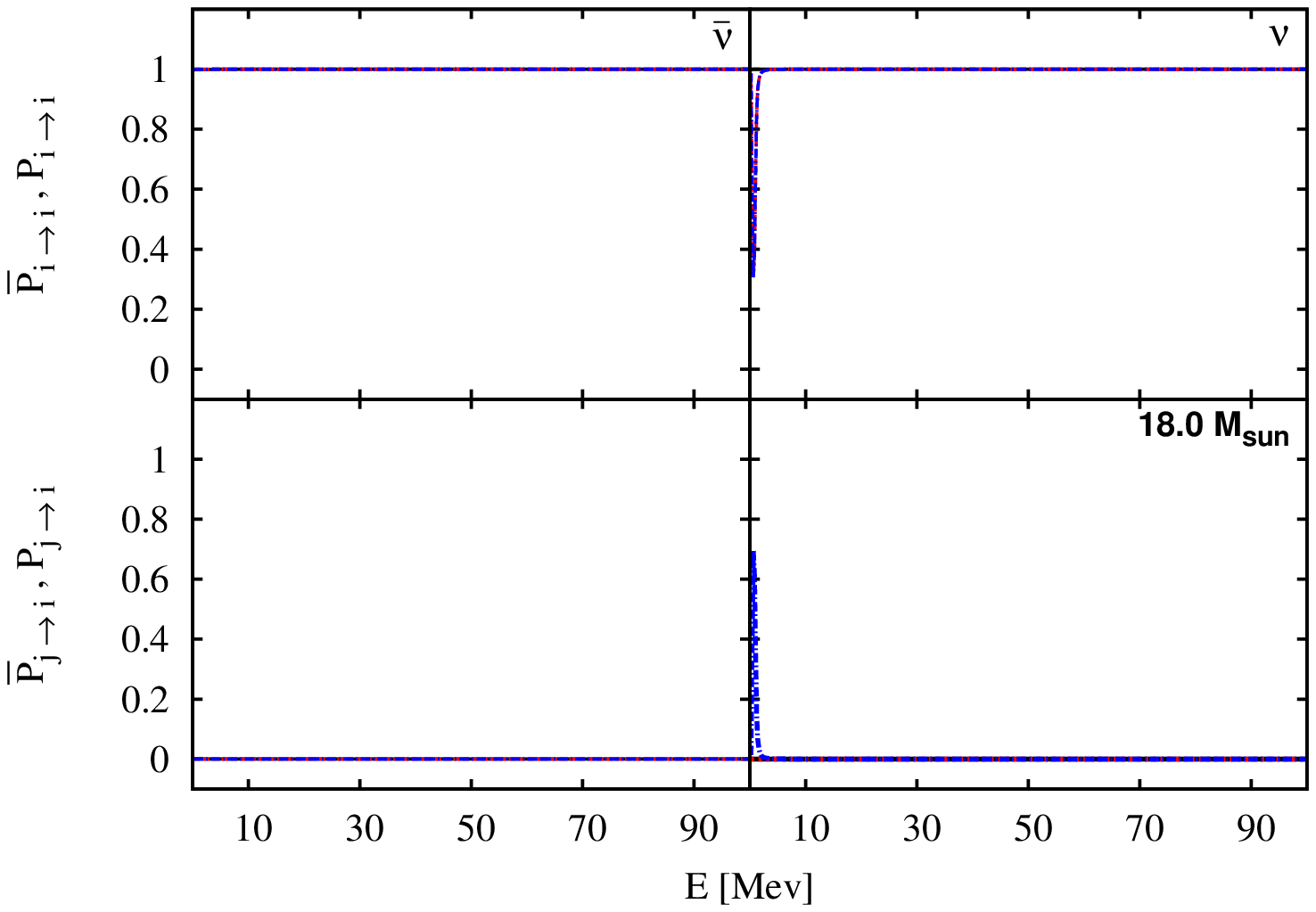}
\caption{\label{fig:moNH_3sec} (color online).   
As Fig.~\ref{fig:coNH_3sec} but for the outer region.  Normal
Hierarchy.} 
\end{figure}

The question becomes whether these major differences in the
profiles translate into equally different results for the neutrino
transition probabilities through each progenitor.  Our results 
for the outer region are shown Figures~\ref{fig:moIH_3sec} (IH) 
and \ref{fig:moNH_3sec} (NH).  Surprisingly we see the 
probabilities
for the \Mei{} and the \Met{} models at 3~s are virtually
identical. The only difference are: In the \Mei{} model a
small amplitude undulation can be seen between 50~MeV and
100~MeV for anti-neutrinos in the IH and for the neutrinos in
the NH; The \Met{} progenitor merely offers a small drop
in the survival probabilities at $\sim$1--2~MeV for
anti-neutrinos in the IH and for neutrinos in the NH.  
The drops in survival probabilities for the \Met{} model is
caused by the lower portion of the forward shock just reaching
into the resonant densities for low energy (anti-) neutrinos in
the (inverted) normal hierarchy. Thus making their passage of
the resonant densities diabatic leading to their enhanced flavor
conversion. 
The reason the two sets of results are so similar
is because the neutrino evolution through both profiles is very
close to adiabatic despite the very different gradients of the
density at the H and L resonances of each progenitor. The
adiabaticity of the L resonance for supernova neutrinos has been
long known but the adiabaticity of the H resonance is
dependent upon the mixing angle $\theta_{13}$ and until recently
this angle was unknown. The measurement of a relatively large
value for $\theta_{13}$ means that neutrinos experiencing a
single, non-shock related, H resonance will do so adiabatically
giving no flavor conversion.  

In contrast the \Mten{} model is a lot more interesting.
Starting with the anti-neutrinos in the IH, we see an incomplete
spectral split between states \bone{} and \bthree{} at 3.5~MeV.
Above 40~MeV is an additional incomplete swap between states
\bone{} and \bthree{}. Furthermore we see sharp changes in the
average survival probability of states \bone{} and \bthree{} at
10~MeV and $\sim$12~MeV.  The anti-neutrino state \btwo{} has a
small dip in the survival probability at $\sim$3~MeV.  A
slightly larger dip is visible for neutrinos in the IH in the
survival probabilities of matter states 1 and 2 at $\sim$3~MeV.
When we turn to the NH, we see that a similar small dip is
present in the survival probabilities of anti-neutrino states
\bone{} and \btwo{} at $\sim$3~MeV.  In the NH the neutrinos now
have the interesting features. We see an incomplete split
between neutrino states 2 and 3 at 3.5~MeV, and above 40~MeV we
see another incomplete swap.  A careful study also reveals that
the average survival probability of neutrino states 2 and 3
increase abruptly at roughly 24~MeV.

In both hierarchies we see phase 
effects \cite{Kneller:2005hf,Dasgupta:2007phase} on top of the
aforementioned swaps. In the IH there are large phase effects in
the anti-neutrino states \bone{} and \bthree{}, but also for the
neutrino states 1 and 2 do we see phase effects at low energies.
In the NH the large phase effects are visible in the neutrino
states 2 and 3, and only smaller effects are visible in the
anti-neutrino states \bone{} and \btwo{} for lower energies.
Contrary to Dasgupta \& Dighe \cite{Dasgupta:2007phase}, who only
see phase effects for neutrinos in the NH and for anti-neutrinos
in the IH, we thus find phase effects for neutrinos and
anti-neutrinos in both hierarchies.  We note that the feature
around $\sim$71~MeV in the IH anti-neutrinos and around
$\sim$73~MeV in the NH neutrinos is an effect of the phase effect
being close to resonance at these energies
(A similar feature is seen in Fig.~2 of \cite{Fogli:2003dw}).
The drops in survival probabilities seen for neutrino states 1 and 2
at roughly 3.5~MeV in both hierarchies, are actually the first
little effects of the shock hitting the MSW L resonance.  The
magnitude of the drops are similar in the two hierarchies, as we
would expect, although the effect is enhanced and somewhat
obscured in the NH, where state 2 also mix with state 3 through
the H resonance.  We are the first ones to follow the shock wave
out to densities relevant for the MSW L resonance, and calculate
the flavor probabilities in this region. We will go into further
detail with these findings in Sec.~\ref{sec:time_evolv}.

\subsubsection{Full profile traversal}{\label{sec:3s_full}}
\begin{figure}[t!]
\includegraphics[width=\linewidth]{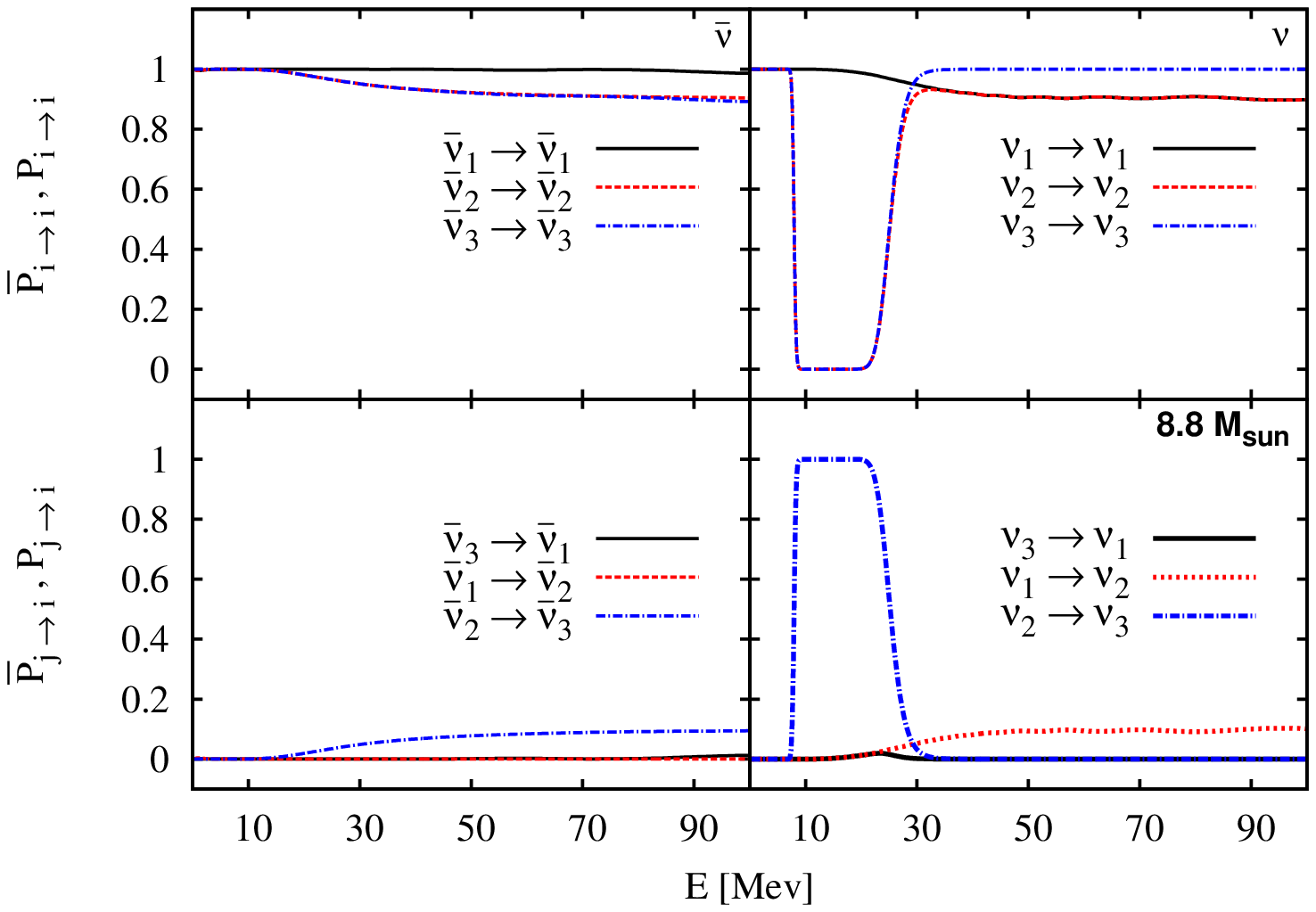}
\includegraphics[width=\linewidth]{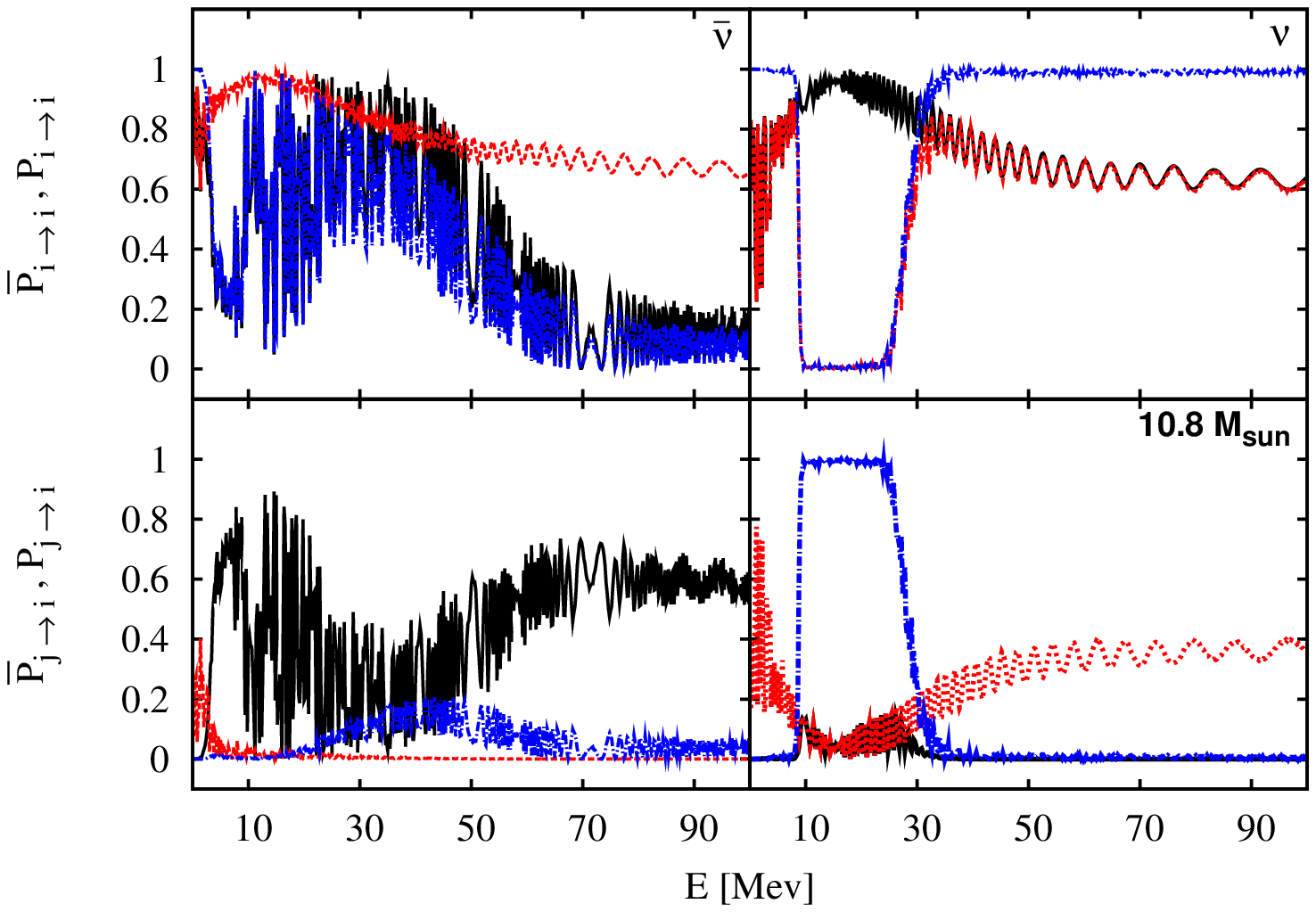}
\includegraphics[width=\linewidth]{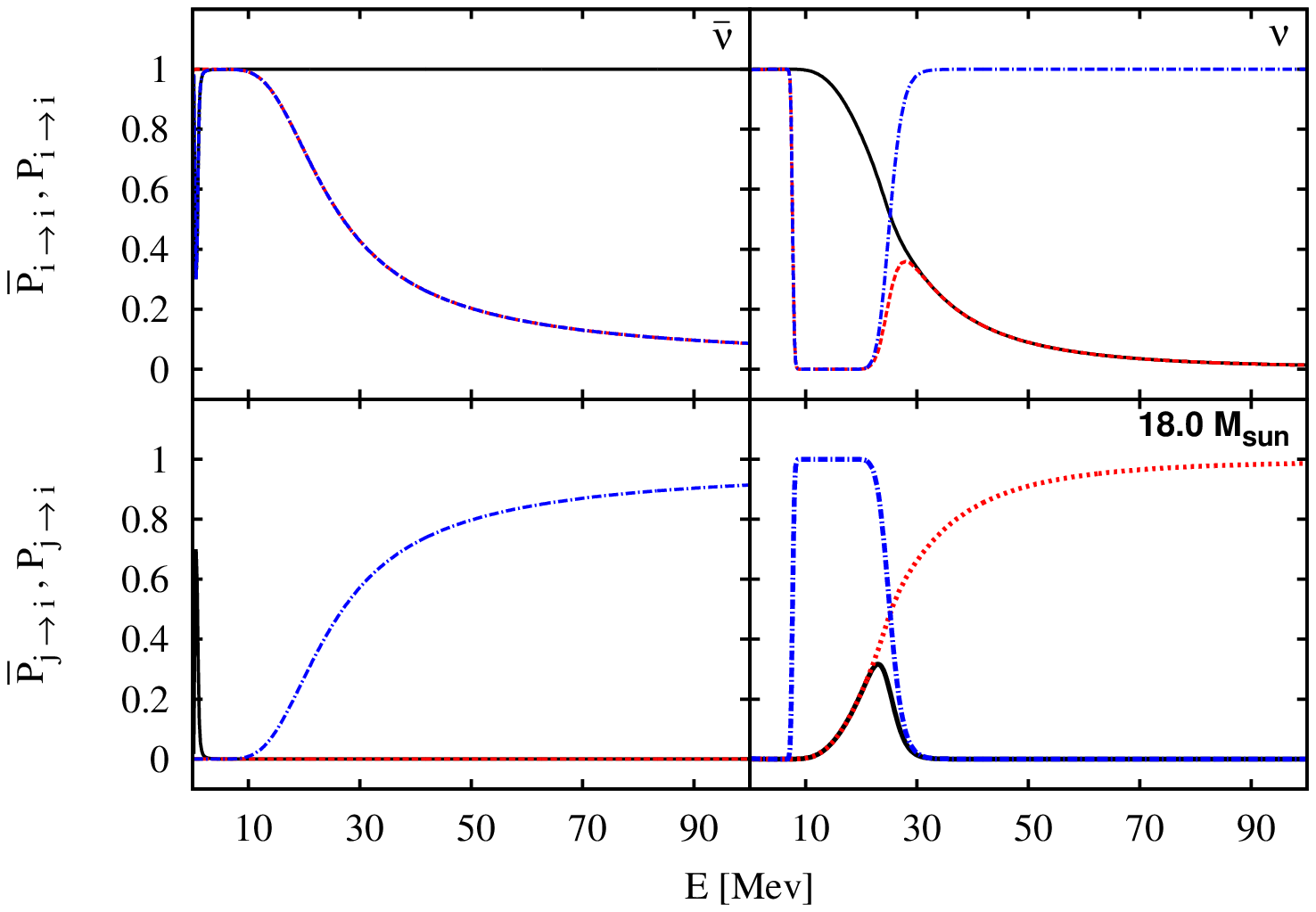}
\caption{\label{fig:nmIH_3sec} (color online).  As
Fig.~\ref{fig:coIH_3sec} but for the full profile traversal.
Inverted Hierarchy.} \end{figure}
\begin{figure}[t!]
\includegraphics[width=\linewidth]{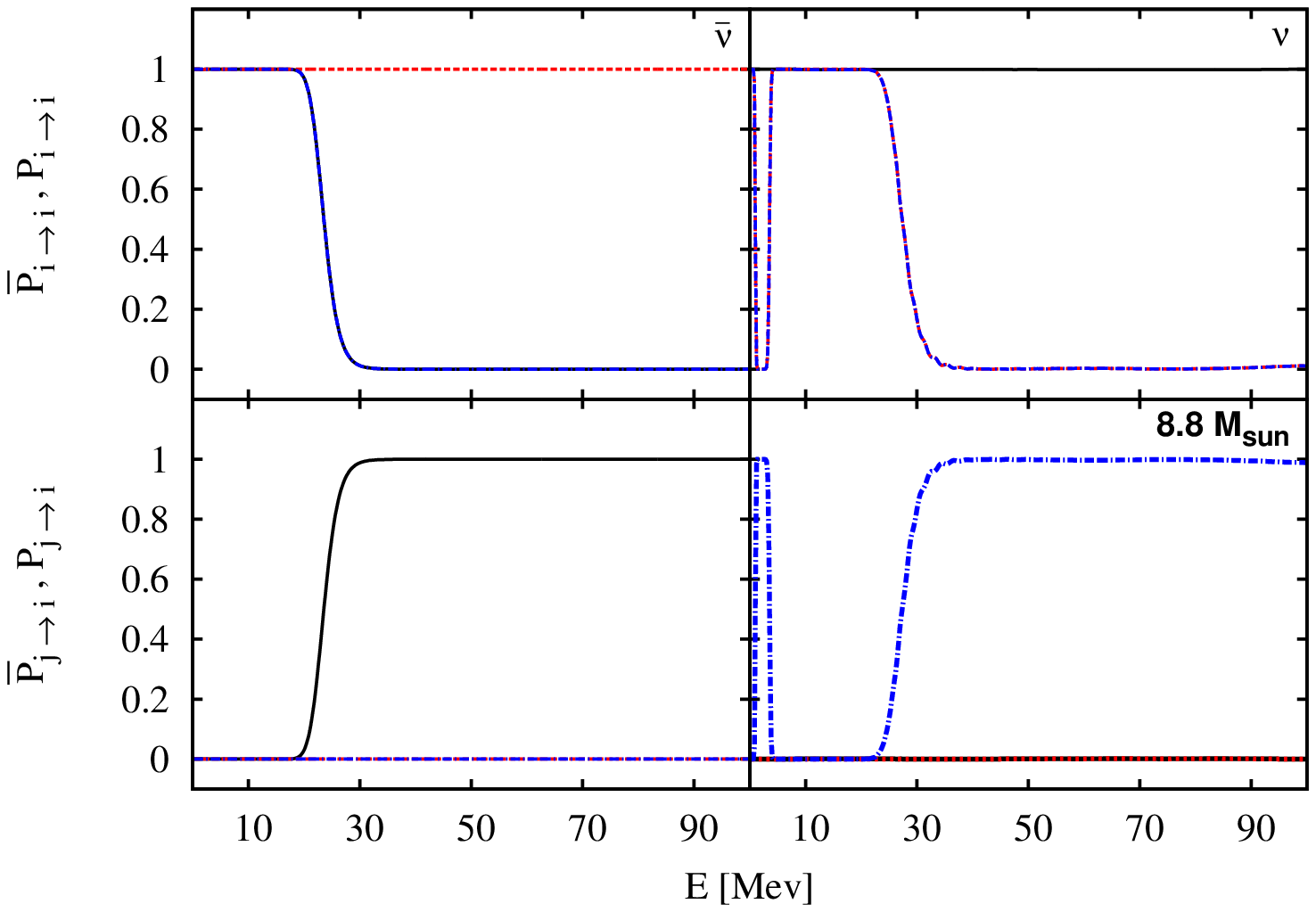}
\includegraphics[width=\linewidth]{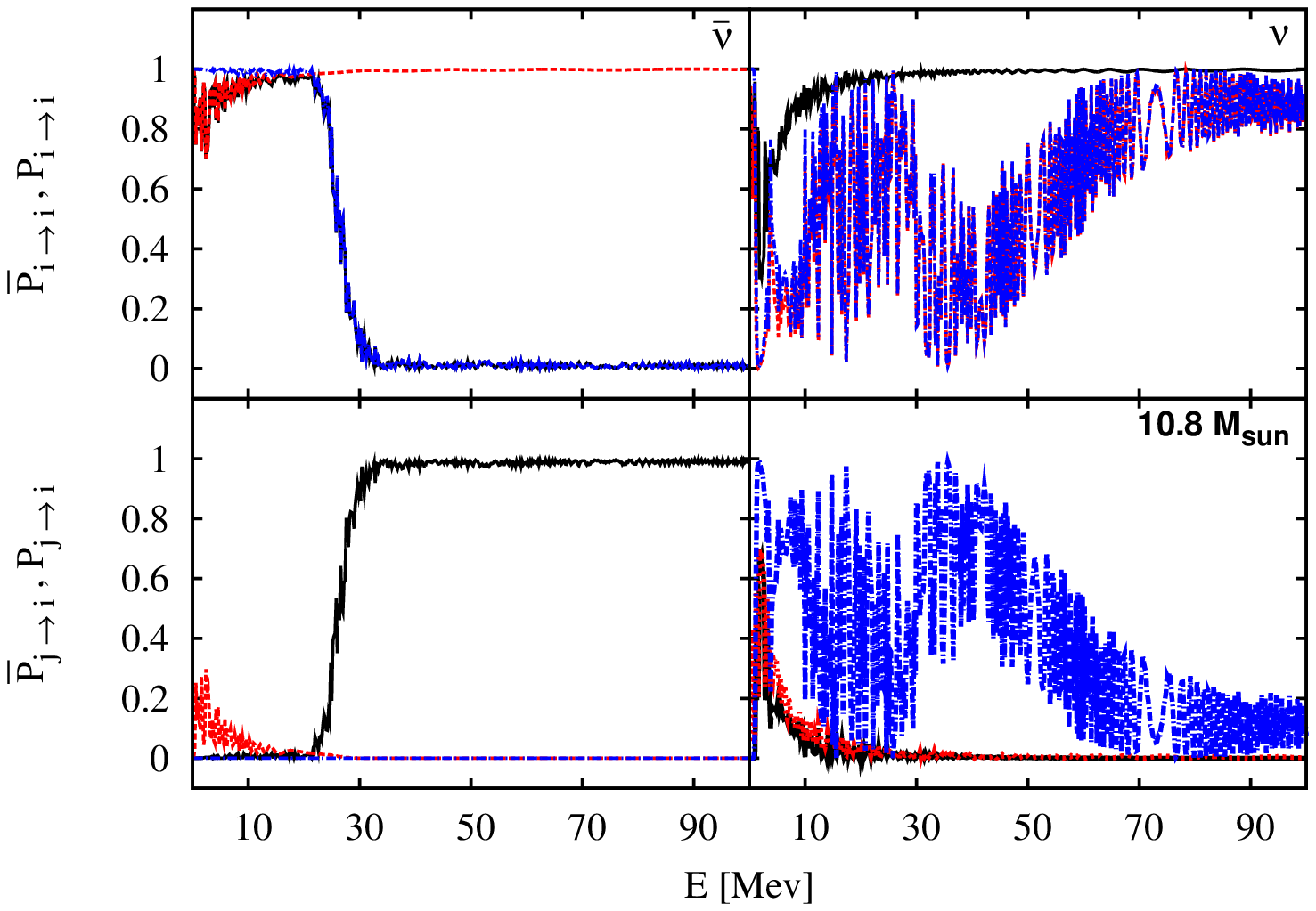}
\includegraphics[width=\linewidth]{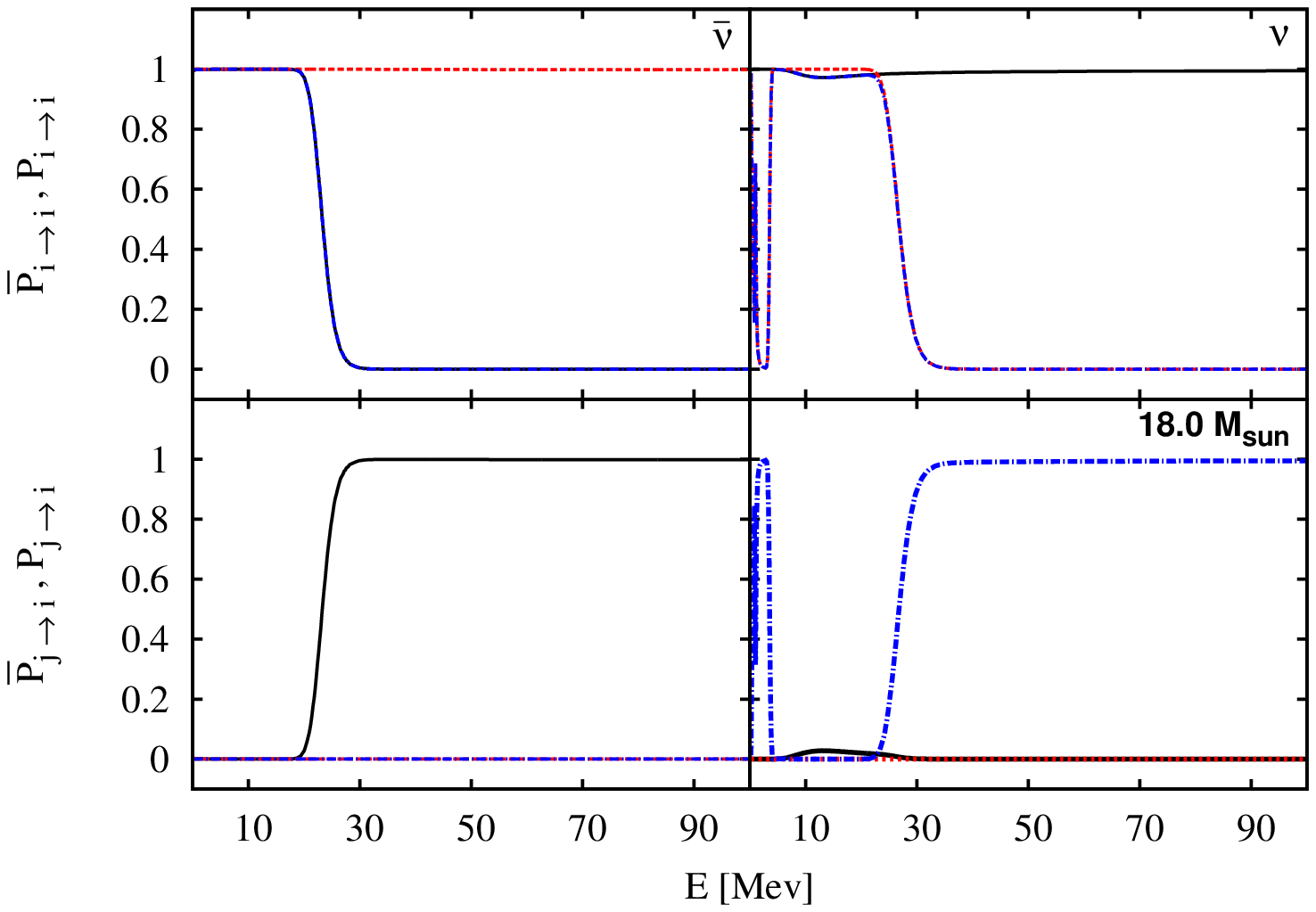}
\caption{\label{fig:nmNH_3sec} (color online).   As
Fig.~\ref{fig:coNH_3sec} but for the full profile traversal.
Normal Hierarchy.} 
\end{figure}
The two previous calculations showed separately how the
collective and MSW effects generate features in the spectrum.
Now we put the two together and show in
Figures~\ref{fig:nmIH_3sec} and \ref{fig:nmNH_3sec} the results
of calculations covering the entire density profile from PNS
surface to the end.

In a broad sense the results display phenomena which are
consistent in most of the profiles we have investigated. In a
full calculation the splits and swaps induced in the inner and
outer regions respectively are now combined in a superposition.
All the features present in the inner region can be recovered,
as well as the features from the outer region. The clean
superposition is evident in the cases of the \Mei{} and \Met{}
models where very little happens in the outer region, which can
be seen by comparing Figures~\ref{fig:nmIH_3sec} and
\ref{fig:nmNH_3sec} with Figures~\ref{fig:coIH_3sec},
\ref{fig:coNH_3sec}, \ref{fig:moIH_3sec} and
\ref{fig:moNH_3sec}.

A slightly different story is visible in the middle panels for
the \Mten{} model in Figures~\ref{fig:nmIH_3sec} and
\ref{fig:nmNH_3sec}.  We see the superposition but new
features have also emerged.  In broad terms we can divide the
probabilities into 3 categories: i) those that remain the same
as they were either in the inner region or in the outer region;
ii) those that are a simple superposition - below a certain
energy they follow the trend seen in the outer region
calculation while above that energy they follow the trends of
the inner region; iii) the complex cases that are neither i) nor
ii).

By comparing the middle quartet of Fig.~\ref{fig:nmIH_3sec} (IH)
to the ones of Fig.~\ref{fig:coIH_3sec} and
Fig.~\ref{fig:moIH_3sec} we see that $\bar P_{11}$ is an example
of case i) because it follows its path from the outer region and
nothing happened to $\bar\nu_{1}$ in the inner region. Similarly
$P_{33}$ is another case i) example because it follows its trend
from the inner region, albeit with minimal fluctuations on top,
since very little happens to $\nu_3$ in the outer region. The
three probabilities $\bar P_{22}$, $P_{11}$ and $P_{22}$ are all
case ii) because we see that each of these probabilities follow
the same pattern as in the outer region
(Fig.~\ref{fig:moIH_3sec}) below an energy of 12, 15 and 8~MeV
respectively while above these energies they follow the trend
they developed in the inner region (Fig.~\ref{fig:coIH_3sec}). 
Additionally, for all three
probabilities new small amplitude oscillations have arisen. At
higher energies the oscillations have a larger frequency, than
at lower energies.
Finally $\bar P_{33}$ falls into the more complex, case iii)
category: Below 25~MeV it follows the pattern from the outer
region, above this energy it has the same oscillations it
displayed in the outer region it just falls off slightly
quicker, so that at higher energies the average survival
probability is slightly lower.

Turning to the NH and comparing quartets of 
Fig.~\ref{fig:nmNH_3sec} to Figures~\ref{fig:coNH_3sec} and 
\ref{fig:moNH_3sec} we find that $P_{11}$ is a case i) example
as it follows its trend from the outer region, although
with oscillations of tiny amplitude above 30~MeV. $\bar P_{33}$
too is a case i) as it
follows its trend from the inner region, but with additional
minor random fluctuations superimposed which arise in the outer
region. Both $\bar P_{11}$ and $\bar P_{22}$ 
belong to case ii) as they follow their
pattern from the outer region below 20~MeV, and change to follow
the pattern from the inner region above this energy, albeit with
slight fluctuations. This leaves us with $P_{22}$ and $P_{33}$
belonging to case iii)
who are slightly more complex, and share a common story. They
both display the splits at 2 and 4~MeV that arose in the inner
region, although the latter split is no longer complete, because
an additional incomplete swap is caused in the outer region.
Between 4~MeV and 30~MeV the probabilities follow the pattern
they had in the outer region. Above 30~MeV the spectral split
from the self-interaction in the inner region has swapped the
spectra of $\nu_2$ and $\nu_3$ completely. When they then
propagate through
the outer region the MSW effect swaps them again, but this time
incompletely. Therefore we see an abrupt drop in the
probabilities at 30~MeV, and above this energy the probabilities
appear as a reflected version of the pattern present for the
outer region alone.

So, in summary, we see that the combination of collective and MSW
effects often produces results which are consistent with a
trivial superposition and in other cases the combined effect
results in a new feature. The exact reason needs to be explored
more carefully but at the same time it is not entirely
unexpected or unprecedented. 
The $S$ matrix describing
the calculation for the entire profile is the product of the $S$
matrices computed for the inner region $S_I$ and the outer
region $S_O$: $S = S_O\,S_I$. Both $S_O$ and $S_I$ are of
the form 
\begin{equation} 
S =  
\begin{pmatrix} \alpha_1 & \alpha_2 &
\alpha_3 \\ \frac{-D\alpha_2^{\ast}\alpha_5^{\ast} - \alpha_1
\alpha_4 \alpha_3^{\ast}}{1-|\alpha_3|^2} &
\frac{D\alpha_1^{\ast}\alpha_5^{\ast} - \alpha_2 \alpha_4
\alpha_3^{\ast}}{1-|\alpha_3|^2} & \alpha_4 \\
\frac{-D\alpha_2^{\ast}\alpha_4^{\ast} - \alpha_1 \alpha_5
\alpha_3^{\ast}}{1-|\alpha_3|^2} &
\frac{D\alpha_1^{\ast}\alpha_4^{\ast} - \alpha_2 \alpha_5
\alpha_3^{\ast}}{1-|\alpha_3|^2} & \alpha_5 \\ 
\end{pmatrix}
\end{equation}
where the $\alpha_i$'s are complex numbers with the requirement
that $|\alpha_1|^2 + |\alpha_2|^2 +|\alpha_3|^2 = |\alpha_3|^2 +
|\alpha_4|^2 +|\alpha_5|^2 =1$ and $D$ is the determinant of
unit magnitude. Multiplying two matrices of this form together
leads to a very messy expression in general which only gets worse
when one computes the square amplitudes of the matrix elements
so as to produce the transition probabilities. The only simple
case is the element $S_{13}$ whose product is $S_{13} =
\alpha_{1}^{(O)}\,\alpha_{3}^{(I)} +
\alpha_{2}^{(O)}\,\alpha_{4}^{(I)} +
\alpha_{3}^{(O)}\,\alpha_{5}^{(I)}$.  The transition probability
$P_{13} = |S_{13}|^2$ is of the form $P_{13} =
P_{11}^{(O)}\,P_{13}^{(I)} + P_{12}^{(O)}\,P_{23}^{(I)} +
P_{13}^{(O)}\,P_{33}^{(I)} + \ldots$ where the $P_{ij}^{(R)}$
are the transition probabilities between states $j$ and $i$ in
region $R$. The term $P_{11}^{(O)}\,P_{13}^{(I)} +
P_{12}^{(O)}\,P_{23}^{(I)} + P_{13}^{(O)}\,P_{33}^{(I)}$ is
exactly what we expect from a straight superposition of
probabilities but in addition there are many extra terms, which
we have not explicitly written out, which depend upon both the
phases and magnitudes of the $\alpha$'s.  These phase terms are
due to `interference' between the collective and MSW
calculations and a sign of this interference is that the new
features found when we combine the calculations are oscillatory
as a function of energy as the interference varies from
constructive to destructive and back again.

\subsubsection{Including turbulence}\label{sec:3s_incTurb}
\begin{figure}[t!]
\includegraphics[width=\linewidth]{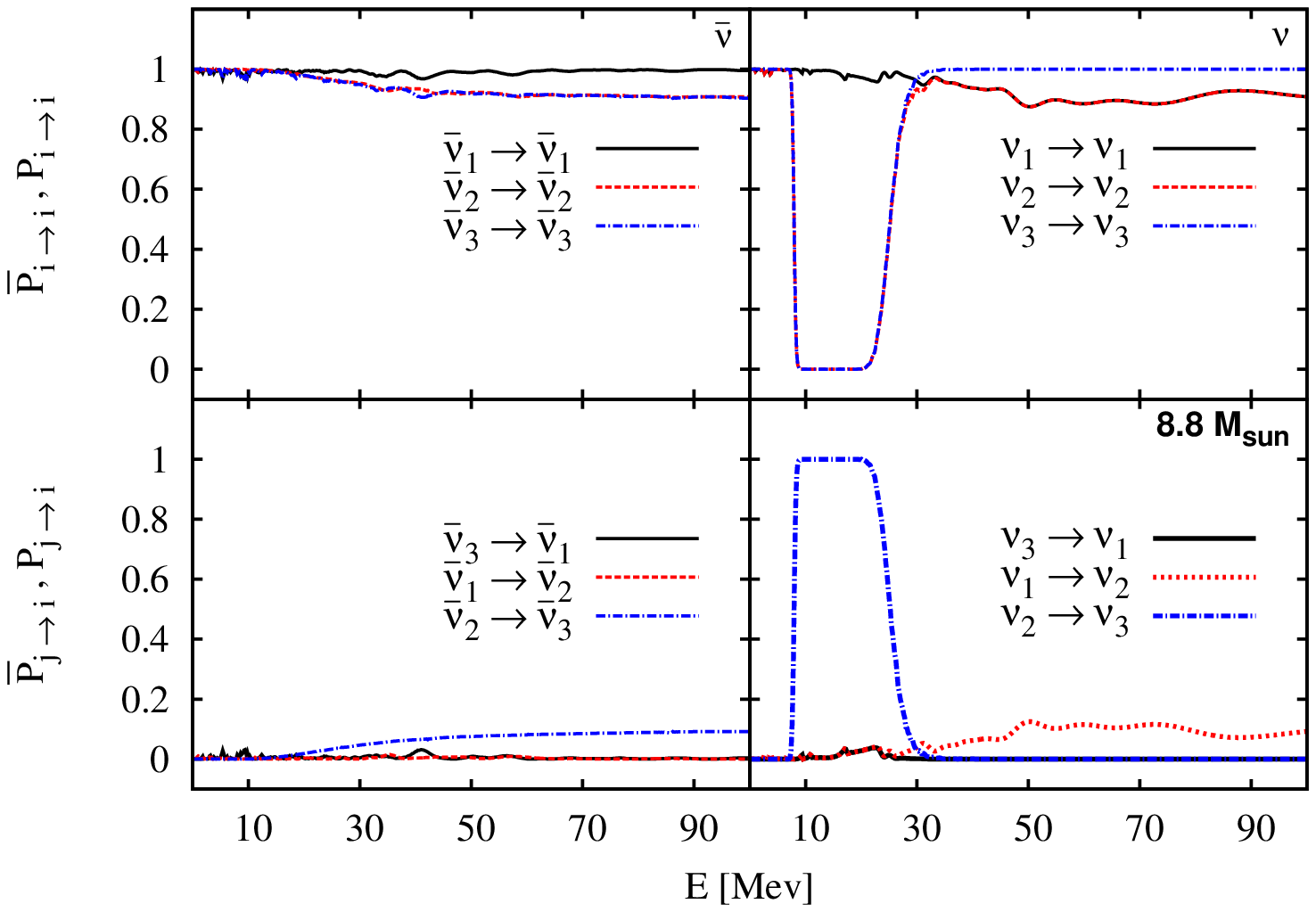}
\includegraphics[width=\linewidth]{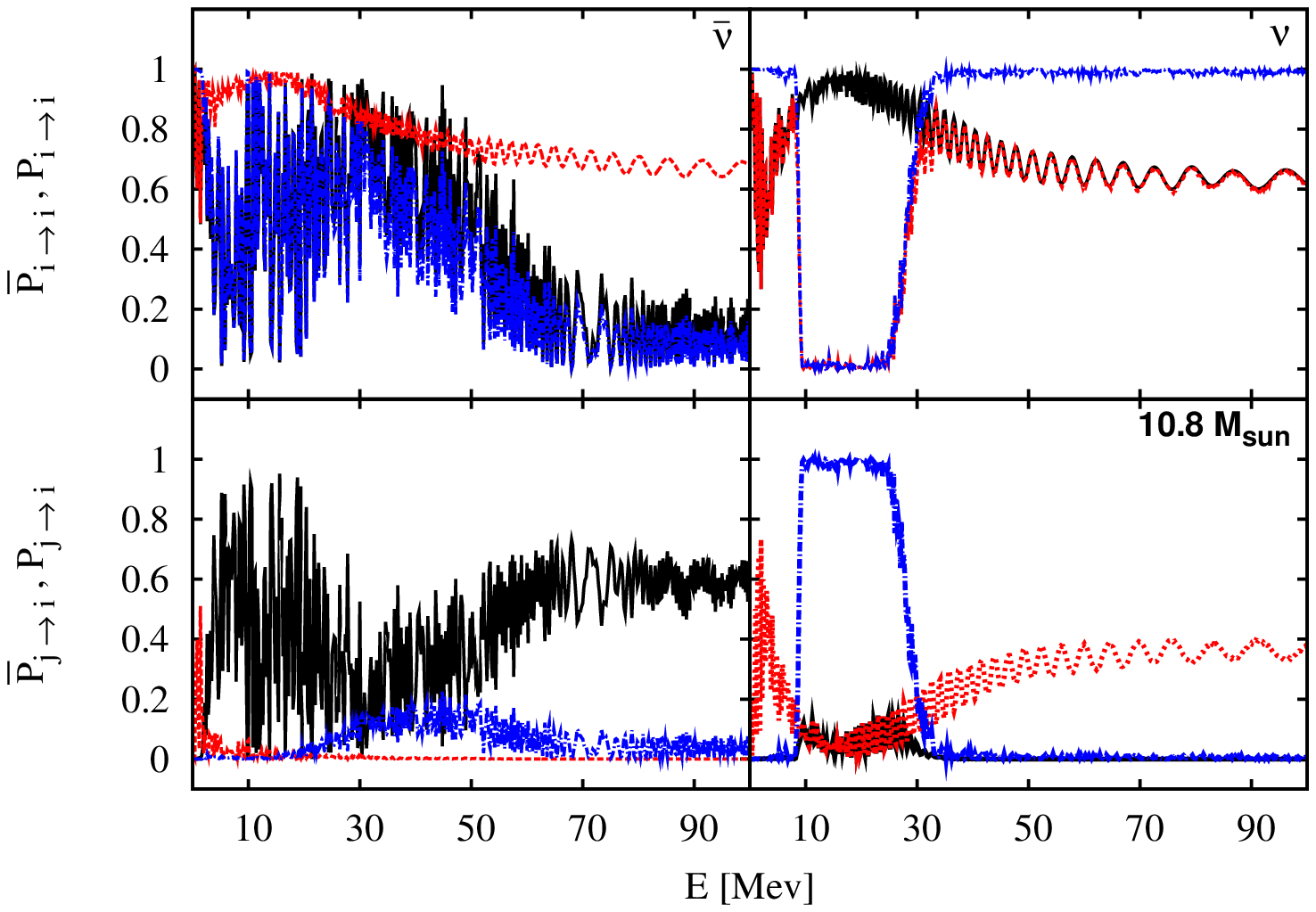}
\includegraphics[width=\linewidth]{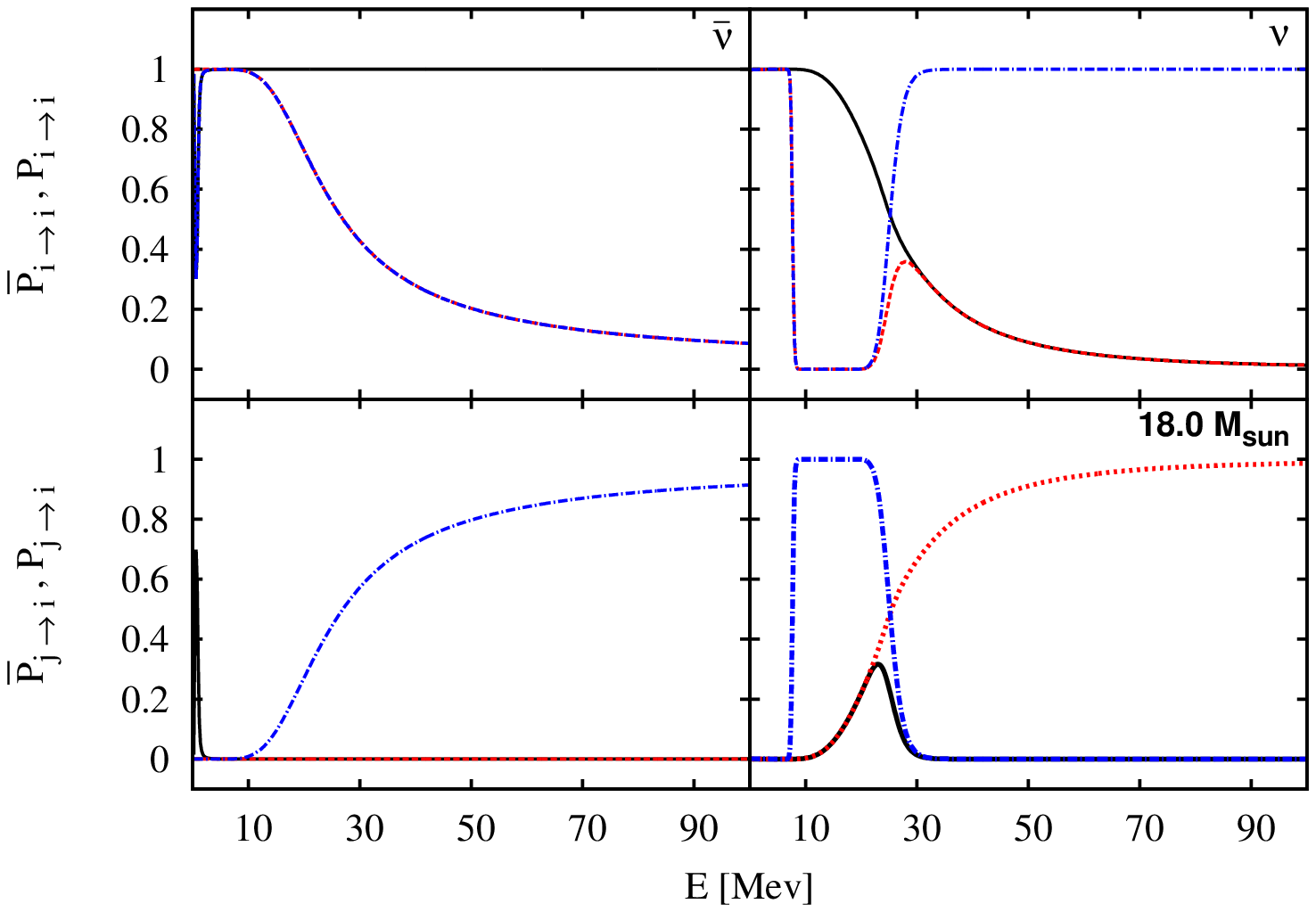}
\caption{\label{fig:tnmIH_3sec} (color online).  As
Fig.~\ref{fig:coIH_3sec} but for the full profile traversal with
10\% turbulence added. Inverted Hierarchy.} \end{figure}
\begin{figure}[t!]
\includegraphics[width=\linewidth]{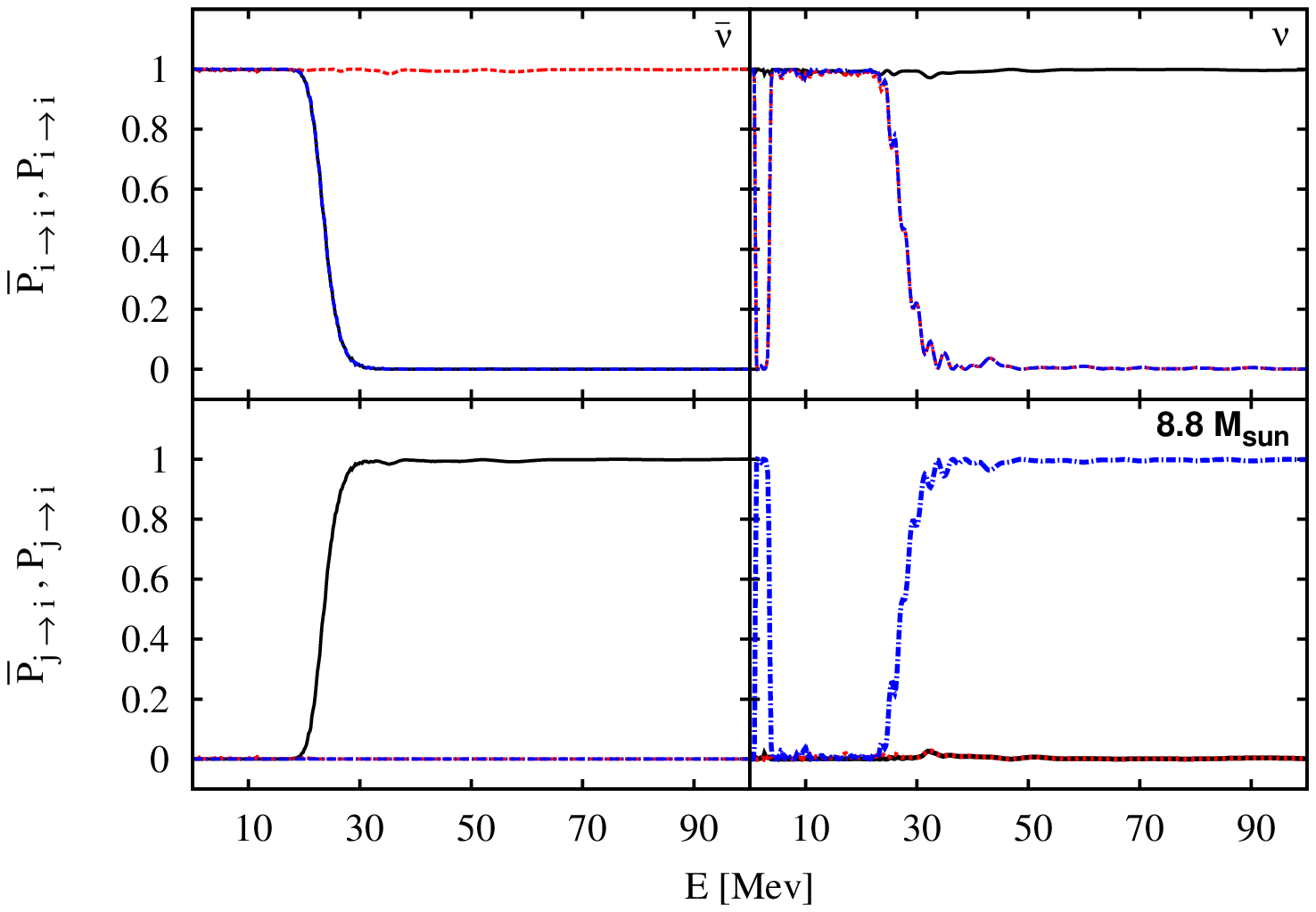}
\includegraphics[width=\linewidth]{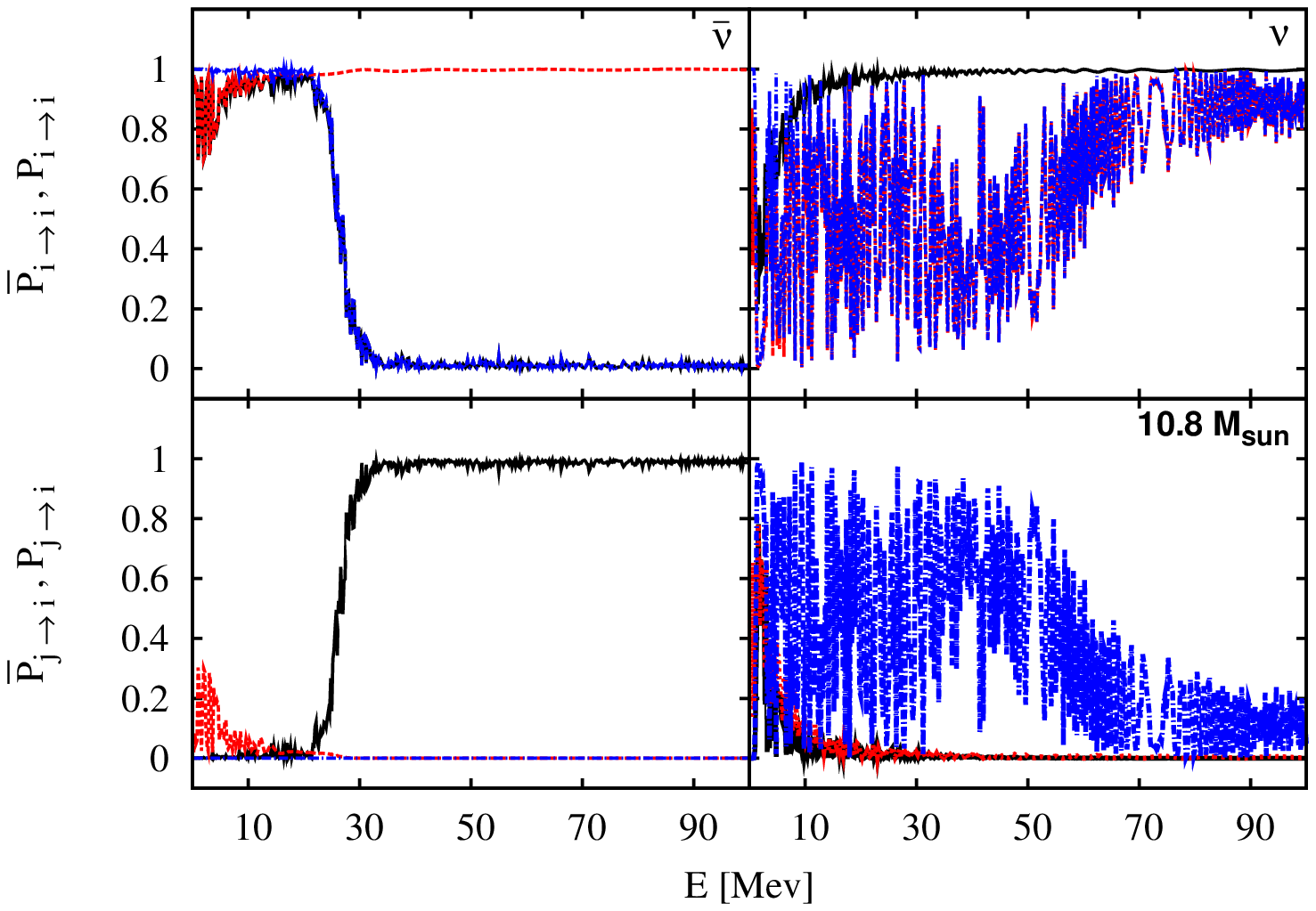}
\includegraphics[width=\linewidth]{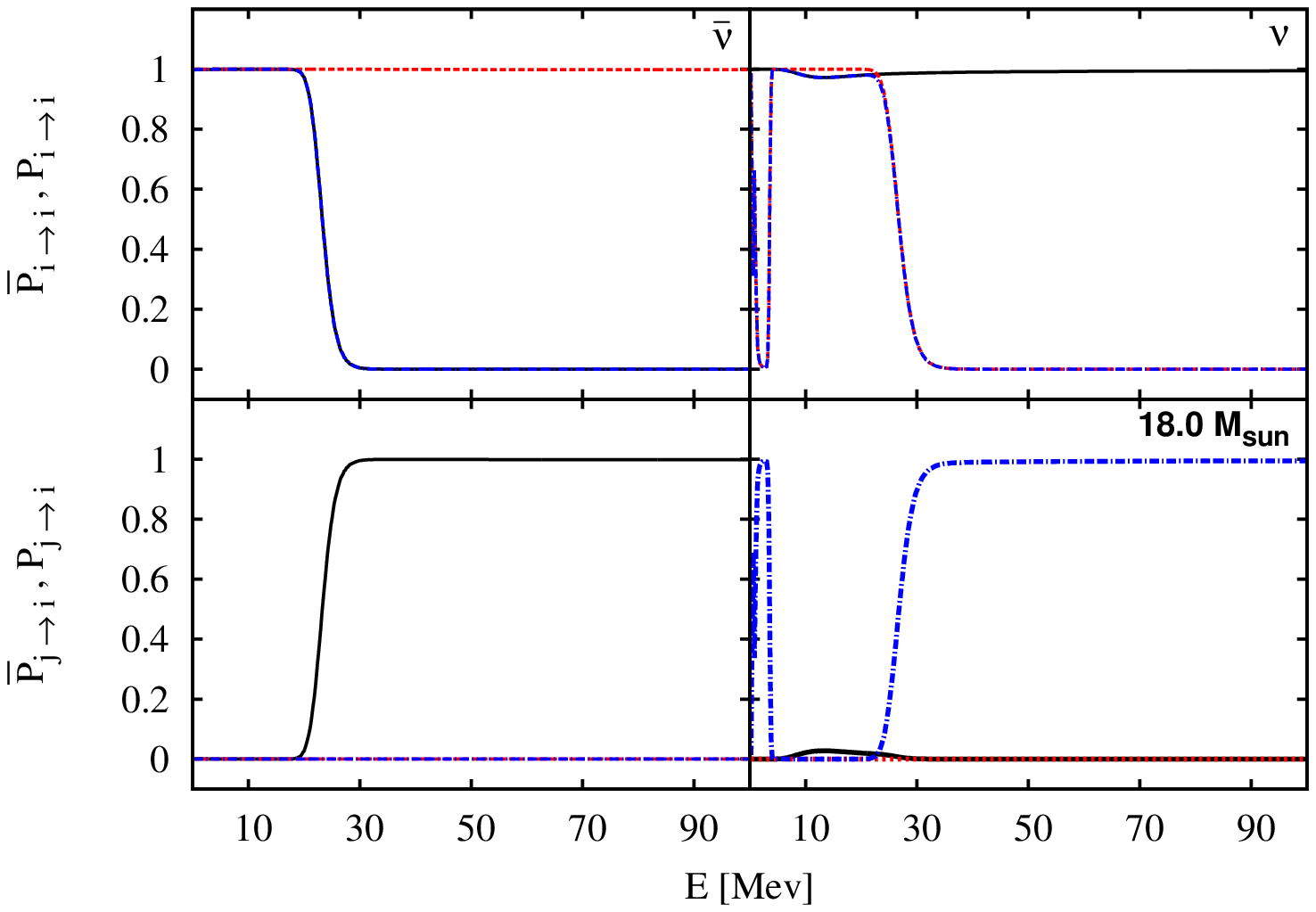}
\caption{\label{fig:tnmNH_3sec} (color online).   As
Fig.~\ref{fig:coNH_3sec} but for the full profile traversal with
10\% turbulence added. Normal Hierarchy.} \end{figure}

Finally we consider the results from our fourth set of
calculations where the neutrinos again traverse the full profile
but now with added turbulence.  These are shown in
Figures~\ref{fig:tnmIH_3sec} and \ref{fig:tnmNH_3sec}. From the
way we have included turbulence (see Sec.~\ref{sec:theo_turb})
the results of the \Mei{} and \Met{} are more directly
comparable but quite generally adding 10\% ($C_*=0.1$)
turbulence to our density profiles at 3~s does not lead to
dramatic changes. 
When we look closely, though, we see that minor alterations have
occurred. Focusing our attention on $\bar P_{22}$ for the \Mten{}
model in Fig.~\ref{fig:tnmIH_3sec}, we see that at 100~MeV the
endpoint of the red dashed line bends the opposite way of what
it does in Fig.~\ref{fig:nmIH_3sec}. This indicates that
turbulence in this case acts as to introduce an additional phase
effect.
The clear transition points at 30~MeV that were visible in
$P_{22}$, $P_{33}$ and $P_{32}$ in Fig.~\ref{fig:nmNH_3sec} for
the \Mten{} model have been obscured by the bigger amplitudes of
the phase effects introduced by the addition of turbulence. 
Generally the effect of adding a moderate amount of turbulence 
to this profile can be
summarized as an increase in the amplitudes of the phase effect
oscillations, and a slight shift in the position of these.

\subsubsection{Larger Turbulence}   \label{sec:turb_larger}
\begin{figure*}[t!]
\includegraphics[width=0.49\linewidth]{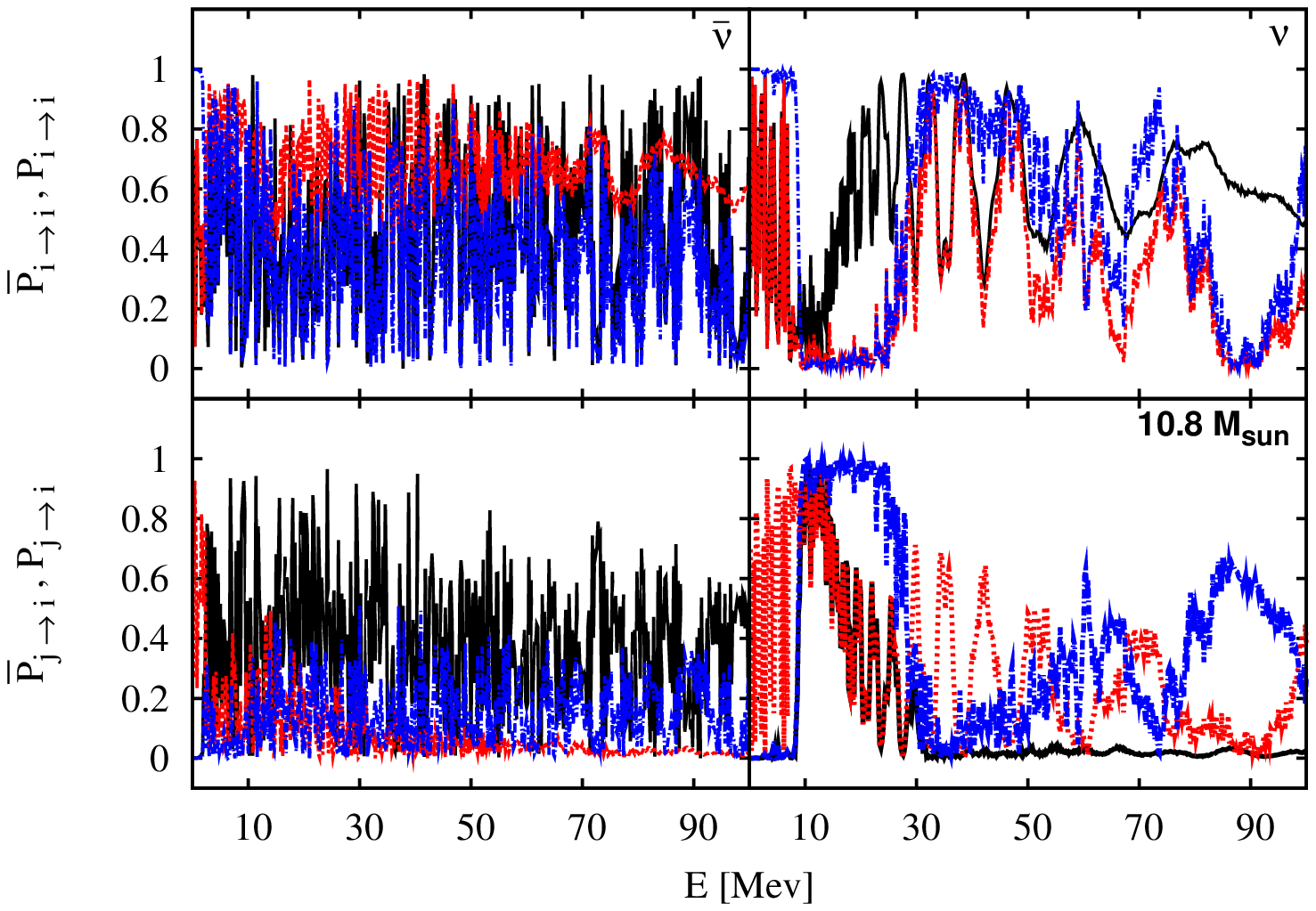}
\includegraphics[width=0.49\linewidth]{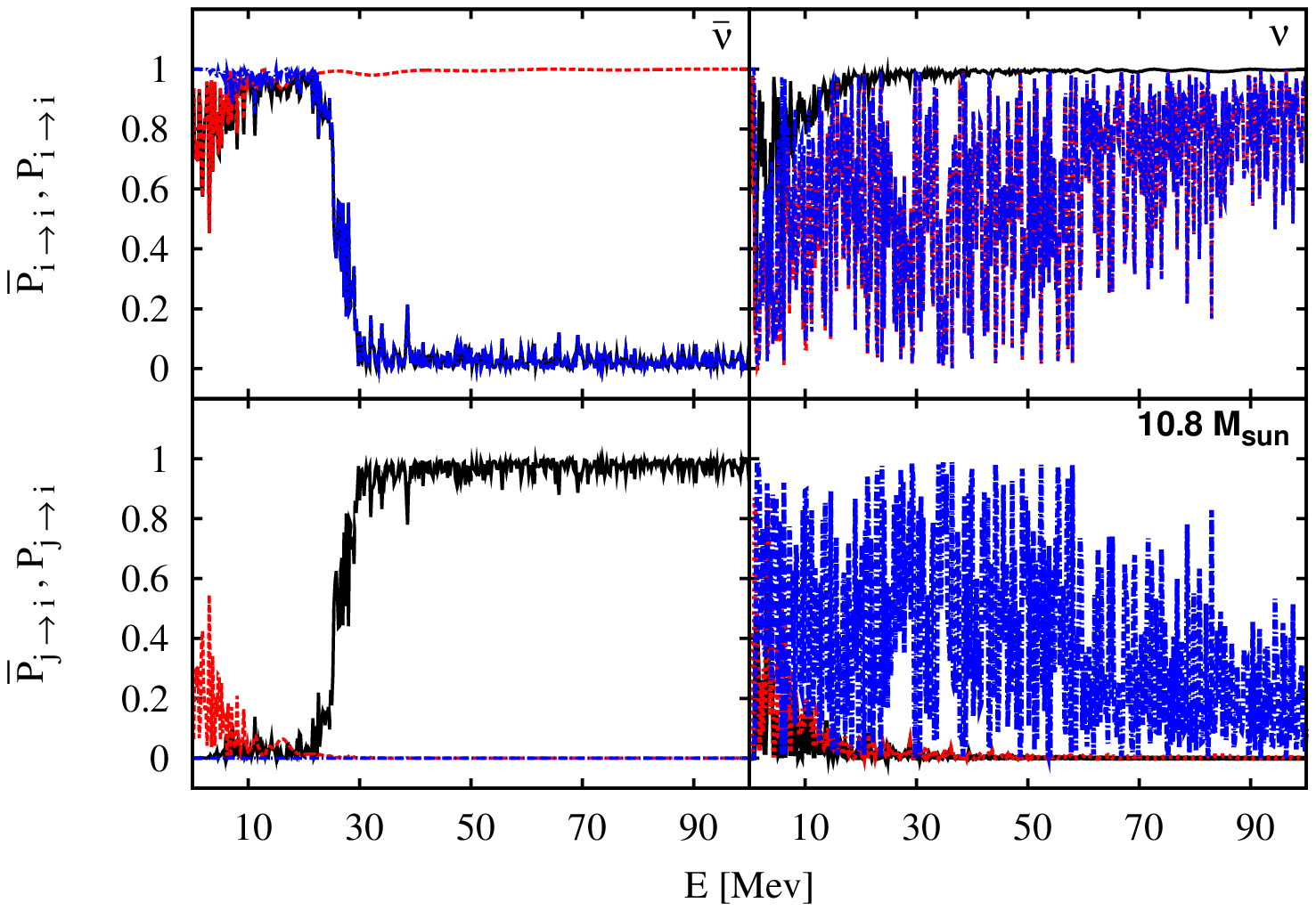}
\caption{\label{fig:30perctTurb} (color online).  Matter
survival and transition probabilities at 2.8~s with 30\%
turbulence for the \Mten{} model.  In the four panels on the
left we show the IH and in the four panels on the right we show
the NH. Lines and layout as before.\\}
\includegraphics[width=0.49\linewidth]{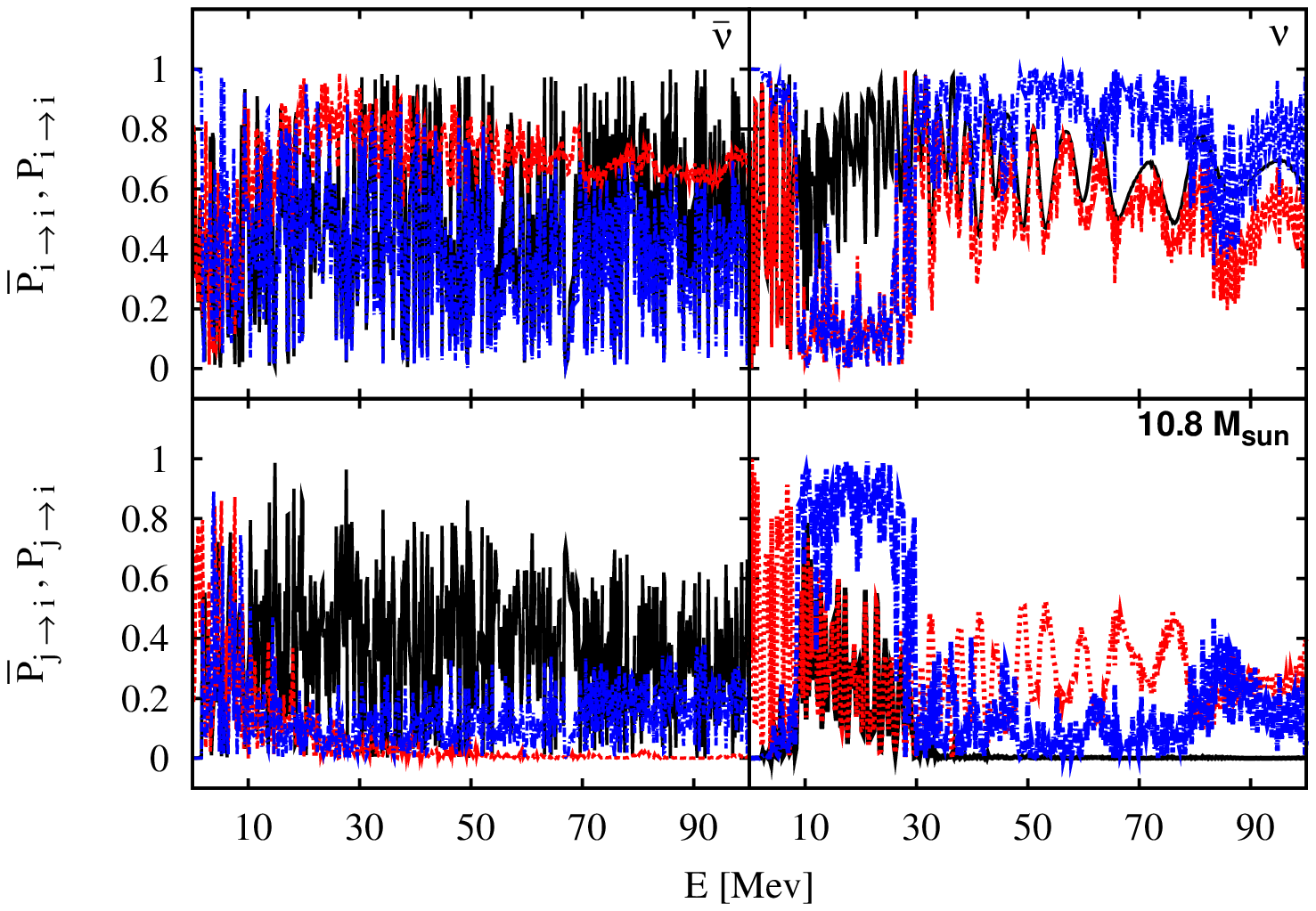}
\includegraphics[width=0.49\linewidth]{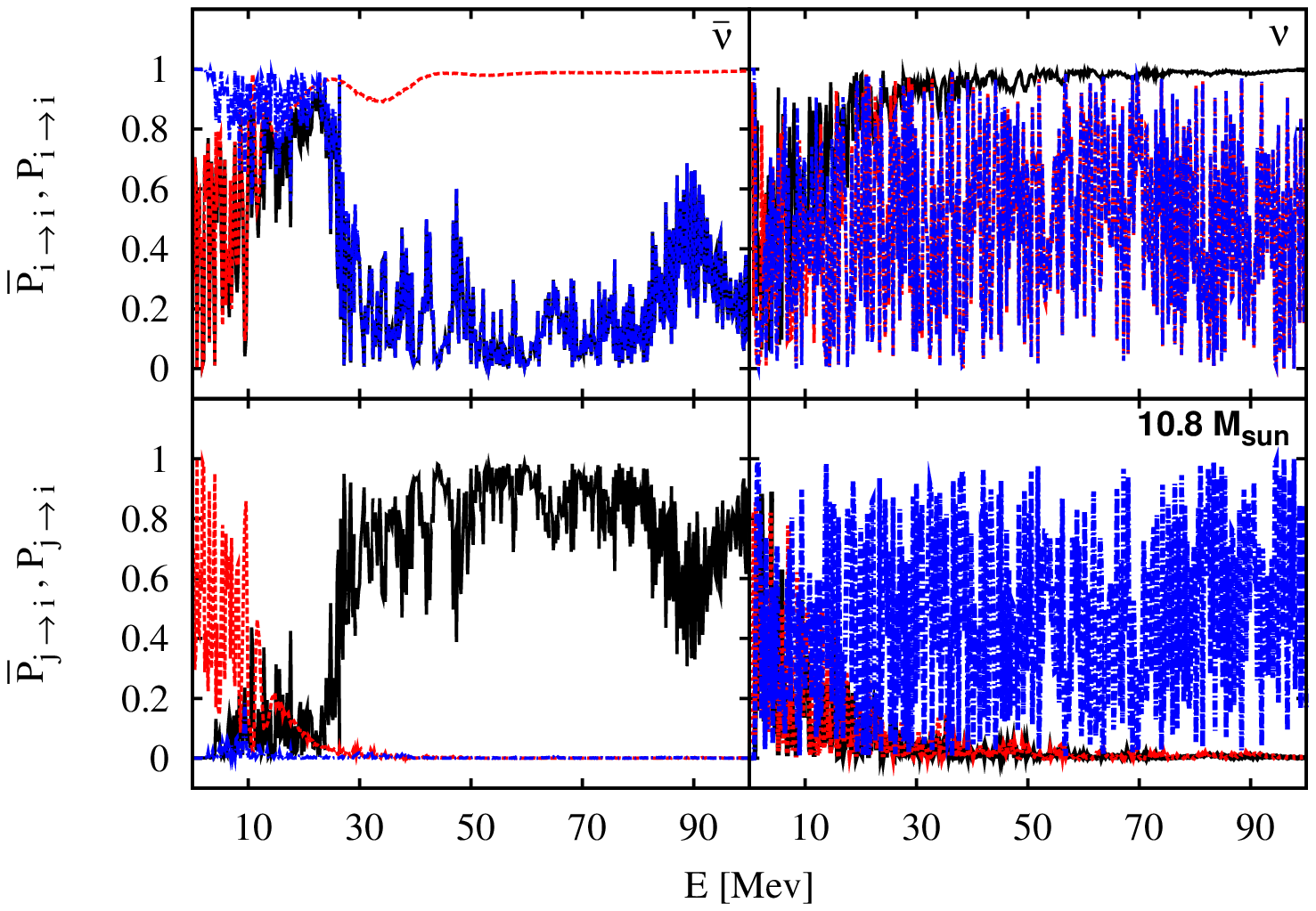}
\caption{\label{fig:50perctTurb} (color online).  As
Fig.~\ref{fig:30perctTurb} but with 50\% turbulence instead of
30\%.} \end{figure*}
So far we have investigated a turbulence amplitude of 10\% and 
seen in the previous section that it has rather limited effect.
The reason is that we are using a value of $\theta_{13}$ in line
with the present measurements which tends to make H resonances
more adiabatic \cite{2013arXiv1302.3825K}. Turbulence
amplitudes of 10\% are the most one might expect to occur in ONeMg
supernovae such as the \Mei{} model because even spherical
simulations of these supernovae successfully explode. But in
contrast, for more massive progenitors spherical models do not
explode and multi-dimensional physics of some kind is necessary.
Such circumstances would naturally lead to aspherical explosions
and the generation of large amplitude turbulence. Therefore we
extended the investigation for the 3~sec profile to 30\% and
50\% turbulence ($C_* = 0.3$ and $0.5$). For these investigations 
we primarily focused on the \Mten{} model, but we have also done 
calculations for the \Met{} model with 30 and 50\% turbulence.

It is evident from Figures~\ref{fig:30perctTurb} and
\ref{fig:50perctTurb} that a number of the main superposed
collective and MSW features remain but the amplitude of the
phase effect oscillations increase, and we also see a shift in
some of the high frequency oscillations. The most interesting
part of  Figures~\ref{fig:30perctTurb} and \ref{fig:50perctTurb}
is that for both levels of turbulence it is always possible to
identify the two spectral splits in neutrino states 2 and 3 in
the IH, and the single split in anti-neutrino states \bone{} and
\bthree{} in the NH, although as a general rule the amplitude of
all fluctuations increases with turbulence increasing from 10\%
to 30\% and 50\%.  This increase in fluctuation amplitude makes
it impossible to identify the trends visible in the full profile
results for anti-neutrinos in the IH, and it gradually obscures
the split in the NH between neutrino states 2 and 3, until not
even the high energy trend is visible at 50\% turbulence.
In the IH additional large amplitude, low frequency oscillations
arises at high energies in the neutrino probabilities as the
turbulence is increased.

By comparing the upper left panel of the right quartet in
Fig.~\ref{fig:30perctTurb} (anti-neutrino survival
probabilities) with the one in Fig.~\ref{fig:50perctTurb} we
see that although the survival probability of both anti-neutrino
states \bone{} and \bthree{} drops above 30 MeV, then in
Fig.~\ref{fig:30perctTurb} they drop to zero but in
Fig.~\ref{fig:50perctTurb} the average probabilities only drops
to about 0.2. From the transition probabilities (lower left
panels) we see that the probability for anti-neutrino state
\bthree{} to go into state \bone{} has fallen from 100\%
to an average of 80\% when the turbulence was increased. Thus
instead of completely converting states \bone{} and
\bthree{} into one another, we now have a small admixture
of the original state.  At the PNS an $\bar\nu_e$ is created in
matter state $\bar\nu_3$ in the NH but it will travel
predominantly as mass state $\bar\nu_1$ in vacuum. From the
transition probabilities we see that this means 80\% of the
$\bar\nu_3$ state gets converted into $\bar\nu_1$, which means
it stays as an $\bar\nu_e$. Naturally this entails that about
20\% of the energy spectrum from neutrinos initially created as
$\bar\nu_x$ will be mixed in with about 80\% of the energy
spectrum of neutrinos initially created as $\bar\nu_e$ (aka
$\bar\nu_3$) to form the final energy spectrum of the state we
would observe as $\bar\nu_e$ (aka $\bar\nu_1$).

As this conversion is happening at the higher energies, above
30~MeV, we get the higher energy non-electron flavor
contribution added to the spectrum, increasing the number of
higher energy neutrinos, which with current detector technology
is the easiest to observe.

\begin{figure}[t!]
\includegraphics[width=\linewidth]{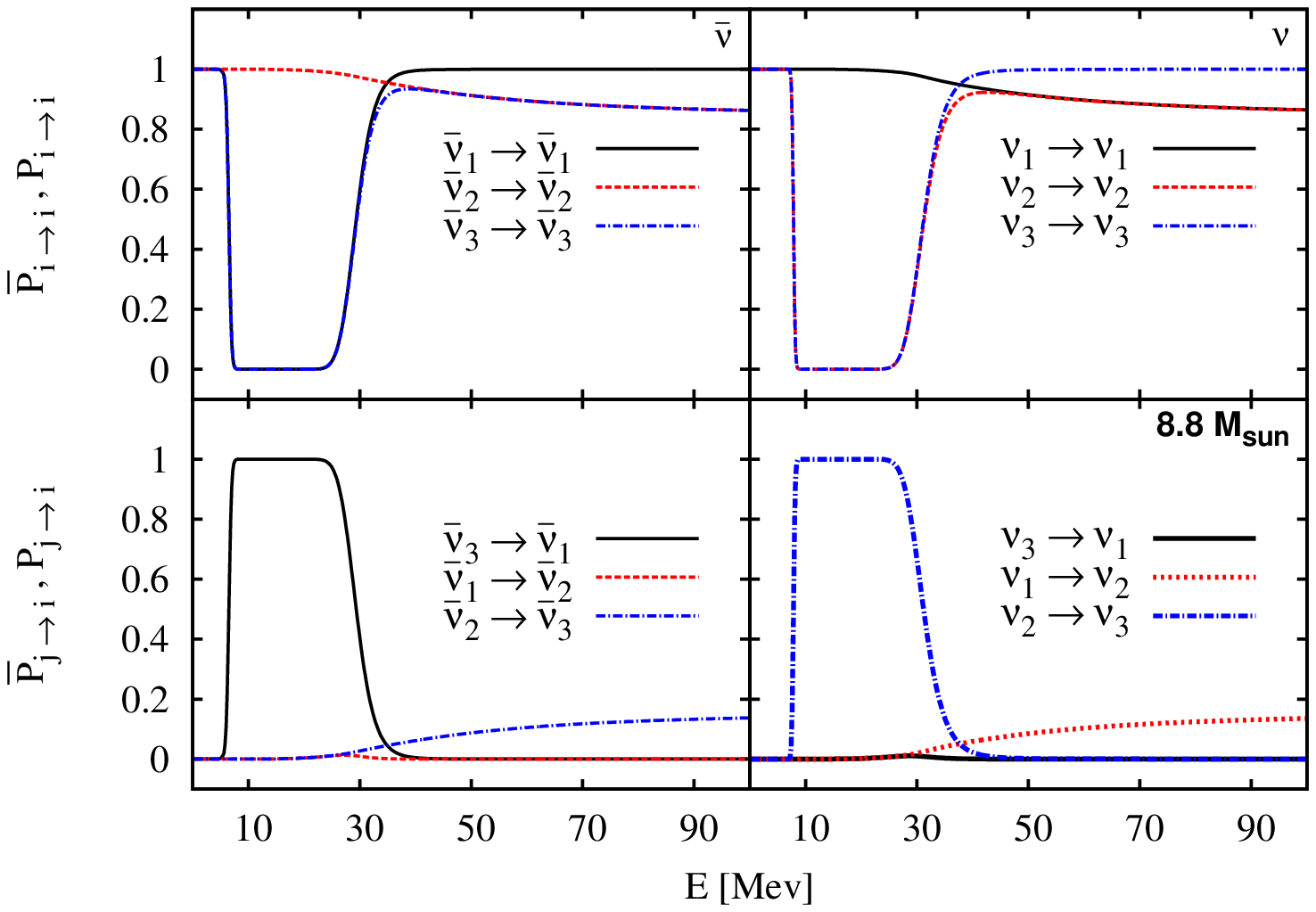}
\includegraphics[width=\linewidth]{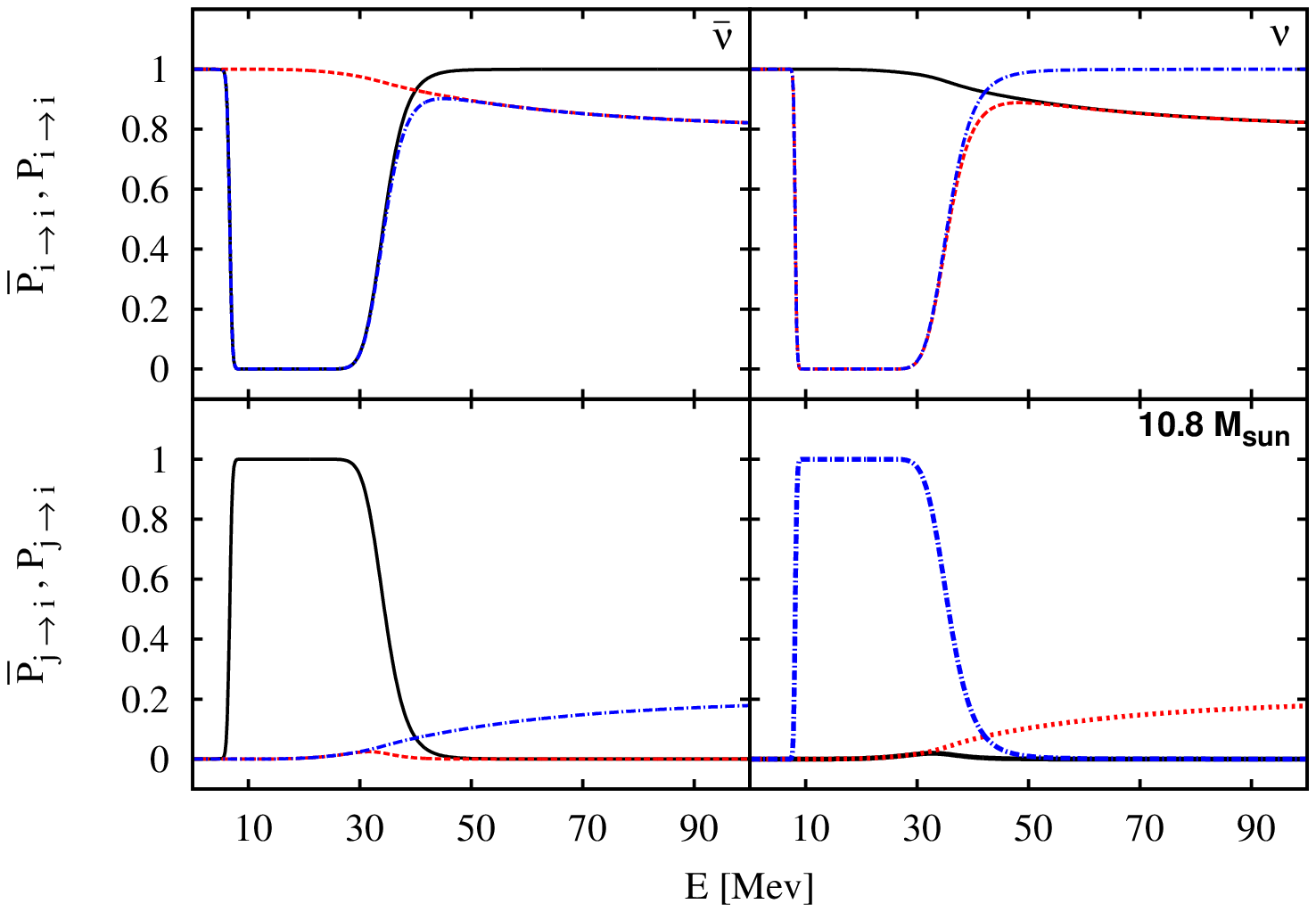}
\includegraphics[width=\linewidth]{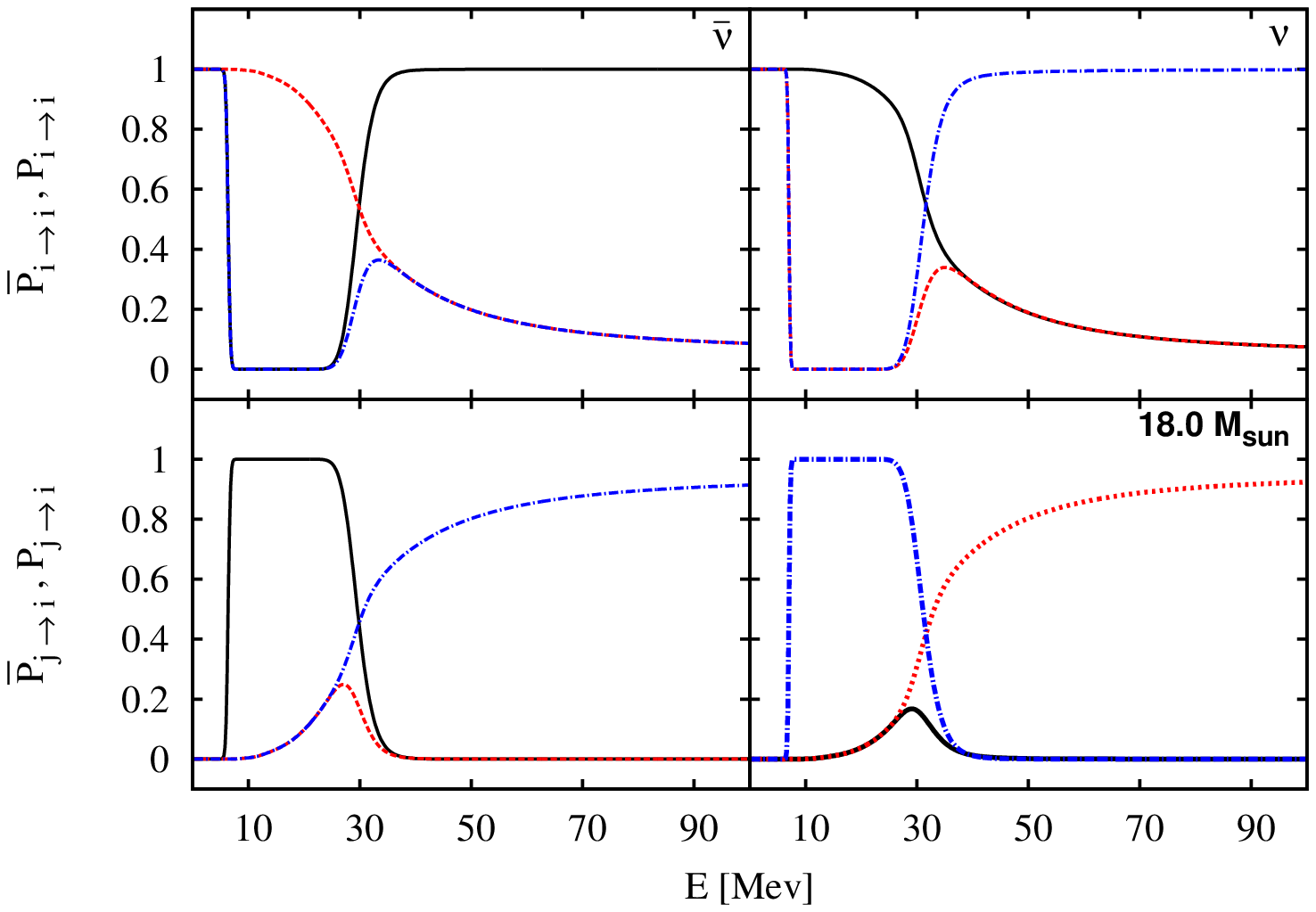}
\caption{\label{fig:coIH_1sec} (color online).  As
Fig.~\ref{fig:coIH_3sec} but for 1~s pb.  Results from the inner
region in the IH.  Lines and layout as before.} 
\end{figure}

\begin{figure}[t!]
\includegraphics[width=\linewidth]{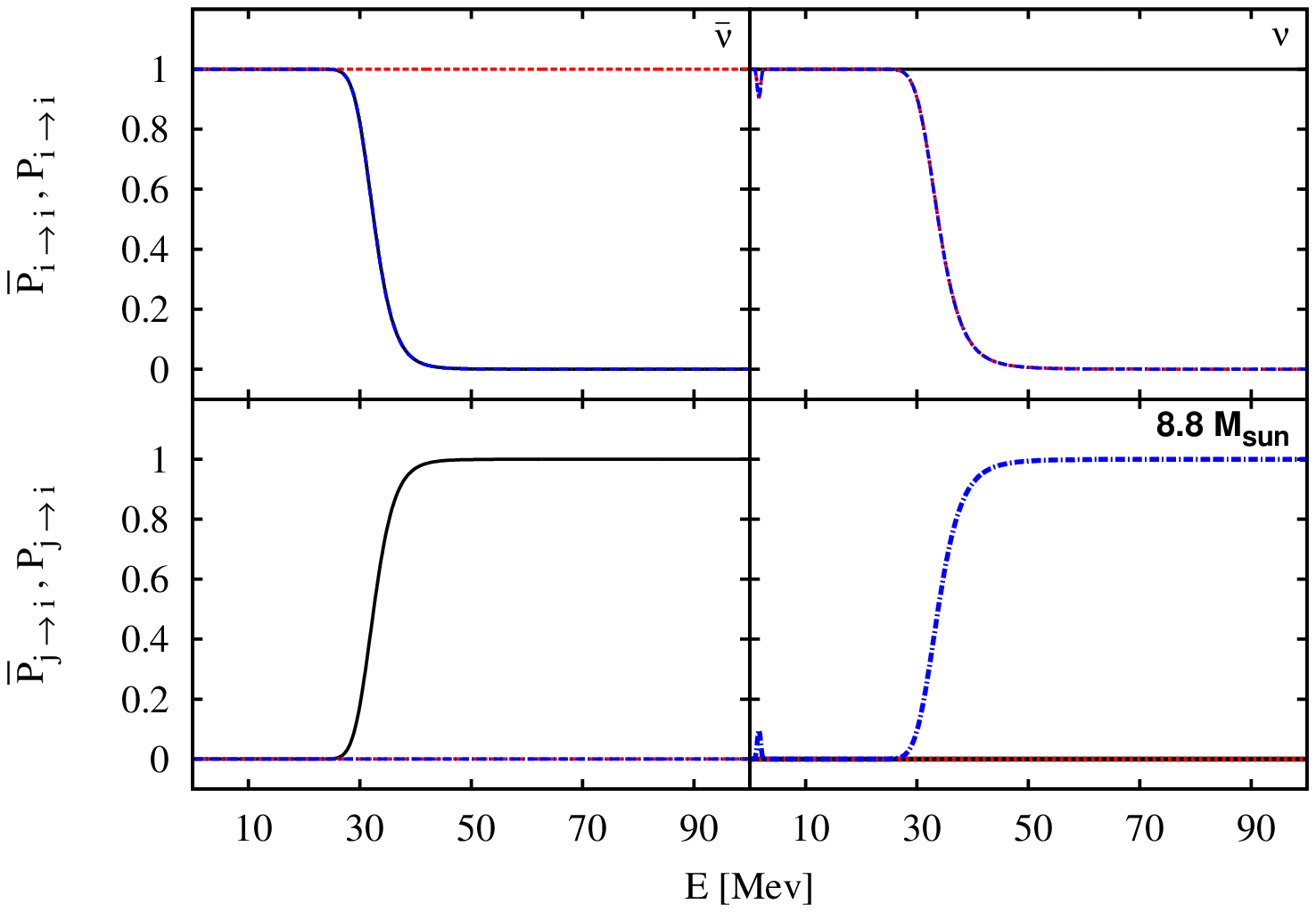}
\includegraphics[width=\linewidth]{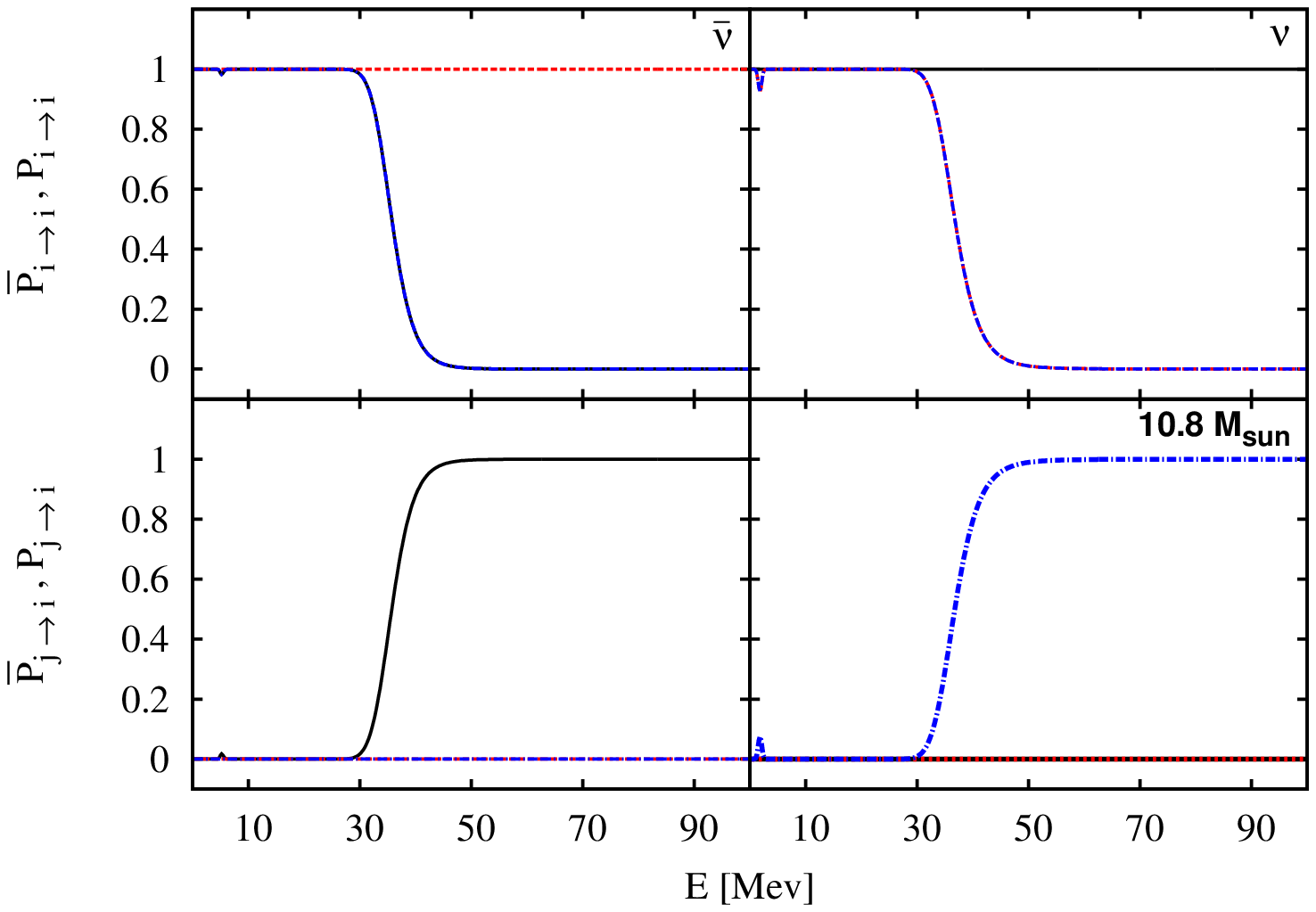}
\includegraphics[width=\linewidth]{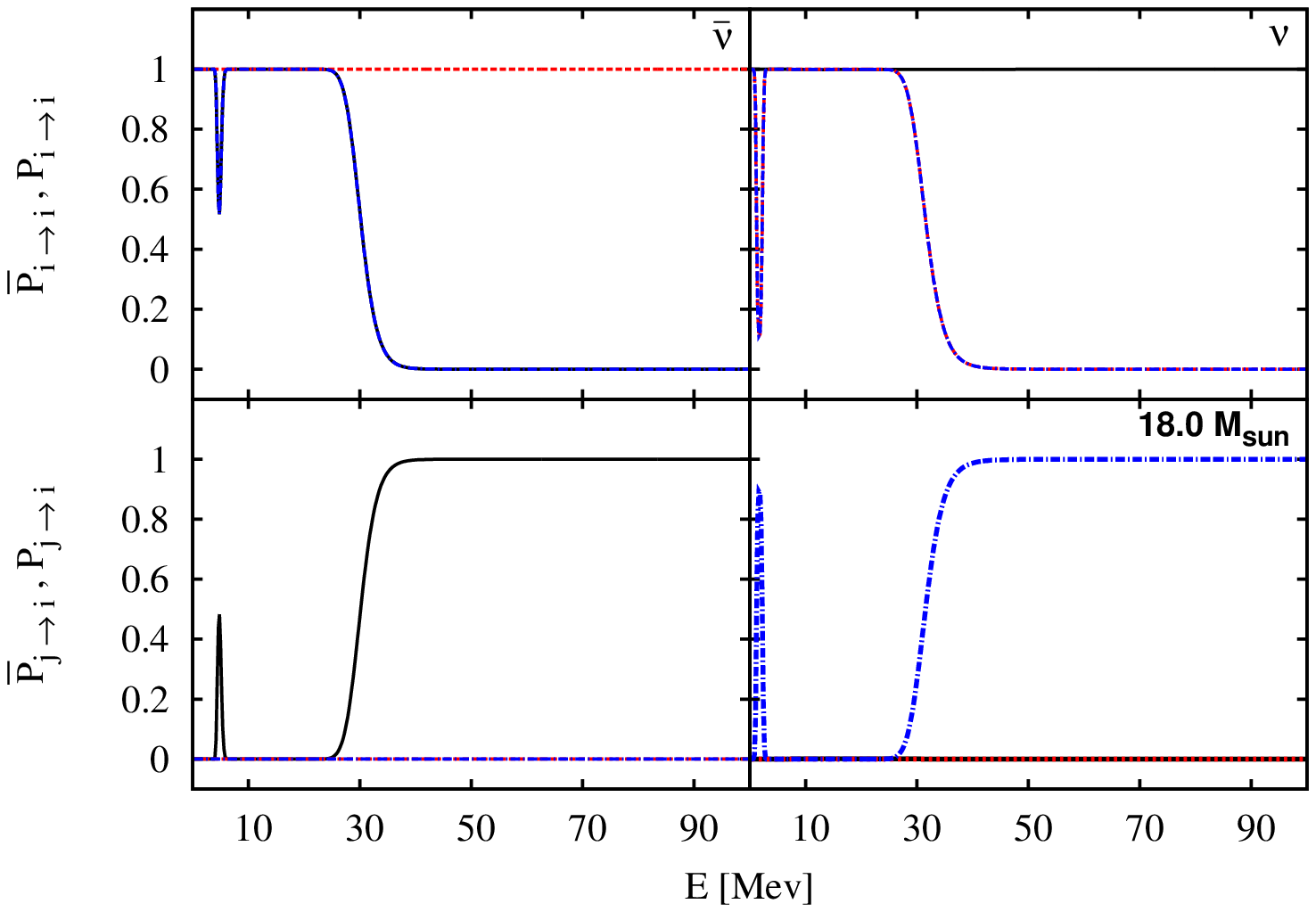}
\caption{\label{fig:coNH_1sec} (color online).   As
Fig.~\ref{fig:coNH_3sec} but for 1~s pb.  Results from the inner
region in the NH.  Lines and layout as before.} 
\end{figure}

\subsection{Profiles at 1~second}         \label{sec:1s_results}
\subsubsection{Inner region, 70 -- 1000 km: Collective dominated} 
Now that we understand how the various neutrino flavor transformation
effects combine we can turn our attention to the density profiles
at other times.  We begin with density profiles at 1~s~pb,
and the calculation results for the inner, collective dominated, region
are shown in Figures~\ref{fig:coIH_1sec} and
\ref{fig:coNH_1sec}. The line colors and styles are the same
as in Figures~\ref{fig:coIH_3sec} and \ref{fig:coNH_3sec}.

Let us focus on the anti-neutrinos first. For the IH
(Fig.~\ref{fig:coIH_1sec}) a common trait in all three models
is a spectral split at a very low energy ($\sim$~5--7~MeV) and a
second split at an intermediate energy ($\sim$~30--35~MeV) in
the anti-neutrino states \bone{} and \bthree{}. From the top
left panels of each quartet we see that between these two
energies the survival probabilities of the two states drop to
zero. For anti-neutrino states \bthree{} and \btwo{} we also see
a decrease in survival probability at energies above
$\sim$~35--40~MeV. From the transition probability plots we
infer that at these energies the anti-neutrino states \btwo{}
and \bthree{} have made an incomplete swap. In the case of the
two lighter models this incomplete swap is very small, but for
the \Met{}  model the incomplete swap is significant and
approaches a full swap, and starts at a slightly
lower energy.

In the NH (Fig.~\ref{fig:coNH_1sec}) we again see a common 
feature across the progenitors: a complete swap of anti-neutrino
states \bone{} and \bthree{} at an energy of $\sim$~30--35~MeV. 
For neutrinos we see the same feature
in the same energy range, but now between matter states 2 and 3.
They also show additional dips in the survival
probabilities at energies of 2--5~MeV for matter states 2 and 3
for all three models. The \Met{} model show a similar feature in
the anti-neutrino states \btwo{} and \bthree{} at the same
energy.

In the IH neutrinos show multiple interesting features as
evident from the right panels of Fig.~\ref{fig:coIH_1sec}. The
most prominent features are two spectral splits at
$\sim$~7--8~MeV and $\sim$~31--36~MeV between matter states 2
and 3, leading to a full swap of the two states in the energy
region in-between. This is not unlike the splits seen between
states \bone{} and \bthree{} in the anti-neutrinos. As in the
anti-neutrino case the \Met{} progenitor has an additional swap
in the neutrinos. This time the swap occurs at an energy of
$\sim$~30~MeV and between matter states 1 and 2.

Although the quartets for the 8.8 and 10.8~$M_\odot$ models
appear quite similar they are in fact different. This is most
easily seen at the crossing point between $P_{11}$ and $P_{33}$
near 35~MeV in the upper right panel for the 8.8~$M_\odot$
model, which are found at 40~MeV for the \Mten{} model.

The similarity of the probabilities across the three progenitors
is again striking. The explanation follows the same lines as
for the 3~s results. The ratios between the fluxes of $\nu_e$
and $\nu_x$, as well as between $\bar\nu_e$ and $\bar\nu_x$, are
as similar for the three progenitors at 1~s as they were at
3~s, which can be easily computed from the luminosities and
mean energies given i table~\ref{tab:LandE}. The overall flux
ratios are about 10\% higher at 1~s than at 3~s so we would
expect the collective interaction to be stronger which is what
we see. The larger interaction strength explains why we see full
swaps between anti-neutrino states \bone{} and \bthree{} in the
IH at 1~s and why those swaps have disappeared at 3~s.
Furthermore, at 1~s the density profiles of the \Mten{} and
the \Met{} models are almost identical out to $\sim$700~km
(upper panel of Fig.~\ref{fig:rhos}). The
\Mei{} density profile follows the other two closely out to
100~km where after it falls off much quicker. The similarity of
the density profiles thus corroborate the similarity of the
results for the three progenitors.

Comparing in more detail the results from the calculations of 
the inner region at 1~s to the ones at 3~s,
we find that the spectral split between
\bone{} and \bthree{} in the NH is present at both 1 and 3~s
but has moved from an energy of 30--36~MeV at 1 s to
24--26~MeV at 3~seconds 
(Compare Figures~\ref{fig:coIH_1sec} and \ref{fig:coNH_1sec} with 
Figures~\ref{fig:coIH_3sec} and \ref{fig:coNH_3sec}).  
The split between neutrino states 2
and 3 in the NH has likewise moved down in energy from 
32--37~MeV at 1~s to 27--30~MeV at 3~s, and of the two 
splits present in neutrino states 2 and 3 the lower one at 
7--8~MeV remains in place, but the one at higher energies has 
moved down from 31--36~MeV to 24--28~MeV from 1~s to 3~s.
The explanation of the movement in energy of the spectral splits
is unclear, but the movement continues over time as is shown in
Sect.~\ref{sec:time_evolv}.
Counter intuitively the small dip in the survival probabilities 
$P_{22}$ and $P_{33}$ at $\sim2$~MeV grow into a full double 
split from 1 to 3~s although the fluxes fall off by 10\% and
the interaction strength would be expected to diminish too.
In the IH the incomplete swaps in the anti-neutrino states 
\btwo{} and \bthree{} above 30~MeV persist from 1 to 3~s.\\

From the top panel of Fig.~\ref{fig:rhos} we clearly see that
within 1000~km the 1~s density profiles of models \Mten{} and
\Met{} have values at least a magnitude larger than the resonant
densities for the MSW effect. We therefore conclude that all the
effects we observe in these two cases must be caused by
collective effects.  Furthermore, since the traversal of the
high density MSW resonance that do take place is adiabatic for
the \Mei{} model, and the probability plots show such
similarities to the ones for the \Mten{} model, we conclude that
the changes in probabilities are also in this case caused by
collective effects.

\subsubsection{Outer region, 1000 km -- profile end: MSW
dominated}
At this relatively early time we see 
no significant evolution of the matter states as they travel
through the outer region.  The survival probabilities 
stays mainly at unity. 
Therefore we do not include figures of the results of our 
calculations for this region.
Only for the \Mten{} model do we see a
dip in the survival probabilities for the anti-neutrino states
\bone{} and \bthree{} at an energy of $\sim$~1~MeV in the IH,
and at the same energy for the neutrinos in the NH where matter
states 2 and 3 transition into one another.

The lack of interesting effects in this region is easily
understood by consulting the top panel of Fig.~\ref{fig:rhos}:
The shocks and the contact discontinuity are at densities well
above the resonant densities for the MSW effect. The H and L
resonant densities are therefore traversed adiabatically
by the neutrinos, if the density profile reaches the relevant
density levels. This is unlike the profile at 3 seconds (lower 
panel of Fig.~\ref{fig:rhos}) where the shocks are in the 
resonant density area, and MSW effects can be seen in the 
probabilities. 
Comparing the results at 1~s with the results 
from 3~s (Sec.~\ref{sec:3s_results} Figures~\ref{fig:moIH_3sec} 
and \ref{fig:moNH_3sec}) for the outer MSW dominated region, 
clearly shows how the MSW impact rises over 
time as the shocks move into the relevant density regions. 
Hardly any effect of the MSW is present at 1~s, but at 3~s 
we see changes in the probabilities from both the H and L MSW 
resonances.  This will be discussed further in 
Sec.~\ref{sec:time_evolv}.

\begin{figure}[t!]
\includegraphics[width=\linewidth]{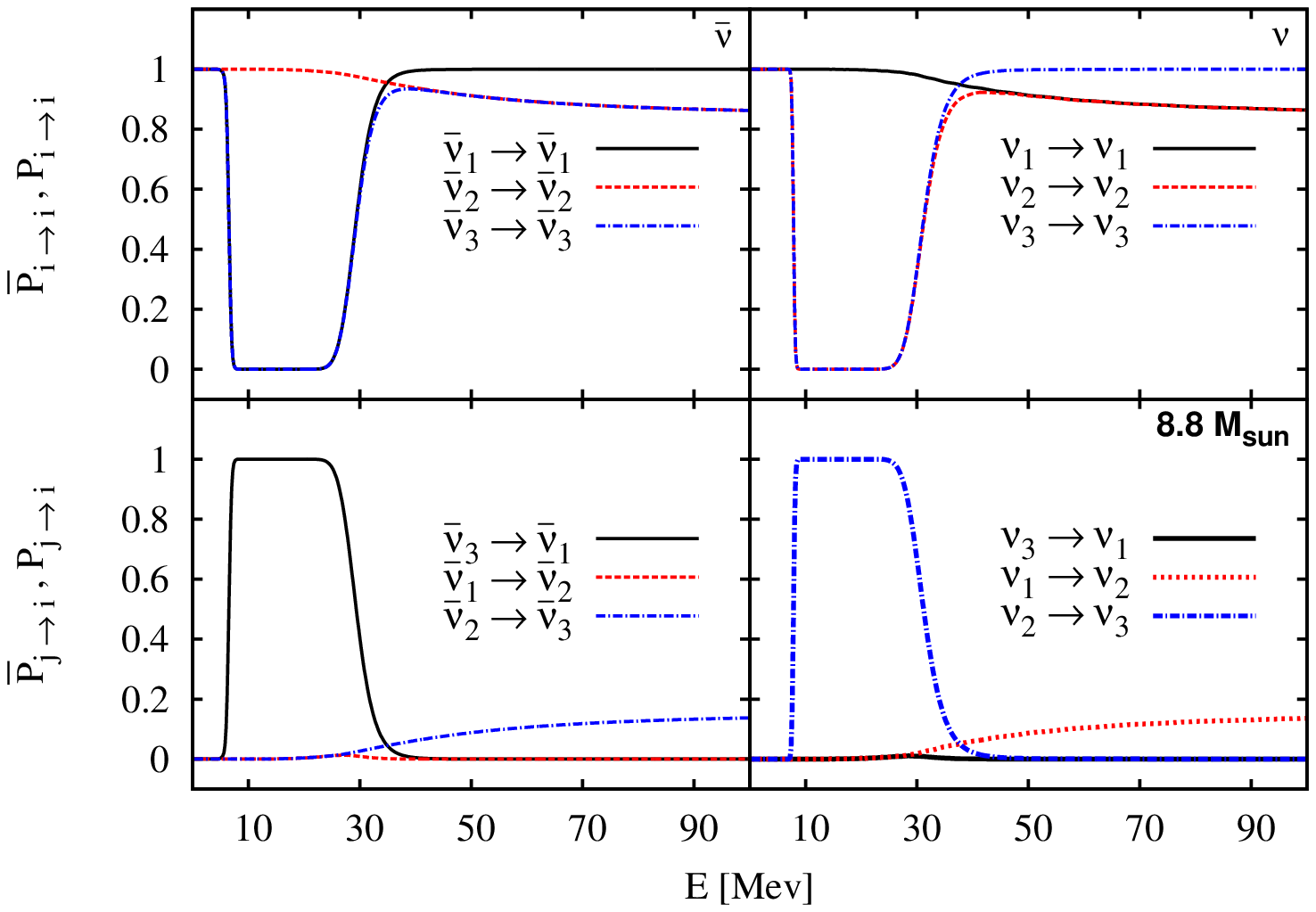}
\includegraphics[width=\linewidth]{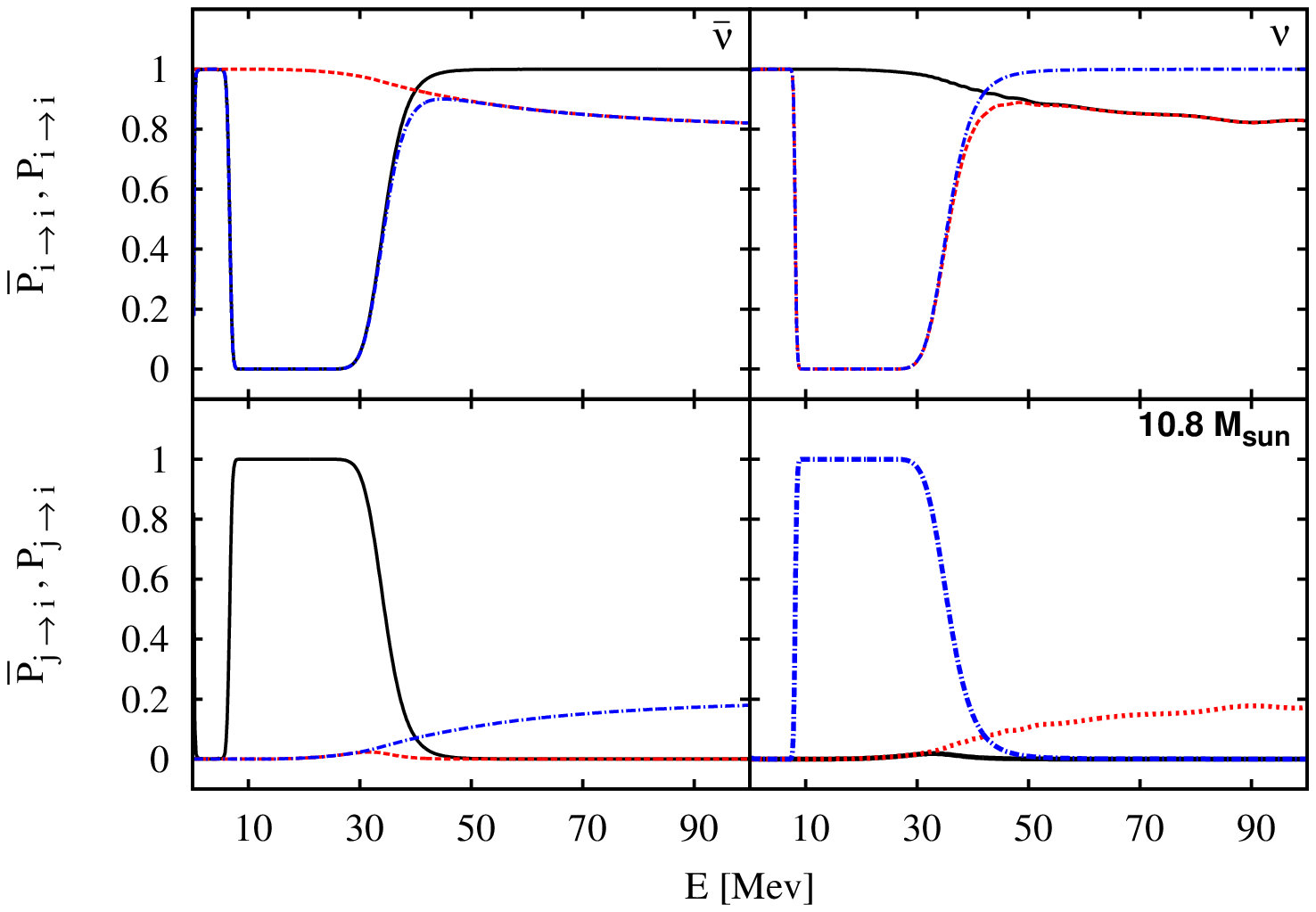}
\includegraphics[width=\linewidth]{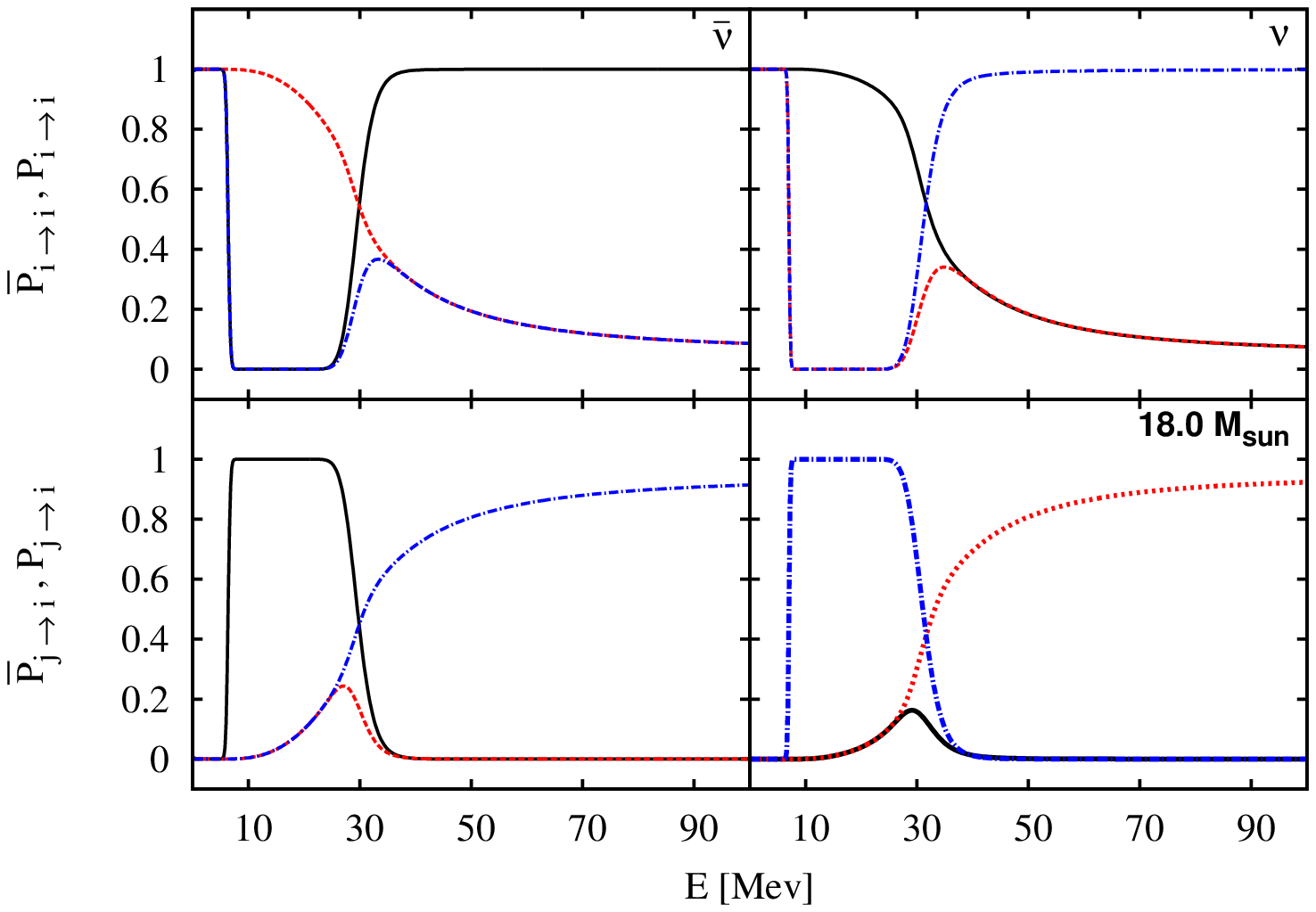}
\caption{\label{fig:nmIH_1sec} (color online).  As
Fig.~\ref{fig:coIH_1sec} but for the full profile traversal.
Inverted Hierarchy.} 
\end{figure}
\begin{figure}[t!]
\includegraphics[width=\linewidth]{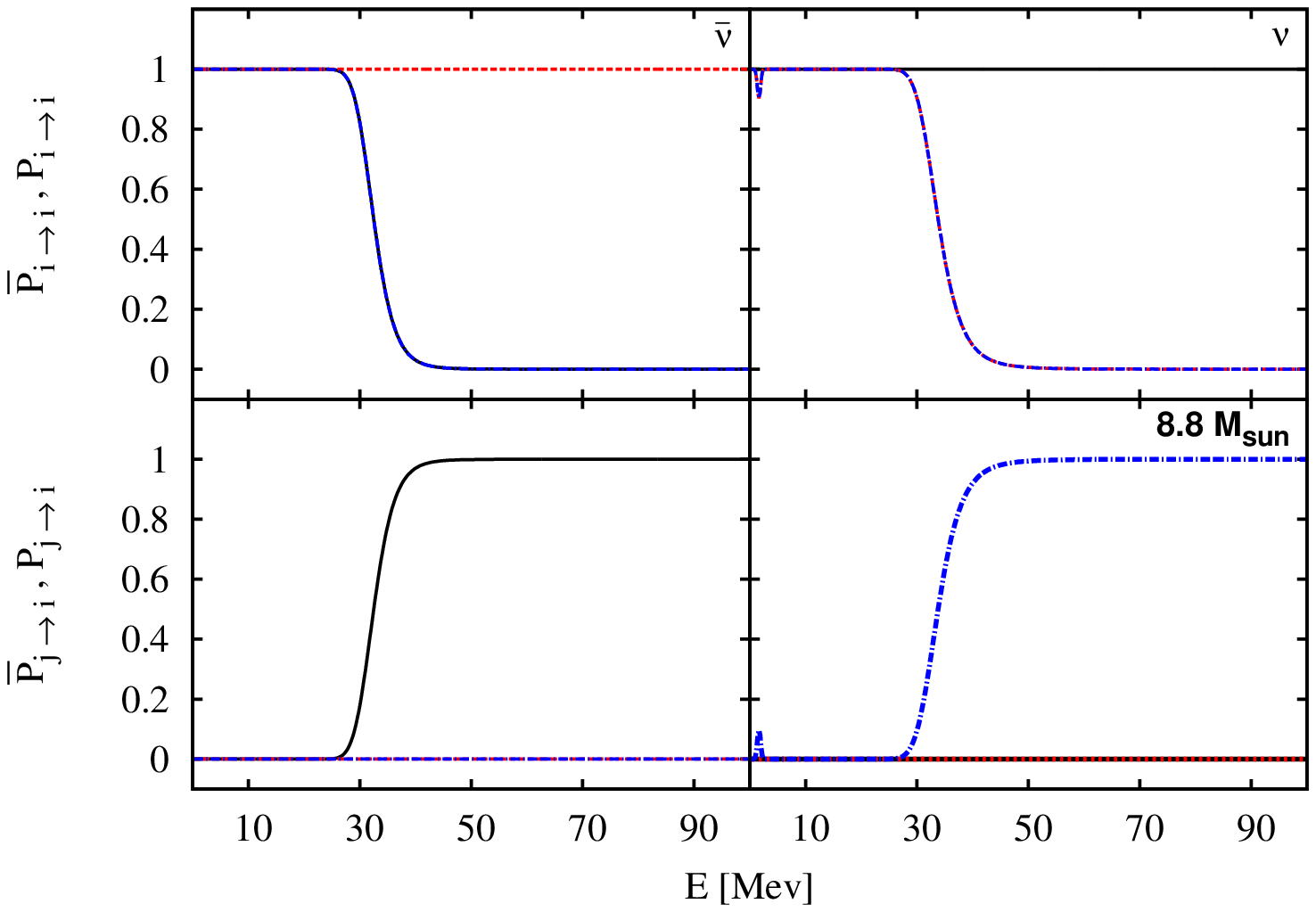}
\includegraphics[width=\linewidth]{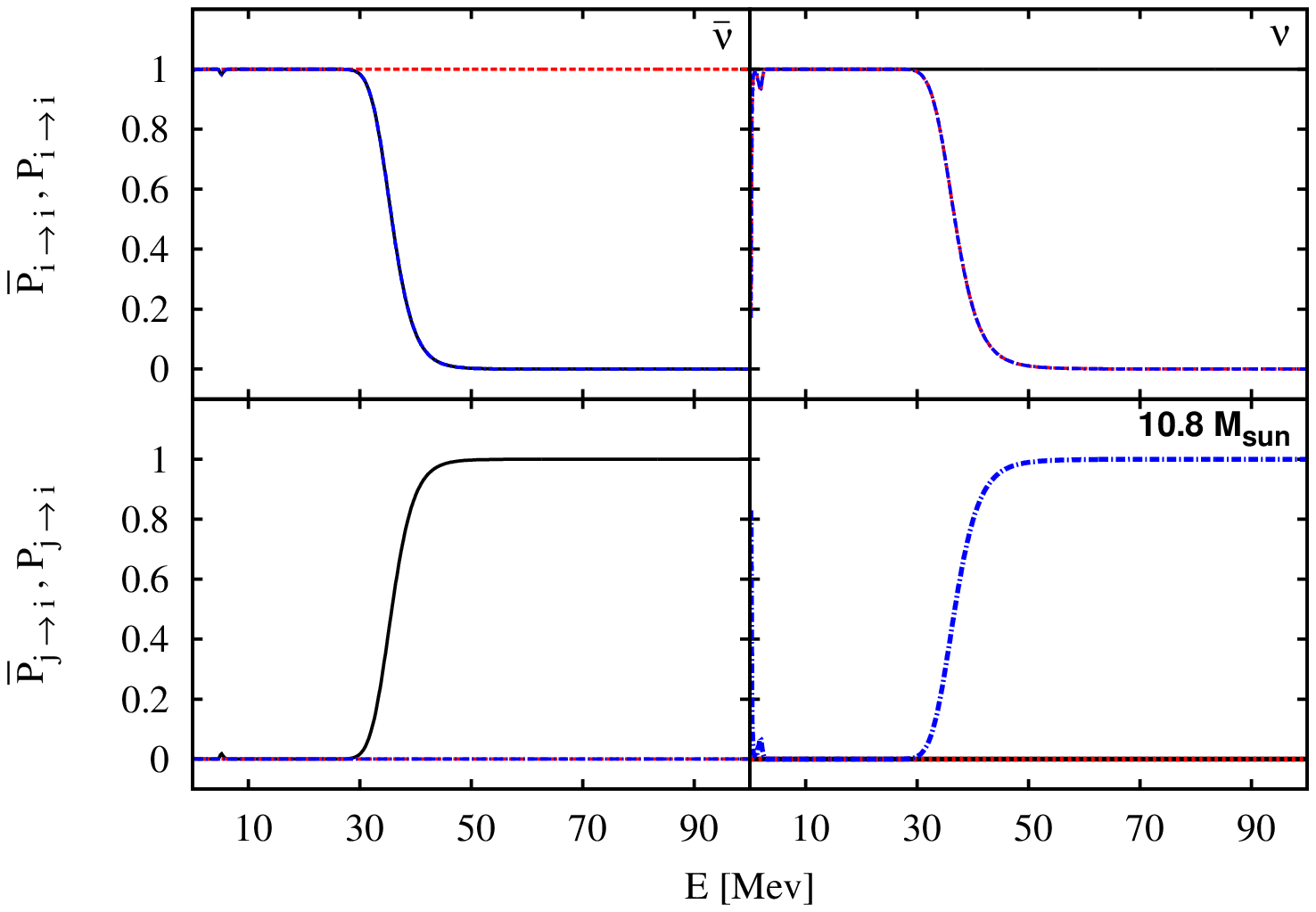}
\includegraphics[width=\linewidth]{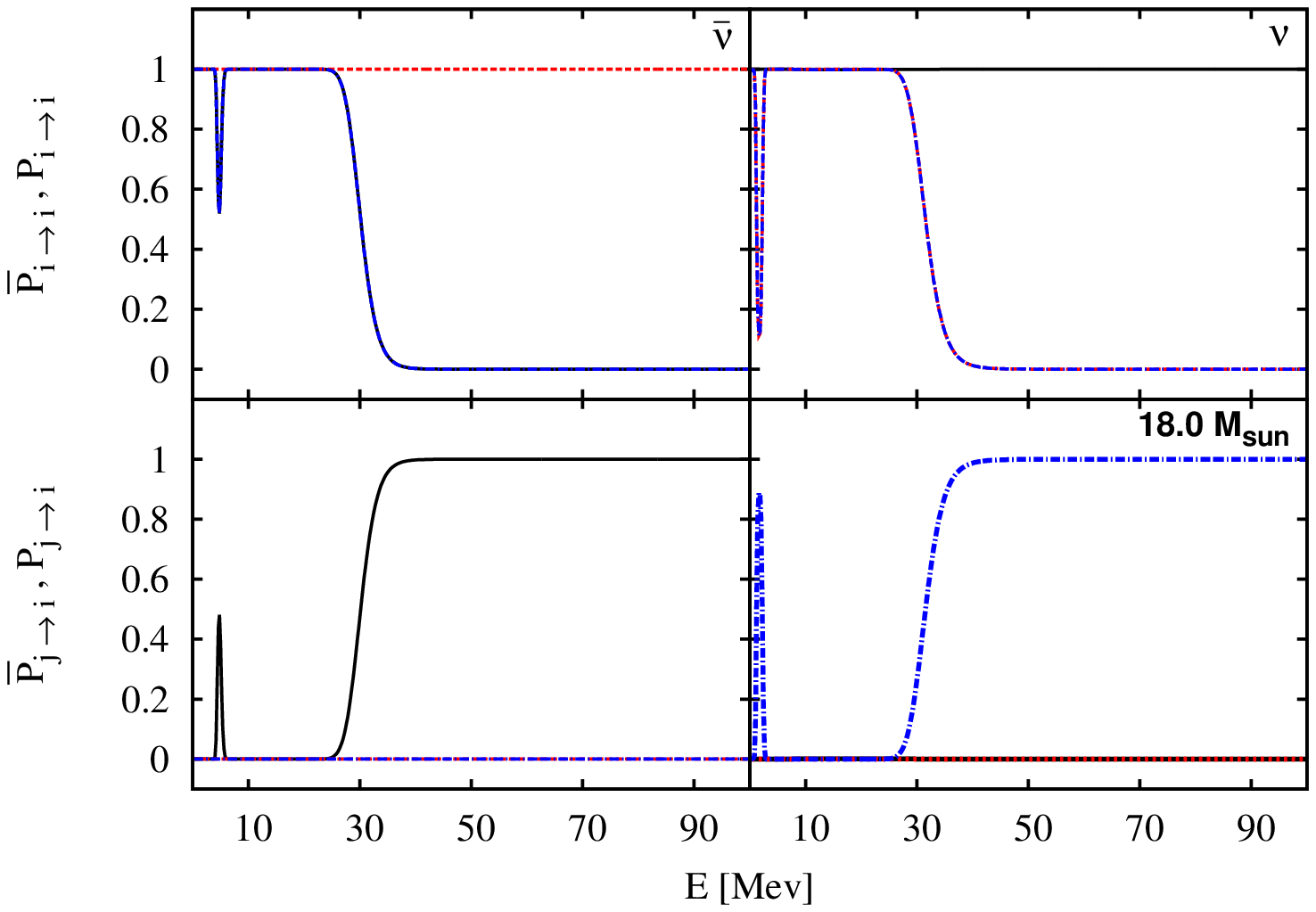}
\caption{\label{fig:nmNH_1sec} (color online).   As
Fig.~\ref{fig:coNH_1sec} but for the full profile traversal.
Normal Hierarchy.} 
\end{figure}

\subsubsection{Full profile traversal} 
For the full profile
traversal we observe that the probabilities are a superposition
of the ones from the inner and outer region, the same result as
we found for the 3~sec profiles. The probabilities for the full
calculations are shown in Figures~\ref{fig:nmIH_1sec} (IH) and
\ref{fig:nmNH_1sec} (NH). Since nothing significant happens in
the outer region they reflect the flavor conversions
occurring in the inner region.

\begin{figure}[t!]
\includegraphics[width=\linewidth]{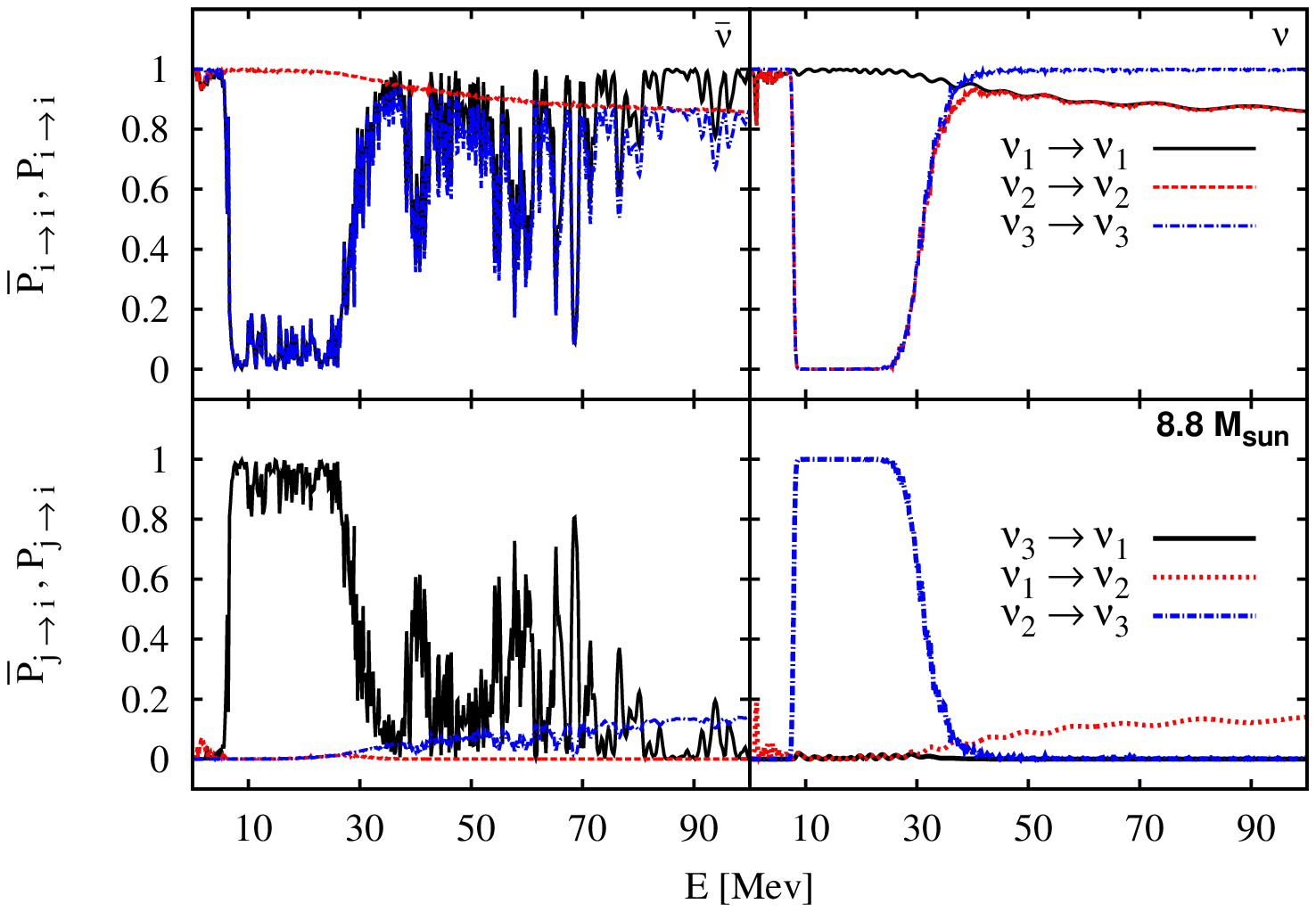}
\includegraphics[width=\linewidth]{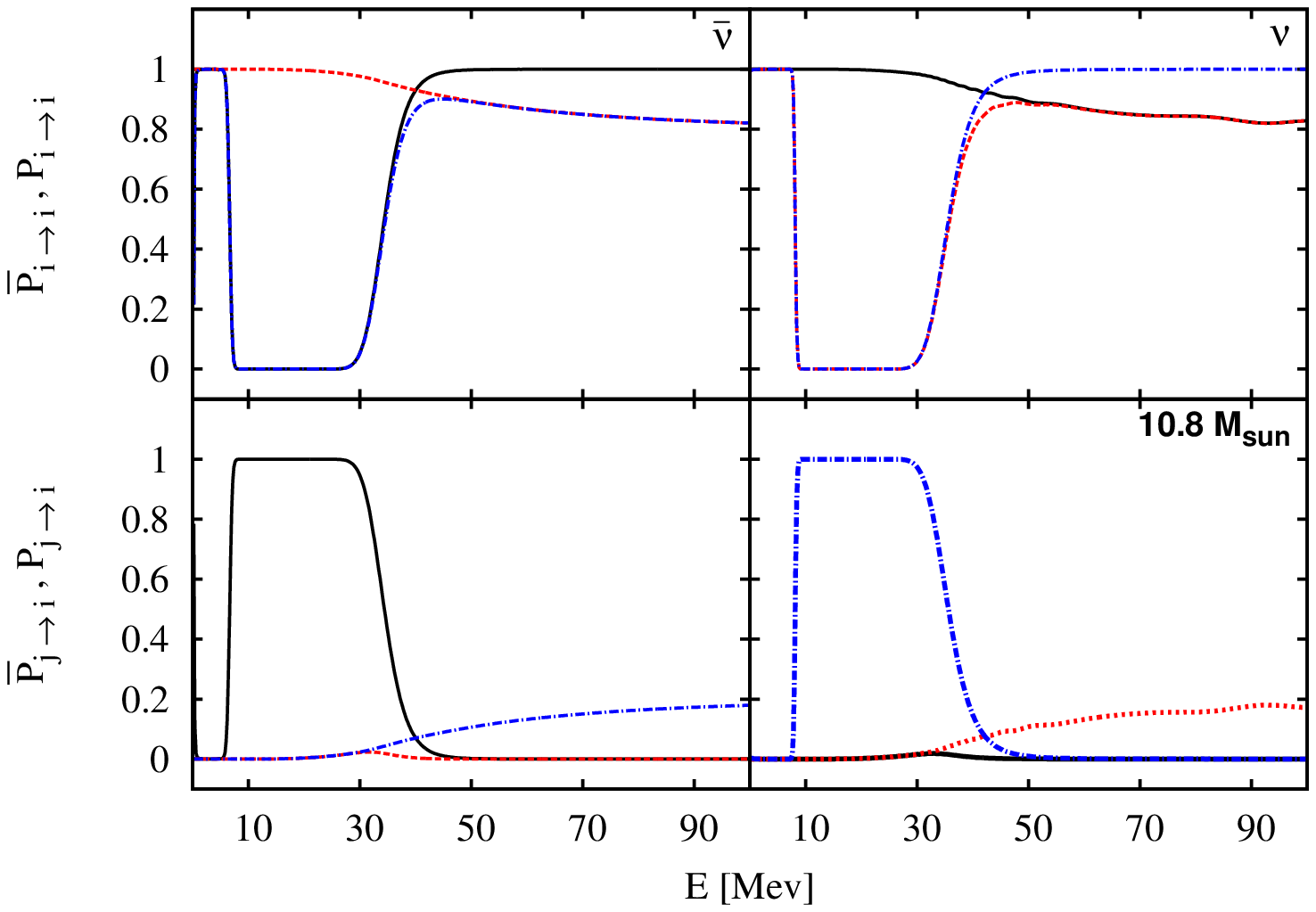}
\includegraphics[width=\linewidth]{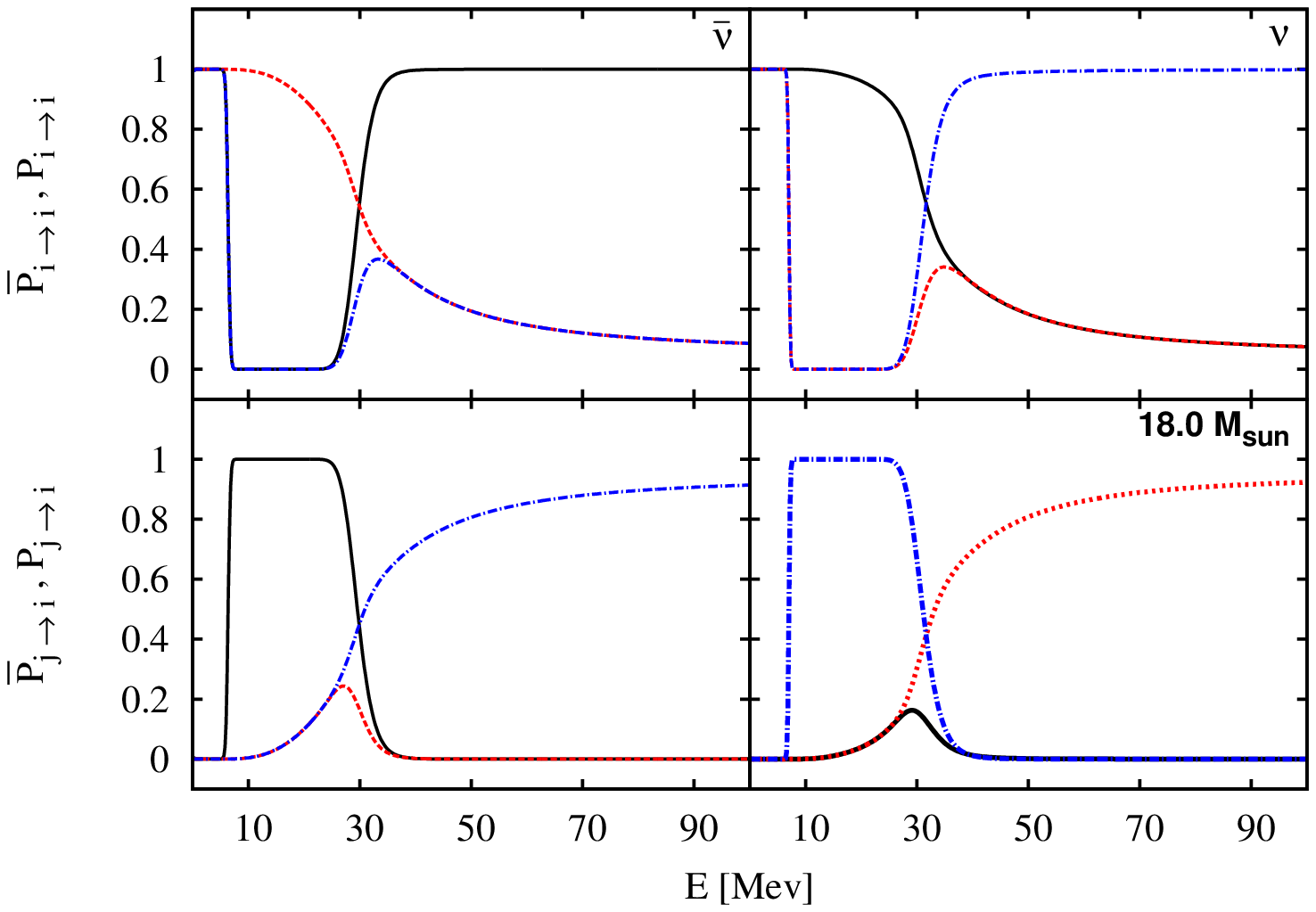}
\caption{\label{fig:tnmIH_1sec} (color online).  As
Fig.~\ref{fig:coIH_1sec} but for the full profile traversal with
10\% turbulence.  Inverted Hierarchy.} \end{figure}

\begin{figure}[t!]
\includegraphics[width=\linewidth]{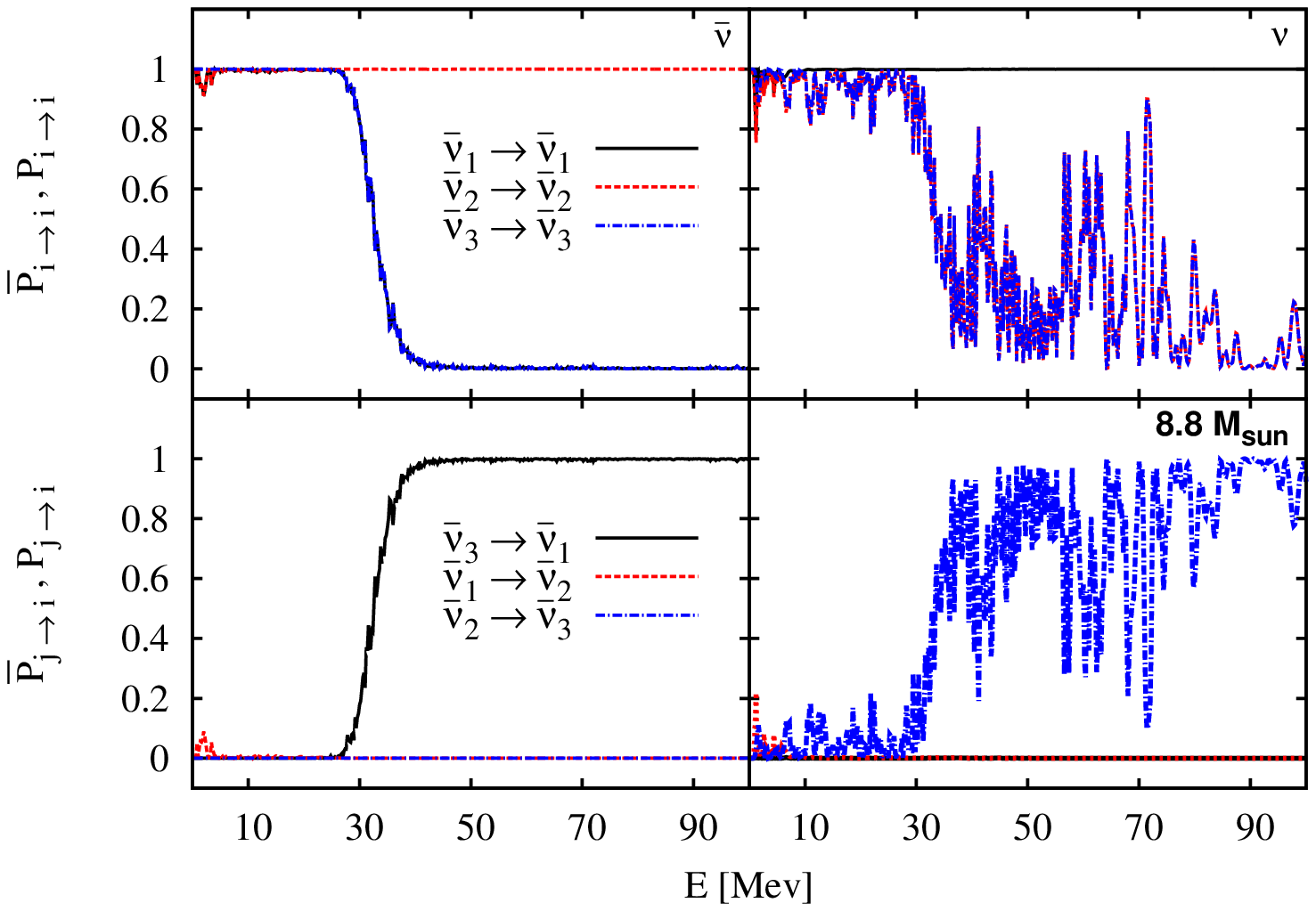}
\includegraphics[width=\linewidth]{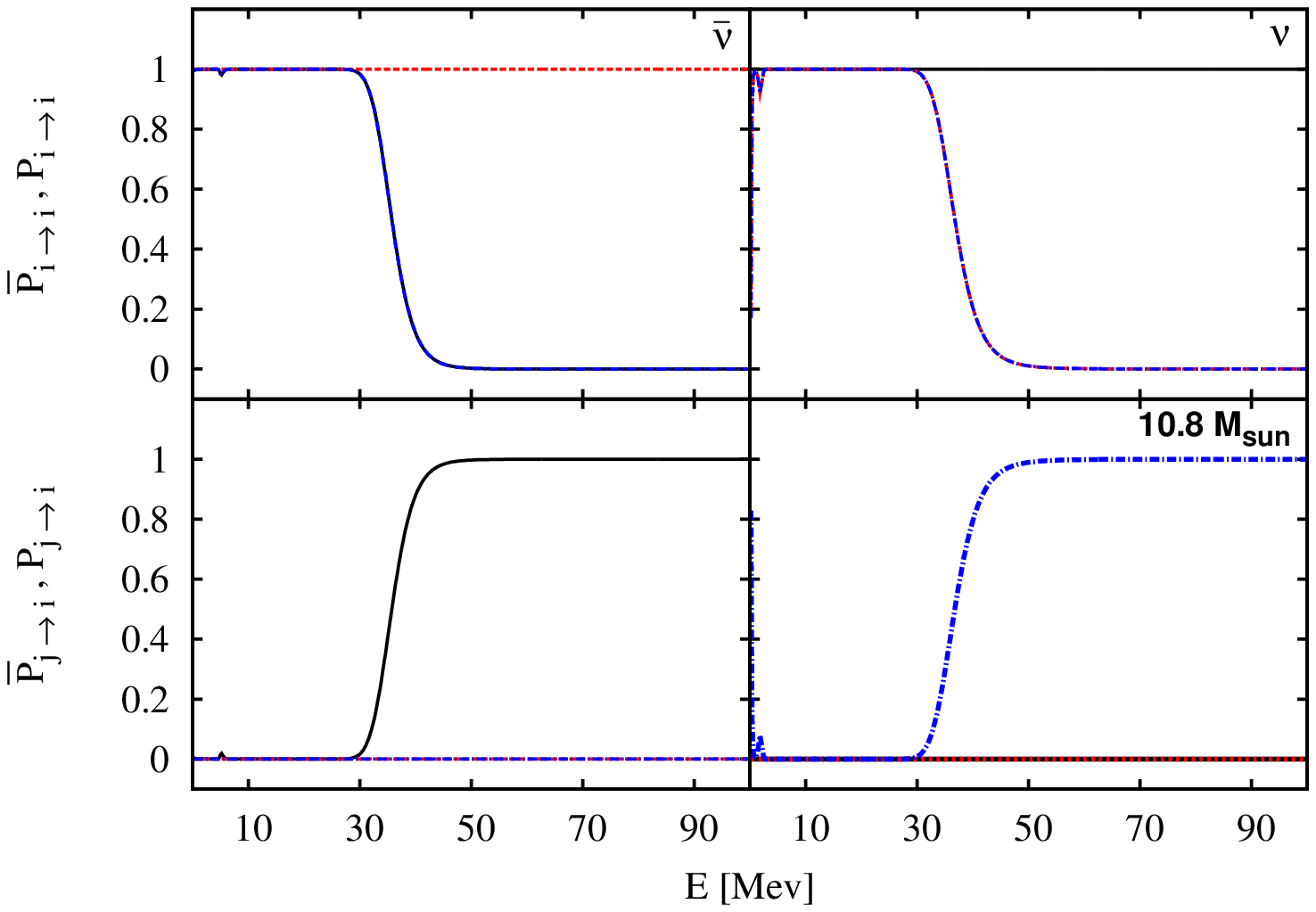}
\includegraphics[width=\linewidth]{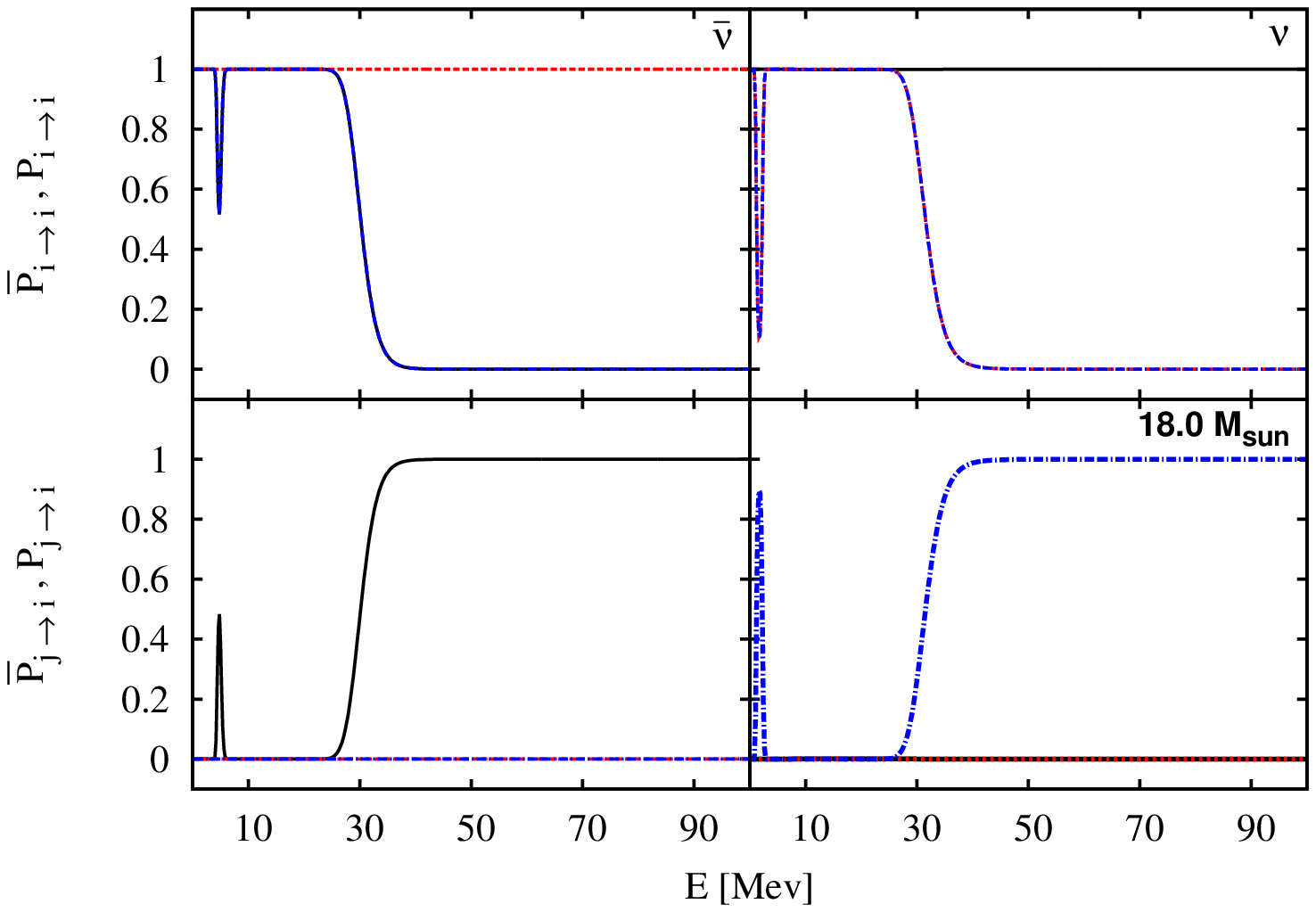}
\caption{\label{fig:tnmNH_1sec} (color online).   As
Fig.~\ref{fig:coNH_1sec} but for the full profile traversal with
10\% turbulence.  Normal Hierarchy.} 
\end{figure}

The differences between results from 1 and 3~s in each of
the inner and outer regions separately are naturally also 
reflected in the full calculations.

\subsubsection{Including turbulence} 
Finally in
Figures~\ref{fig:tnmIH_1sec} and \ref{fig:tnmNH_1sec} we show
the results of calculations where turbulence has been included
as the neutrino traverses the full profile. For the \Mten{} and
the \Met{} progenitor the addition of 10\% turbulence to this
profile has no impact on the survival and transition
probabilities as can be seen by comparing
Fig.~\ref{fig:nmIH_1sec} with Fig.~\ref{fig:tnmIH_1sec}, and
Fig.~\ref{fig:nmNH_1sec} with Fig.~\ref{fig:tnmNH_1sec}. For the
\Mei{} progenitor the story is quite different. We see distinct
imprints in 9 out of 12 survival probabilities. The unchanged
probabilities are $\bar P_{22,\textrm{IH}}$, $\bar
P_{22,\textrm{NH}}$ and $P_{33,\textrm{NH}}$, while
$P_{11,\textrm{IH}}$, $P_{22,\textrm{IH}}$,
$P_{33,\textrm{IH}}$, $\bar P_{11,\textrm{NH}}$ and $\bar P_{33,
\textrm{IH}}$ show only small amplitude oscillations as changes.
Finally $\bar P_{11,\textrm{IH}}$, $\bar P_{33,\textrm{NH}}$,
$P_{22,\textrm{NH}}$ and $P_{33,\textrm{NH}}$ all show large
amplitude oscillatory behavior on top of their still clearly
visible spectral splits.

The addition of turbulence has such a
profound effect on the \Mei{} model because the density profile
at 1~s is relatively flat (see the upper panel of
Fig.~\ref{fig:rhos}). The addition of even low amplitude
turbulence effectively creates a multitude of H
resonances compared to the case without turbulence. By comparing
the panels for the \Mei{} model in Figures~\ref{fig:nmIH_1sec}
and \ref{fig:tnmIH_1sec}, and Figures~\ref{fig:nmNH_1sec} and
\ref{fig:tnmNH_1sec}, we see that, as expected, the new resonances
impact primarily the anti-neutrino states \bone{} and \bthree{}
in the IH, and the neutrino states 2 and 3 in the NH. In
both cases the effect is seen most strongly for the higher
energies since the density profile lingers in the H
resonance region corresponding to the higher neutrino energies.
Additionally, at the lowest energies we see that neutrino states
1 and 2 are affected by the MSW L resonance since it has become
diabatic with the turbulence induced resonances. A strong phase
effect from the multiple resonances is visible in all the
affected channels.

\section{Time evolution of features}    \label{sec:time_evolv}
In Sections~\ref{sec:3s_results} and \ref{sec:1s_results} we saw
how features in the neutrino survival probabilities arose
and subsided as time progressed. We saw how the self-interaction
induced double split between neutrino states 2 and 3 in the IH grew
more narrow from 1 s to 3 s because the higher energy
spectral split moved down in energy.  Similarly we saw in the NH
that the self-interaction splits between states \bone{} and
\bthree{}, and between states 2 and 3 also moved down in energy.
Hardly any imprint of the MSW effect was present at 1~s thus it
is hard to discuss the evolution of MSW imprints from the 1 and
3~s profiles alone. However, as will become apparent below,
the MSW imprints evolve significantly when we look at later
times.  

\begin{figure}[b!]
\includegraphics[width=\linewidth]{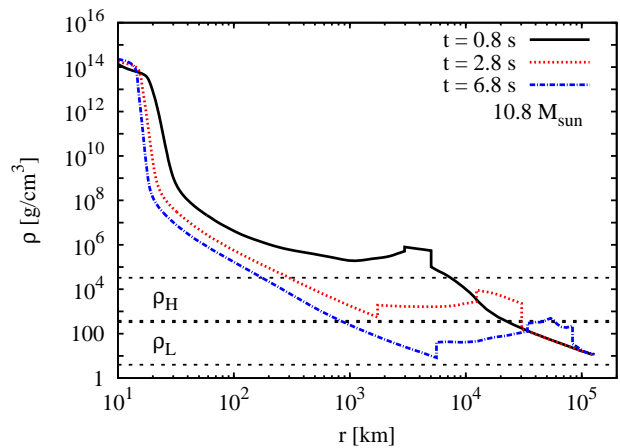}
\caption{\label{fig:rhoProfiles} (color online).  Density
profiles at 3 different times for our 10.8~$M_\odot$ model.}
\end{figure}

If observed, one might hope that such time dependence of the
features imposed by both neutrino self-interactions and the MSW
effect (as well as turbulence) will lead to a better handle on
and understanding of the neutrino flavor evolution.  In this
section we therefore more thoroughly discuss the time evolution
of some collective and MSW induced features.
The movement in energy of the spectral splits is associated
with the evolution of the density profiles. As time passes the
shock front moves progressively further out, and thus into lower
density regions, bringing higher energy neutrinos into
resonance, see Figures~\ref{fig:rhoProfiles} and
\ref{fig:shockMove_dens}. The MSW induced
spectral splits therefore move up in energy.
The self-interaction induced splits change due to the decreasing
neutrino density close to the PNS, allowing less and less time
for collective oscillations to develop. Additionally the
neutrino luminosities and mean energies decrease over the same
time frame causing the interaction strength to diminish.

All these changes to the density profiles, neutrino
luminosities and mean energies lead to a complicated evolution of
the flavor evolution that one would hope to disentangle from a
real neutrino burst signal. From a careful analysis of which
effects create which features we shall see that the time
evolution of MSW collective effect signatures is unique. We
will focus first on the time evolution of the probabilities that
are caused by the progression of the shock front, then we will
discuss the time evolution of a sample feature of the collective
effect.

\subsection{Progression of the shock front and its impact on
probabilities} 
\begin{figure}
\includegraphics[width=\linewidth]{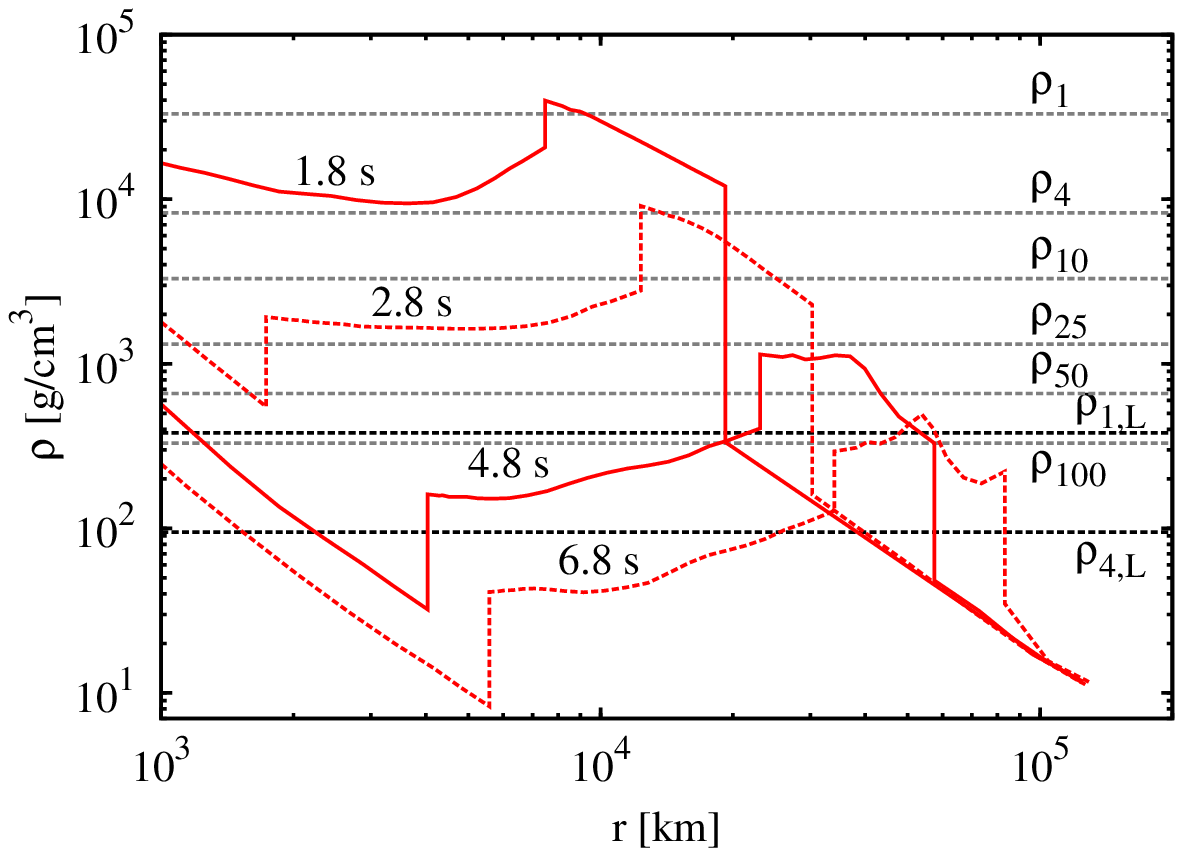}
\caption{\label{fig:shockMove_dens} (color online).  Density
profiles for the \Mten{} model showing the progression of the
forward shock through resonant density layers. The neutrino
energy corresponding to a given resonance is given in MeV.  
The resonant densities are calculated with the assumption
$Y_e=0.5$.\\} 
\includegraphics[width=\linewidth]{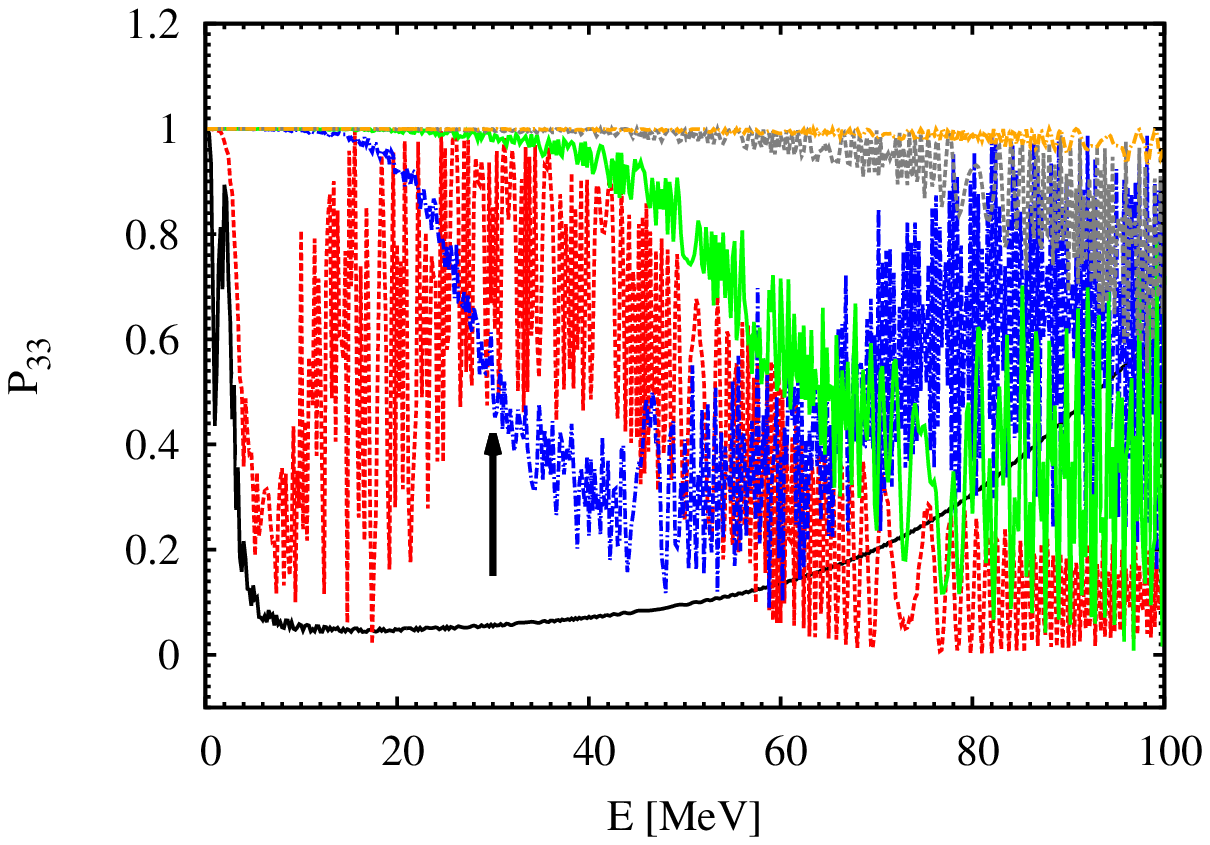}
\caption{\label{fig:shockMove_probImpact} (color online).
Survival probabilities for neutrino matter state 3, $P_{33}$, at
1.8~s (black solid line), 2.8~s (red dashed line), 4.8~s (blue
dot-dashed line), 5.8~s (green solid line), 6.8~s (gray dashed
line) and 7.8~s (yellow dot-dashed line) for our \Mten{} model
in the NH and for the outer regions.  The black arrow at 30~MeV
marks on the 4.8~s profile the drop from unit survival
probability discussed in the text.}
\end{figure}
The shock wave features in the signal are inserted in the outer
region of the supernova envelope therefore we focus upon this
region exclusively for the time being. Including the collective
effects from the inner region would make the probabilities more
complex and make it harder to show the effects we are trying to
illustrate.  Narrowing our focus further we shall consider the
\Mten{} model and the NH where the effect of the MSW H (high
density) resonance is to mix neutrino matter states 2 and 3.  We
shall discuss the evolution of the other two models later. In
Fig.~\ref{fig:shockMove_probImpact} we show the survival
probability of neutrino matter state 3 for several different
times. We will not consider the survival probability of matter
state 2 since it is also entangled with the survival probability
of matter state 1 through the MSW L (low density) resonance.
Multiple features in the $P_{33}$ survival probabilities in
Fig.~\ref{fig:shockMove_probImpact} deserve attention and
further explanation. First and foremost we would like the reader
to focus on the drop from unit survival probability which occurs
at all times, but which appears at low energies for early times
and progresses to higher energies at later times. In
Fig.~\ref{fig:shockMove_probImpact} we have indicated with a
black arrow on the $t=4.8 s$ result (blue dot-dashed line) the
drop feature in question. We see that the midpoint of the drop
moves from about 4~MeV at 2.8~s, to 25~MeV at 4.8~s, to 56~MeV
at 5.8~s. At 6.8~s the shock is beginning to slip out of the
energy range we consider and by 7.8~s it has vanished.  The
probability drop from $P=1$ to $P\sim0$ has a direct relation to
the appearance of the density profile.  For the sake of clarity
we have selected a small sample of the density profiles we are
investigating and show them in Fig.~\ref{fig:shockMove_dens}.
This figure illustrates the progression of the forward shock
through the MSW H resonance region, and we clearly see how the
forward shock moves out and into lower densities, thereby
changing the resonance of the neutrinos with higher and higher
energies from adiabatic to diabatic. If the shock feature can be
found and followed in the neutrino signal then it should be
possible to map that back so as to follow the progression of the
shock front over time through the star. 

The probability $P_{33}$ at 1.8~s (black solid line in
Fig.~\ref{fig:shockMove_probImpact}) shows a particularly
interesting feature at energies below 5 MeV: a double drop. From
the corresponding density profile in
Fig.~\ref{fig:shockMove_dens} we see that the contact
discontinuity at this particular snapshot of the simulation
covers resonant densities corresponding to energies of 1--2~MeV
and the forward shock covers energies of 3--100~MeV. This leaves
a tiny gap of roughly an MeV in which the neutrino transition
probability is not enhanced. This is reflected in the
probability in Fig.~\ref{fig:shockMove_probImpact} where we see
an initial drop in survival probability at 1--2~MeV, followed by
an increase in the survival probability around 3~MeV, only to
drop again at 4~MeV. Part of the reason the initial drop and the
subsequent increase in survival probability is incomplete is
that the flavor conversion resonances have widths which are
proportional to $\tan 2\theta$. Another manifestation of this
effect is the gradual change from a survival probability of 1 to
0 over a range of energies, as can be seen e.g.\ in the $t=4.8
s$ result (blue dot-dashed line) where the drop occurs over
energies of $\sim$~12--44~MeV.
The rapid oscillations in energy that overlay the large scale
trends mentioned above is caused by phase effects, which we have
discussed before.

\begin{figure}
\includegraphics[width=\linewidth]{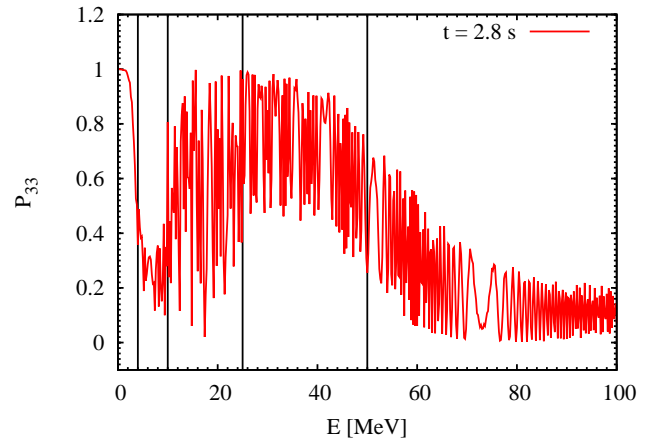}
\caption{\label{fig:threeSec_extraClose_probs} (color online).
Survival probability for neutrino matter state 3, $P_{33}$, at
2.8~s for our \Mten{} model (red solid line) in the NH and for
the outer region.  The vertical lines are the energies for which
the H resonant densities are marked in
Fig.~\ref{fig:shockMove_dens} (gray dashed lines).}
\end{figure}

We now take a closer look at the survival probability $P_{33}$
for the 2.8~s profile with its multitude of features, and
investigate how they relate to features in the corresponding
density profile. 
From Fig.~\ref{fig:shockMove_dens}
we clearly see the discontinuous jump, that is the
reverse shock, spans densities corresponding to the resonant
densities for neutrinos with energies between 17 and 62~MeV.
Similarly the contact discontinuity spans densities
corresponding to energies in the range 3.5 to 12~MeV and the
forward shock covers energies from about 14 to 210~MeV.
Consequently, a neutrino with an energy in one of the following
ranges 3.5--12~MeV, 14--17~MeV or 62--210~MeV will only
experience one diabatic resonant enhancement of its flavor
conversion.  If the neutrino instead possesses an energy in the
range 17--62~MeV then it will experience two diabatic
resonances; first at the reverse shock and secondly at the
forward shock as it gets further out.  A neutrino with an energy
outside of these ranges will not experience any diabatic
enhancement in the conversion probability.  Neutrinos of all
energies, outside the range 62 to 210~MeV, will also experience
at least one adiabatic density resonance.

If we now turn our attention to
Fig.~\ref{fig:threeSec_extraClose_probs}, we see these features
of the density profile reflected in the resonance survival
probability $P_{33}$.  The initial drop in survival probability
at 3.5~MeV is caused by the diabatic crossing of the
contact discontinuity enhancing the conversion of matter state 3
into matter state 2. Above 62~MeV we find the same familiar
approach of $P_{33}$ to 0 caused by a single diabatic resonance,
this time due to the forward shock.  The closely spaced
energy regions of zero and single resonance between 12 and
14~MeV, and 14 and 17~MeV do not show their respective absent or
full conversion due to the width of the resonances, which causes
them to overlap. 
Neutrinos with energies between 17 and 62~MeV have two diabatic
resonances due to the reverse and forward shocks, and in
addition they can have up to 3 adiabatic resonances. The
crossing of the reverse shock will enhance the conversion of
matter state 3 into 2 and the subsequent passage of the forward
shock will cause the neutrinos in matter state 2 to be converted
back into matter state 3, leaving the survival probability at
almost unity.

\begin{figure}[t!]
\includegraphics[width=\linewidth]{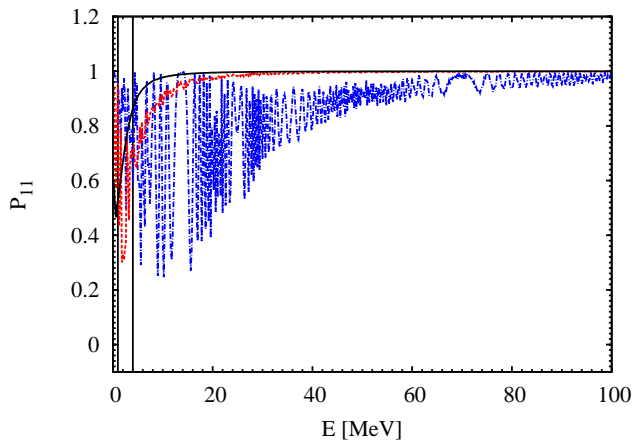}
\caption{\label{fig:threeSec_extraClose_resLprobs} (color
online).  Survival probability for neutrino matter state 1,
$P_{11}$, at 1.8~s (black solid line), 2.8~s (red dashed line)
and 4.8~s (blue dot-dashed line) for our \Mten{} model in the NH
and for the outer region.  The vertical lines are the energies
for which the L resonant densities are marked in
Fig.~\ref{fig:shockMove_dens} (black dashed
lines).} 
\end{figure}

Consulting Fig.~\ref{fig:threeSec_extraClose_resLprobs} allows
us to determine that we expect a minimal impact from the L
resonance. Fig.~\ref{fig:threeSec_extraClose_resLprobs} show the
survival probability of matter state 1 (which mixes with matter
state 2 at the L resonance). From
Fig.~\ref{fig:shockMove_dens} we expect neutrinos with
energies up to 2.5~MeV to be affected by a diabatic L resonance
caused by the forward shock. Due to the width of the resonance
actually neutrinos with up to about 24~MeV will feel the impact
of the L resonance (see the red dashed line in
Fig.~\ref{fig:threeSec_extraClose_resLprobs}). 
At energies above $\sim$17~MeV we see that 96\% or more of the
neutrino state 1 remain in state 1, while the remaining 4\% or
less mix with state 2. We can therefore conclude that the
majority of neutrinos converted from matter state 3 into matter
state 2 at the reverse shock H resonance will be converted back
into matter state 3 at the forward shock. Although the majority
of the neutrinos in matter state 2 remain in matter state 2, the
small mixing with matter state 1 through the L resonance
actually means we can have a tiny admixture of matter state 1
into matter state 3 (coming through matter state 2).  Thus, this
little example shows that it is not possible to completely
separate the effect of the H and the L resonances, although the
contamination is limited. 

On top of the general undulating trend displayed in $P_{33}$ --
the initial drop around 3.5~MeV, the increase around 10~MeV and
the subsequent slow fall off from roughly 40 to 60~MeV -- we
also see high frequency oscillations.  These rapid oscillations
are attributed to the phase effect. Phase effects arise when
neutrinos encounters multiple resonances (diabatic or adiabatic)
\cite{Kneller:2005hf,Dasgupta:2007phase}.  Neutrinos of different
energies will have slightly different path lengths between their
respective resonance points so the phase will not be the same
for each. Obviously the presence of the phase effects will make
identifying split features more difficult but the reader must be
aware that the phase effects typically have such high
``frequency'' - ``periods'' of $50\;{\rm keV}$ or smaller - that
current detectors are not capable of detecting them due to their
comparatively lower energy resolution.

The connection between features of the density profile and
features in the survival probabilities is readily made for all our
profiles.  We chose to focus on the 3~s profile of the the
\Mten{} model in this section because this profile possesses all
three features in the density profile. Although not shown here,
we see the exact same shock induced behavior in the
anti-neutrino state \bthree{} in case of the IH, and $P_{11}$
exhibit a similar behavior in both hierarchies too.  When we
consider the other models we again find a match between profile
features, and their evolution, and transition probability
behavior. In the case of the \Mei{} model the shock moves out so
fast that our relative coarse sampling in time cannot follow it.

\begin{figure}
\includegraphics[width=\linewidth]{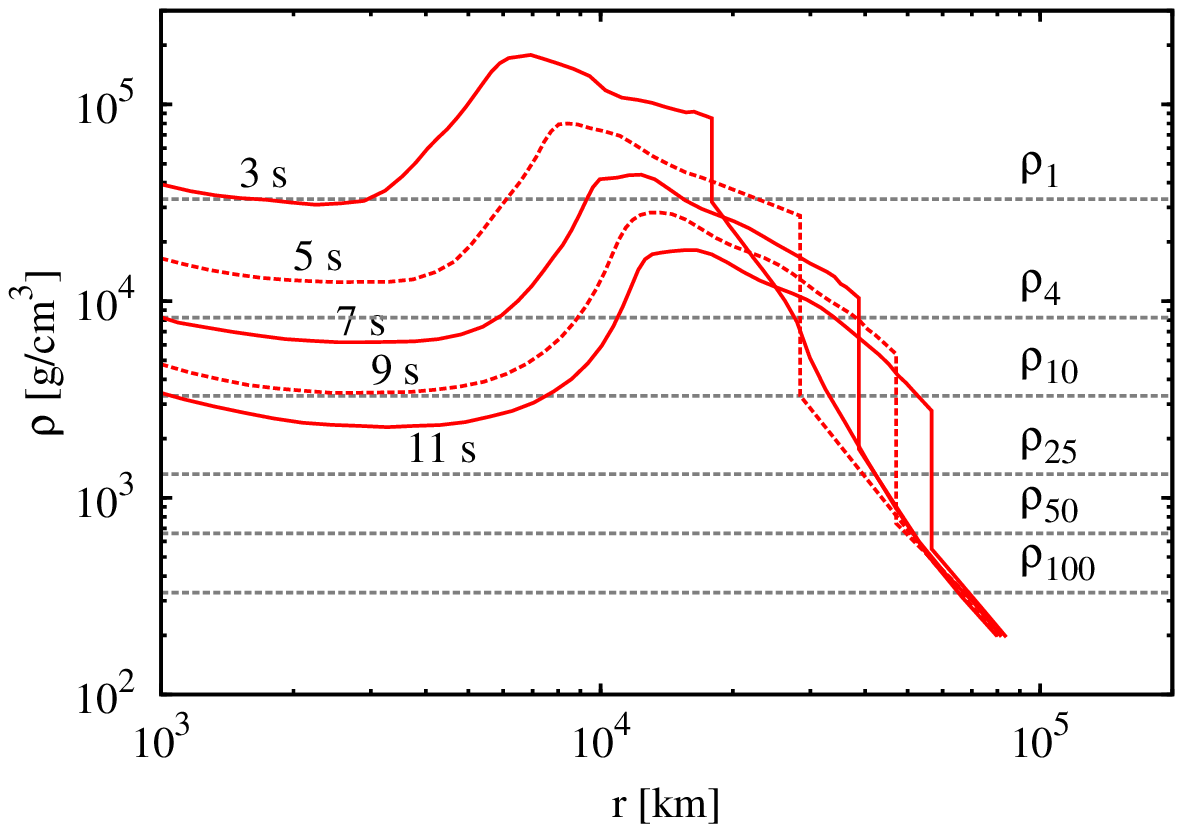}
\caption{\label{fig:shockMove_dens18} (color online).  Density
profiles for the \Met{} model showing the progression of the
forward shock through the MSW H resonant density layers.  The
horizontal gray dashed lines mark the resonant density for a
neutrino  of a given energy (in MeV).\\}
\includegraphics[width=\linewidth]{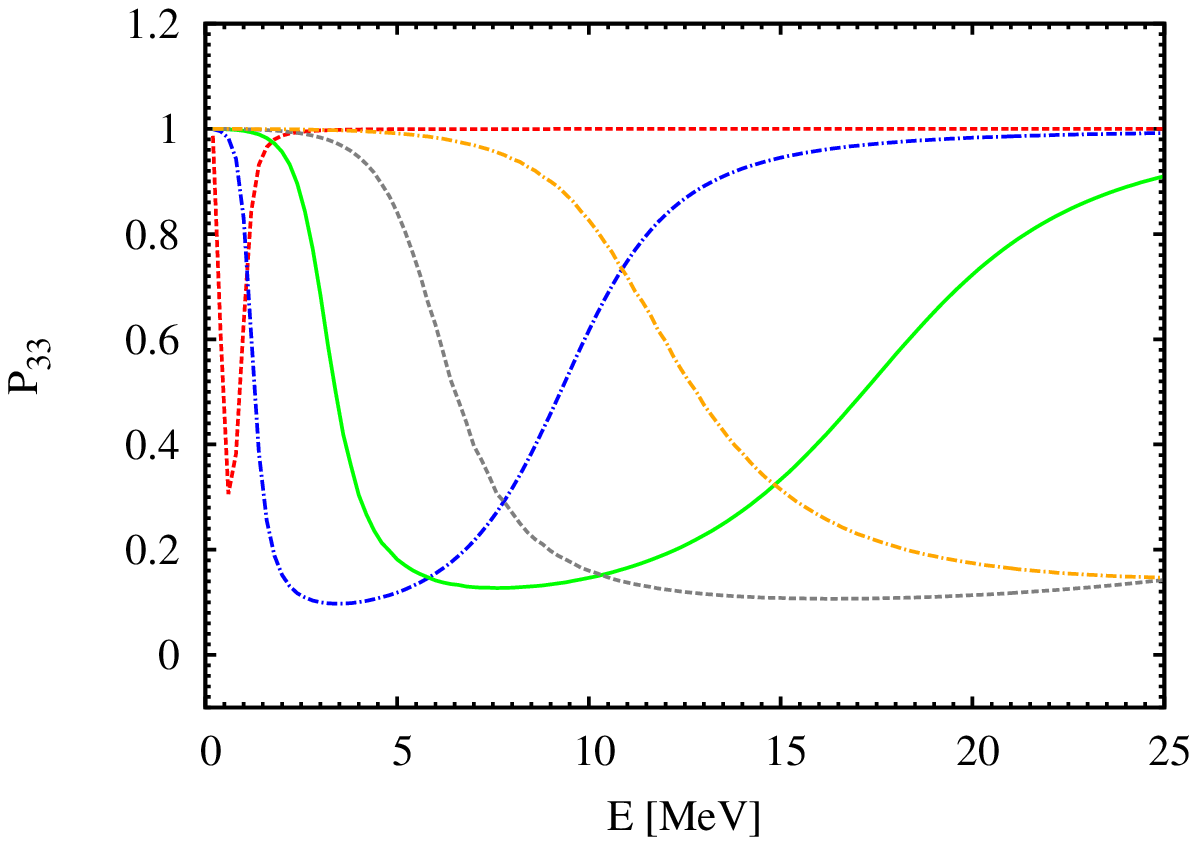}
\caption{\label{fig:shockMove_probImpact18} (color online).
Survival probabilities for neutrino matter state 3, $P_{33}$, at
3~s (red dashed line), 5~s (blue dot-dashed line), 7~s (green
solid line), 9~s (gray dashed line) and 11~s (yellow dot-dashed
line) for our \Met{} model in the NH and for the outer regions.}
\end{figure}

For the \Met{} model we see the effect of the progression of the
shock wave almost as well as we do in the \Mten{} case,
see Figures~\ref{fig:shockMove_dens18} and 
\ref{fig:shockMove_probImpact18}. The
major difference between the \Mten{} and \Met{} models is not
that the outward motion of the shock in the \Met{} progenitor is
slightly slower than the shock progression in the \Mten{}
progenitor, they have similar speeds, but rather the profile of
the \Met{} is so extended that the the shock feature of the
transition probabilities takes much longer to sweep through the
spectrum. For the survival probability $P_{33}$ in
Fig.~\ref{fig:shockMove_probImpact}, we followed the impact of
the shock progression as the drop from $P=1$ to $P\sim 0$ moved
quickly from a few MeV through the entire spectrum and above 100
MeV in a handful of seconds (basically from 2 to 8 seconds). In
contrast, in Fig.~\ref{fig:shockMove_probImpact18} we see how
this same drop in $P_{33}$ for the \Met{} model inches its way
up through the energies; from below 1~MeV at 3~s to 12~MeV at
11~s. (Note the energy scale in
Fig.~\ref{fig:shockMove_probImpact18} ends at 25~MeV not
100~MeV.) In addition to the much slower progression of the
shock feature, the \Met{} model develops neither a contact
discontinuity nor a reverse shock at the times we are looking
at, so the probability plot in
Fig.~\ref{fig:shockMove_probImpact18} is much ``cleaner'' than
Fig.~\ref{fig:shockMove_probImpact} for the \Mten{} model. This
absence of multiple resonances leads to an absence of phase
effects in the \Met{} case.  Finally, for the \Met{} model any
discussion regarding the shock front moving into the L resonant
density region is obviously moot since the L resonance densities
starts at around 380~$g/cm^3$ and from
Fig.~\ref{fig:shockMove_dens18} it is clear that only the very
last couple of points on the density profiles are in this
regime, and that the shock does not reach such low densities at
the times we investigate here.

This very different evolution of MSW signatures in our
progenitors means that
identifying the MSW contribution to the neutrino signal can be
turned around to learn about the progenitor. If the simulations
we have used in this paper are representative then we expect
ONeMg supernovae neutrinos to be briefly affected by shock waves
early in the signal while much more massive progenitors are
affected only after a longer delay and then the effects persist
for much longer if not all of the remaining burst. Identifying
what kind of star exploded makes the comparison of observations
of the supernova via neutrinos, gravitational waves and photons
with a simulation easier in the sense that we can compare like
with like.    

\begin{figure}[b!]
\includegraphics[width=\linewidth]{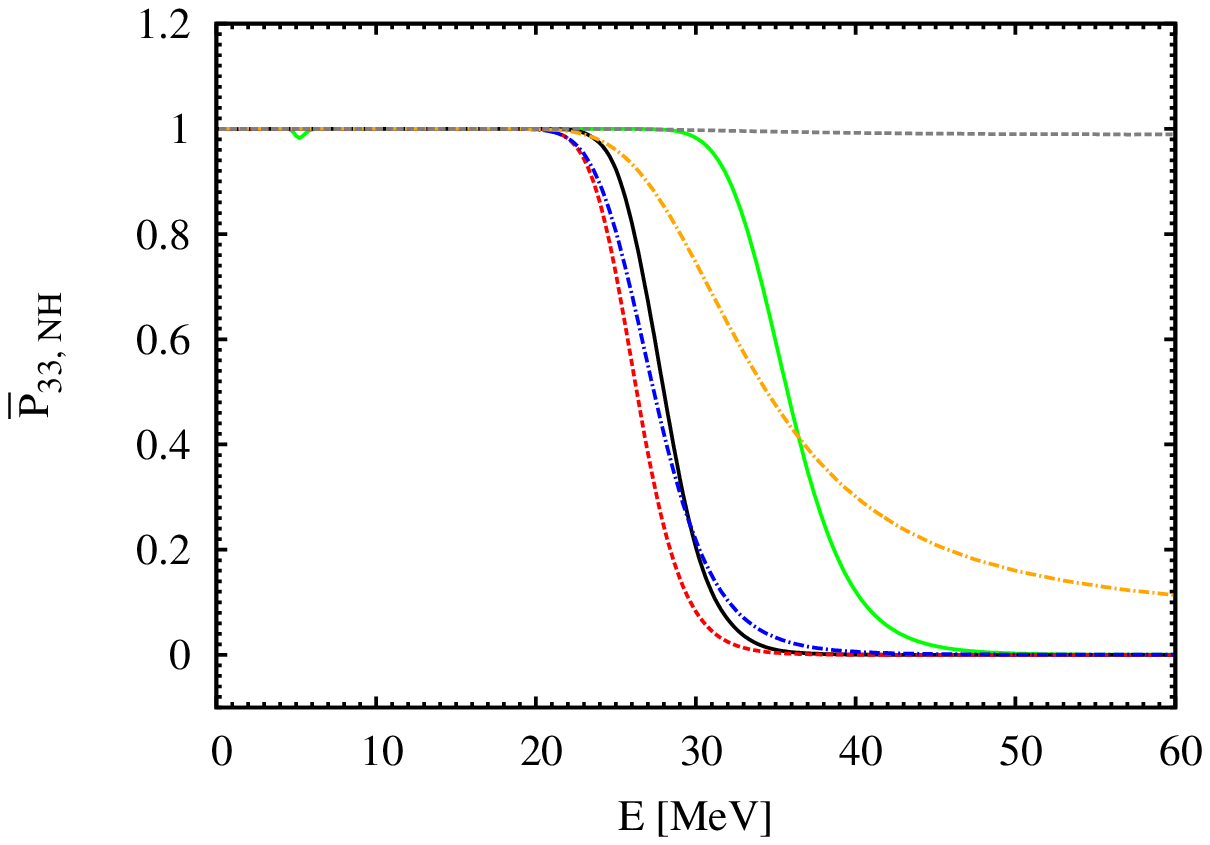}
\caption{\label{fig:aP33_coNH_tEvolv} (color online).  Survival
probabilities for anti-neutrino state \bthree, $\bar P_{33}$, at
0.8~s (green solid line), 1.8~s (black solid line), 2.8~s (red
dashed line), 4.8~s (blue dot-dashed line), 7.8~s (yellow
dot-dashed line) and 10.5~s (gray dashed line) for our \Mten{}
model for the inner regions and the NH.} 
\end{figure}

\begin{figure*}
\includegraphics[width=0.45\linewidth]{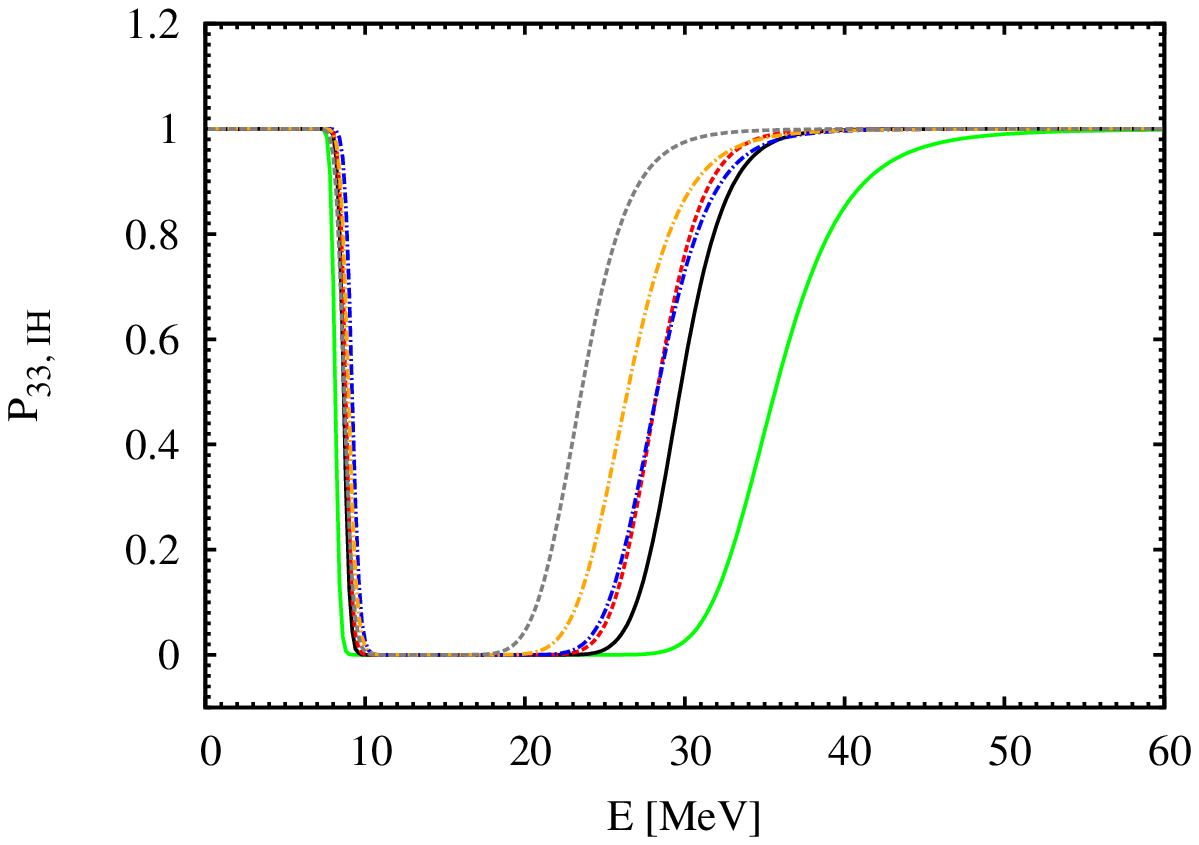}
\includegraphics[width=0.45\linewidth]{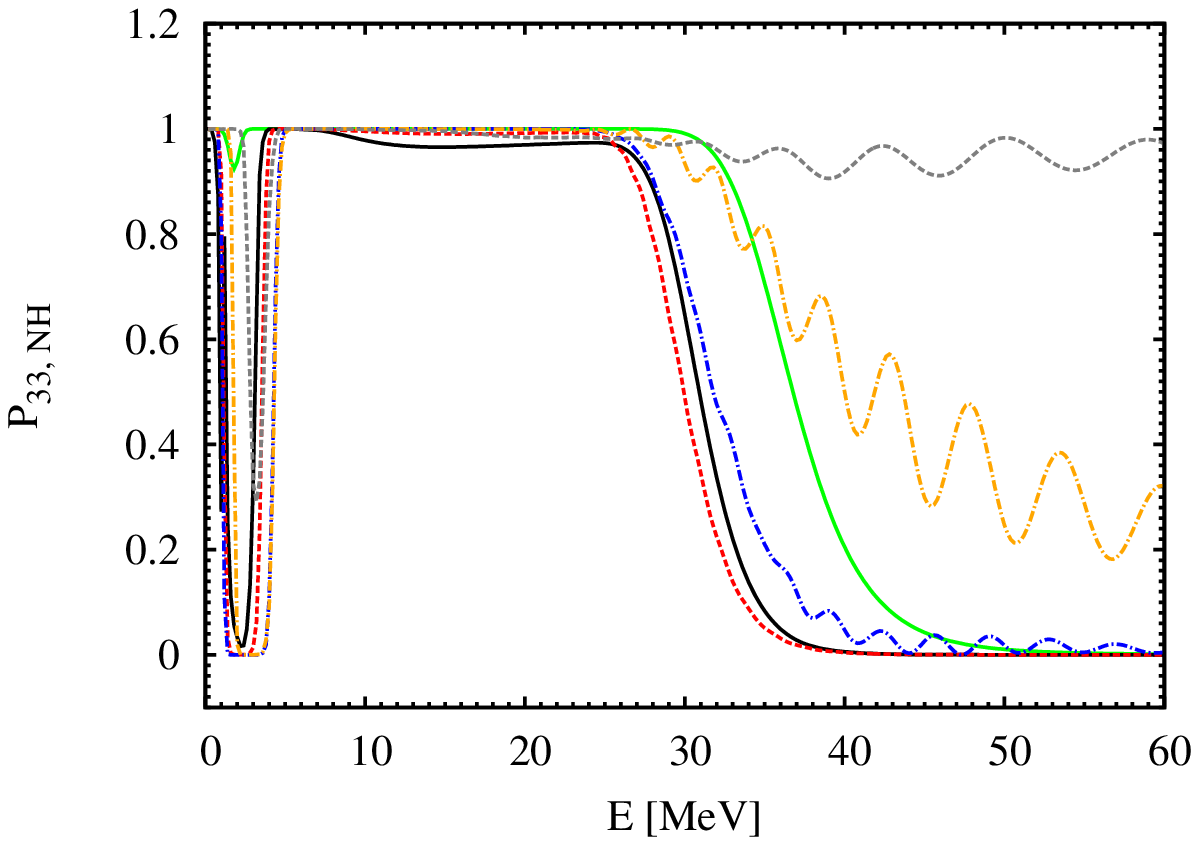}
\caption{\label{fig:P33_co_tEvolv} (color online).  Survival
probabilities for neutrino matter state 3, $P_{33}$, at 0.8~s
(green solid line), 1.8~s (black solid line), 2.8~s (red dashed
line), 4.8~s (blue dot-dashed line), 7.8~s (yellow dot-dashed
line) and 10.5~s (gray dashed line) for our \Mten{} model for
the inner regions.  On the left for the IH and for the NH on the
right.} 
\end{figure*}

\subsection{Time evolution of collective features}
\label{sec:coll_time_evolv}
We now consider the time evolution of a few of the collective
features and again focus our attention for the time being on the
\Mten{} model. For this one progenitor we plot in
Figures~\ref{fig:aP33_coNH_tEvolv} and \ref{fig:P33_co_tEvolv}
the time evolution of $\bar P_{33}$ and $P_{33}$ in the inner
region for a select set of times.
Fig.~\ref{fig:aP33_coNH_tEvolv} shows how the spectral split for
the NH anti-neutrino survival probability $\bar P_{33}$
initially starts at 36~MeV at 0.8~s (green solid line). Then it
moves down in energy until 2.8~s (red dashed line) where it
reaches 26~MeV only to climb back up in energy and become
increasingly softer until the split is completely gone at 10.5~s
(gray dashed line).  Equally clean is the narrowing over time of
the double spectral split in $P_{33}$ in the IH that is shown in
the left panel of Fig.~\ref{fig:P33_co_tEvolv}. The higher
energy split starts at 35.5~MeV at 0.8~s (green solid line) and
gradually becomes steeper as it moves down in energy to end at
23.5~MeV at 10.5~s (gray dashed line). On the same time scale
the spectral split at the lower energy starts at 8~MeV at 0.8~s,
moves up to 9.2~MeV at 4.8~s, and then back down to 8.7~MeV at
10.5~s. 

The evolution of $P_{33}$ in the NH shown in the right panel of
Fig.~\ref{fig:P33_co_tEvolv} is more convoluted. In this case,
initially at 0.8~s (green solid line) there is only one split at
an energy of 37~MeV, and a minuscule dip in the survival
probability at 2~MeV.  As time passes the higher energy split
moves slightly down in energy until 2.8~s (red dashed line)
where after it reverse direction and moves up in energy,
becoming increasingly softer until it has almost disappeared at
10.5~s (gray dashed line). This is exactly as we saw in the
anti-neutrino case but now we also observe at the later times
\begin{figure}[b!]
\includegraphics[width=\linewidth]{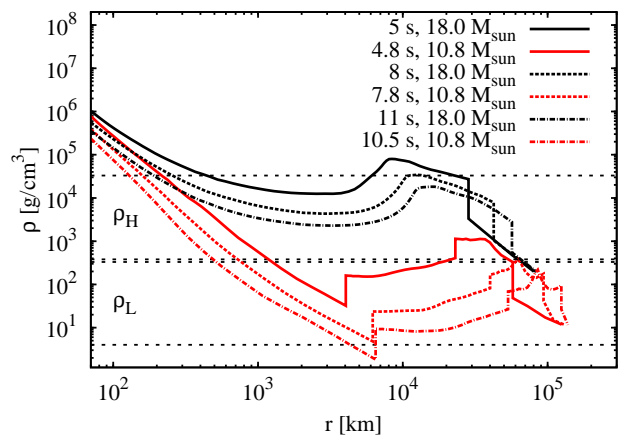}
\caption{\label{fig:rhos_late} (color online).
Density profiles for the \Mten{} progenitor at 4.8~s (solid red), 
7.8~s (dashed red) and 10.5~s (dot-dashed red), as well as for
the \Met{} progenitor at 5~s (solid black), 8~s (dashed black) 
and 11~s (dot-dashed black).}
\end{figure}
oscillations in the survival probability.  Meanwhile, at lower
energies the initial dip at 2--3~MeV grows to a full double
split feature that initially widens slightly and moves up in
energy only to grow narrower again at late times.  In principle
one would expect to see this additional double spectral split
arise over time in an observation. However, the two splits
around 3~MeV are so closely spaced that current detector
technology would not be able to resolve them, especially at such
a low energy. By comparing the right panel of
Fig.~\ref{fig:P33_co_tEvolv} to
Fig.~\ref{fig:shockMove_probImpact} we see that the energy range
over which the probability features changes for the inner region
is the same as the energy range for the outer region changes.
Thus in an observed signal it would be a challenge if not
impossible to disentangle the evolution from the two regions,
let alone their time evolution.

The basic conclusion from these figures is that the collective
features in any signal are not static but rather evolve in a
fashion that does not lend itself to an easy analysis. Spectral
splits due to collective effects can disappear over time and new
ones at different energies emerge. 

\subsection{Late time evolution of the \Mten{} and \Met{} models}
\label{sec:addtProfiles} 
\begin{figure*}
\includegraphics[width=0.49\linewidth]{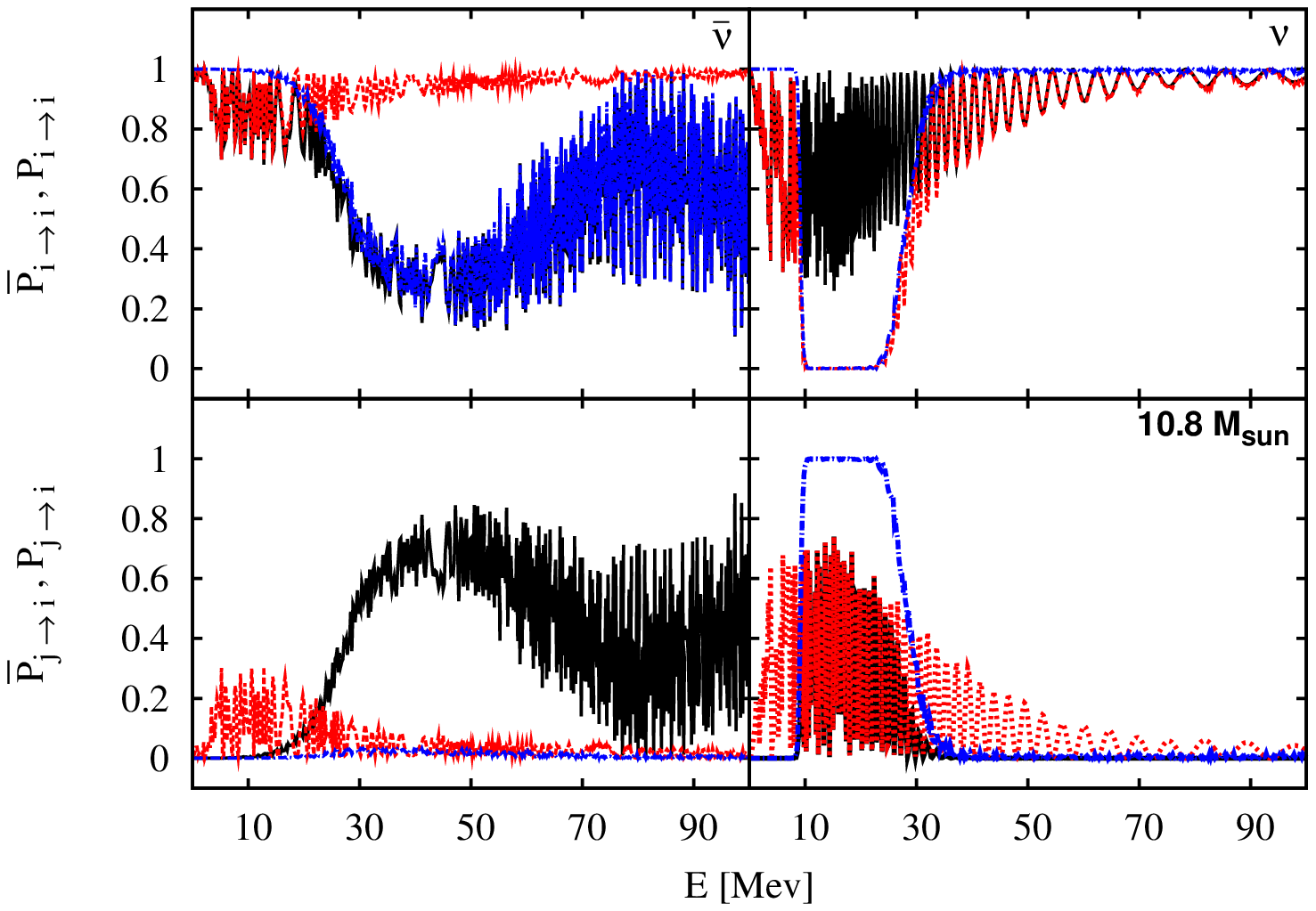}
\includegraphics[width=0.49\linewidth]{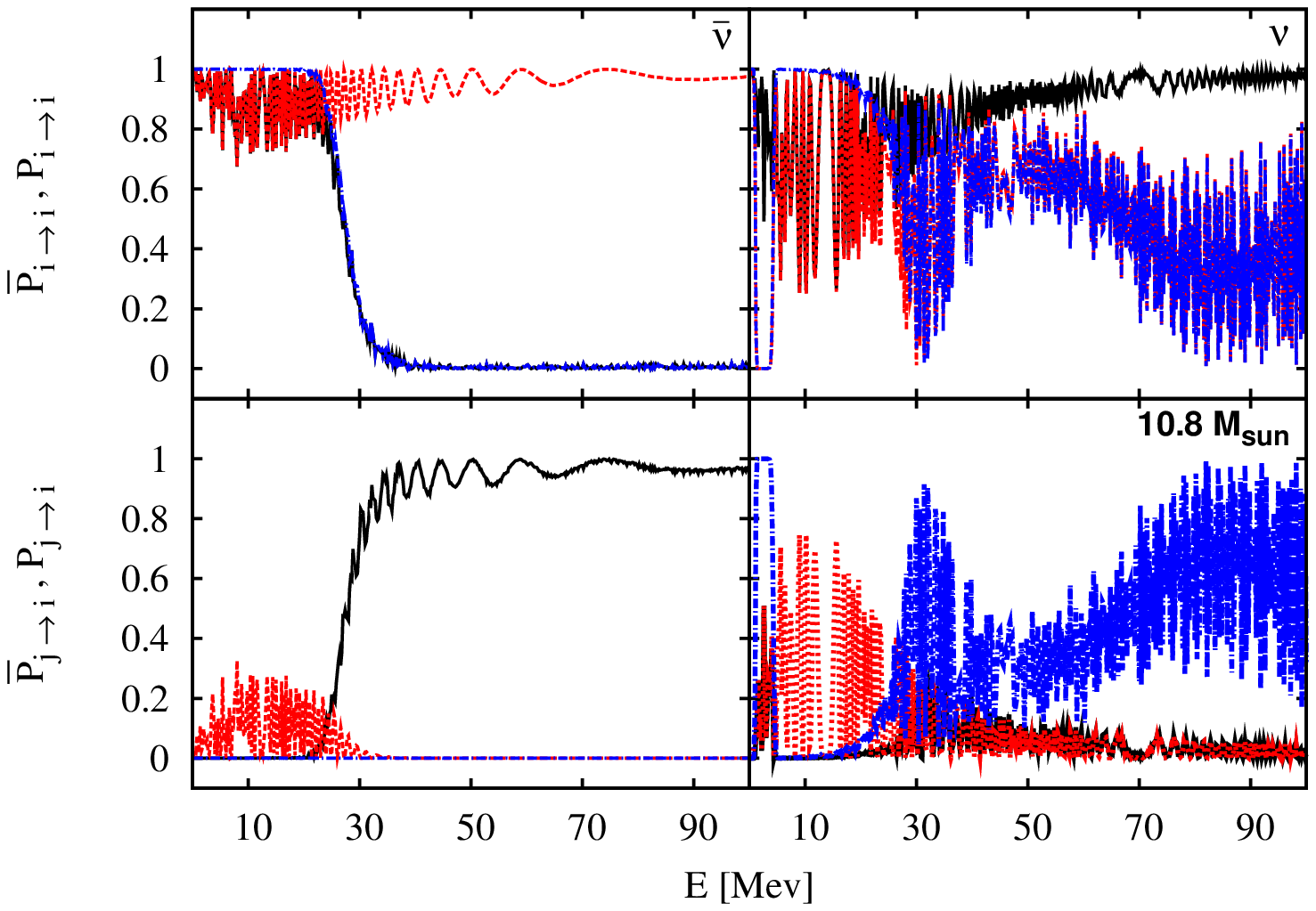}
\includegraphics[width=0.49\linewidth]{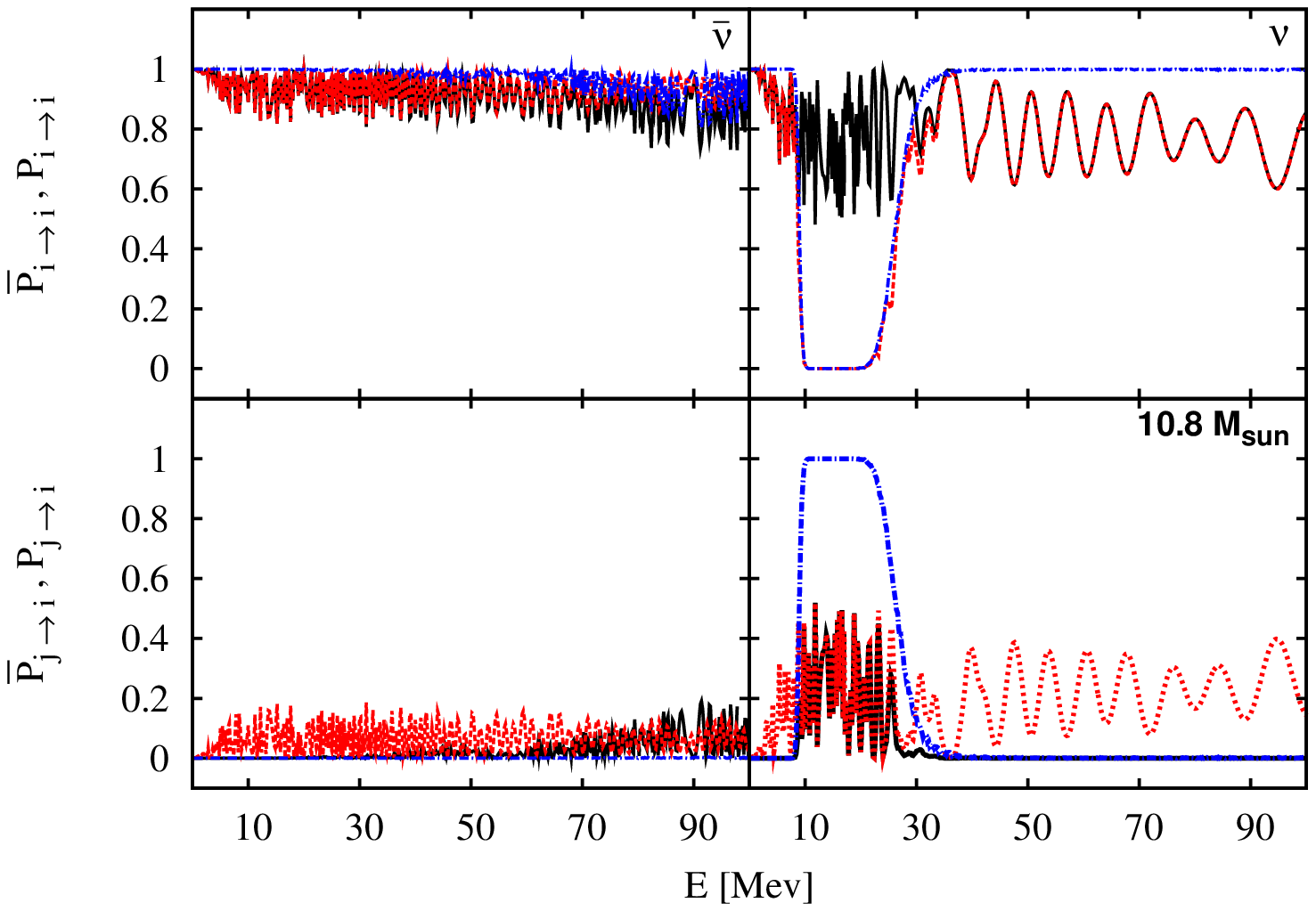}
\includegraphics[width=0.49\linewidth]{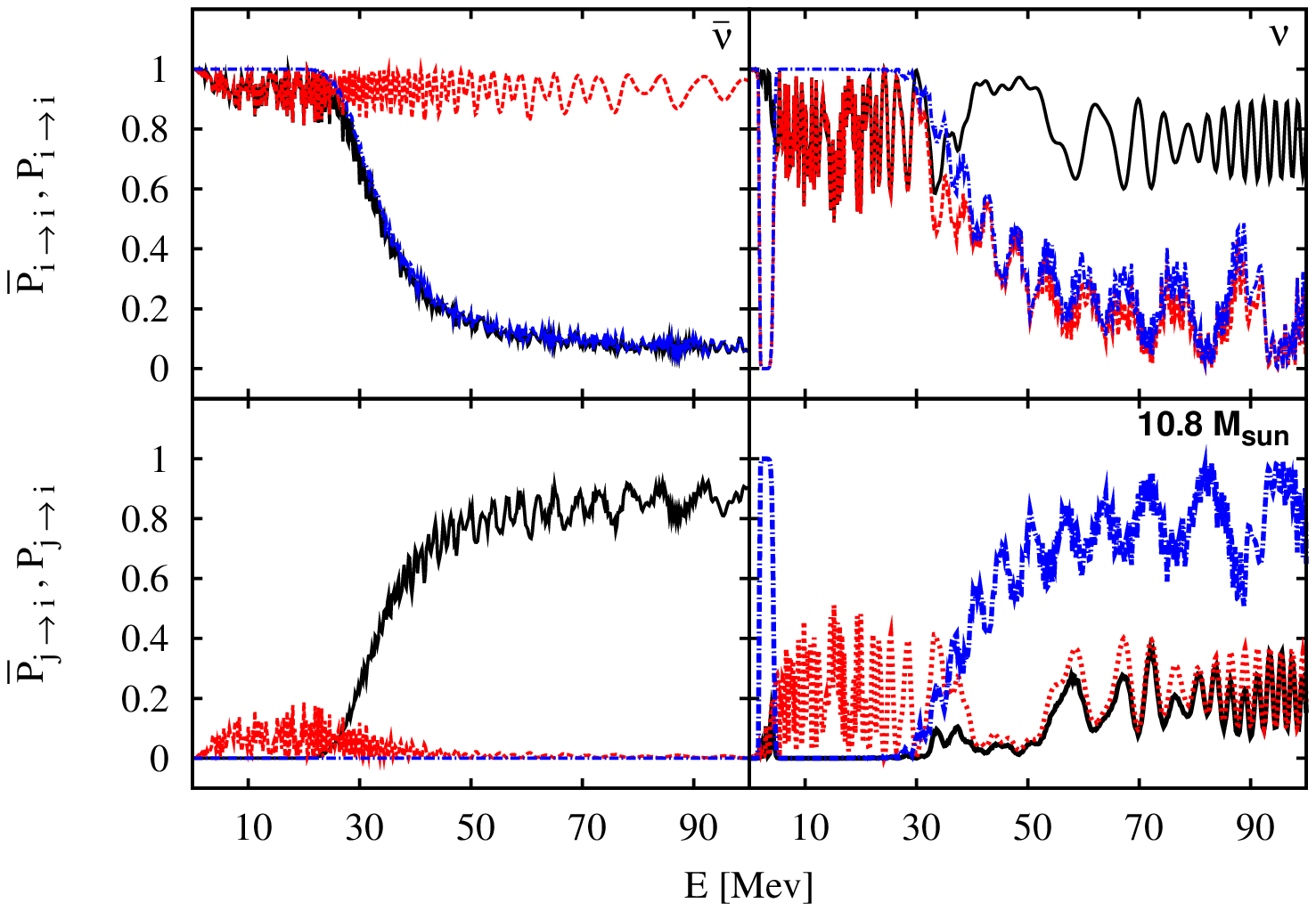}
\includegraphics[width=0.49\linewidth]{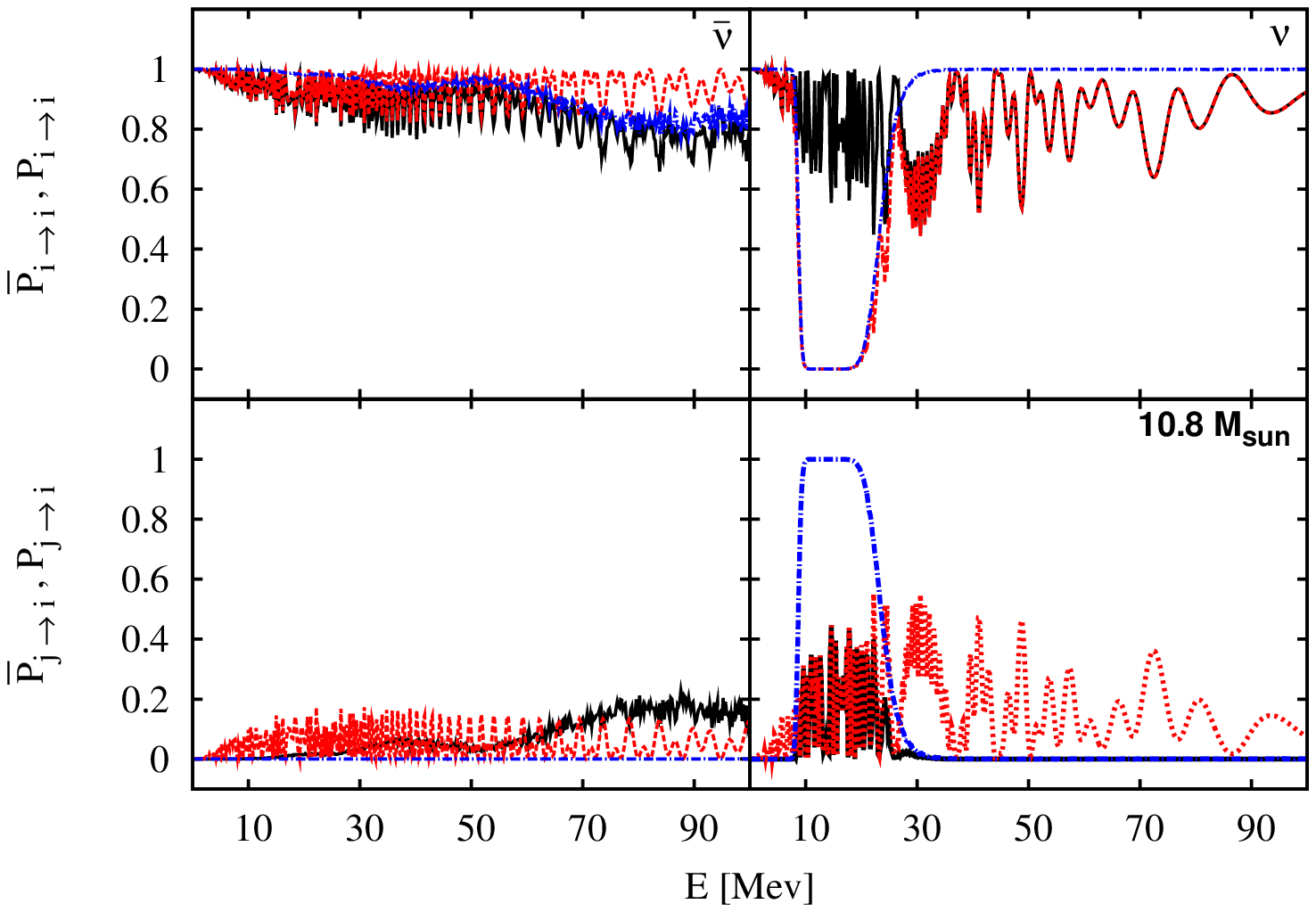}
\includegraphics[width=0.49\linewidth]{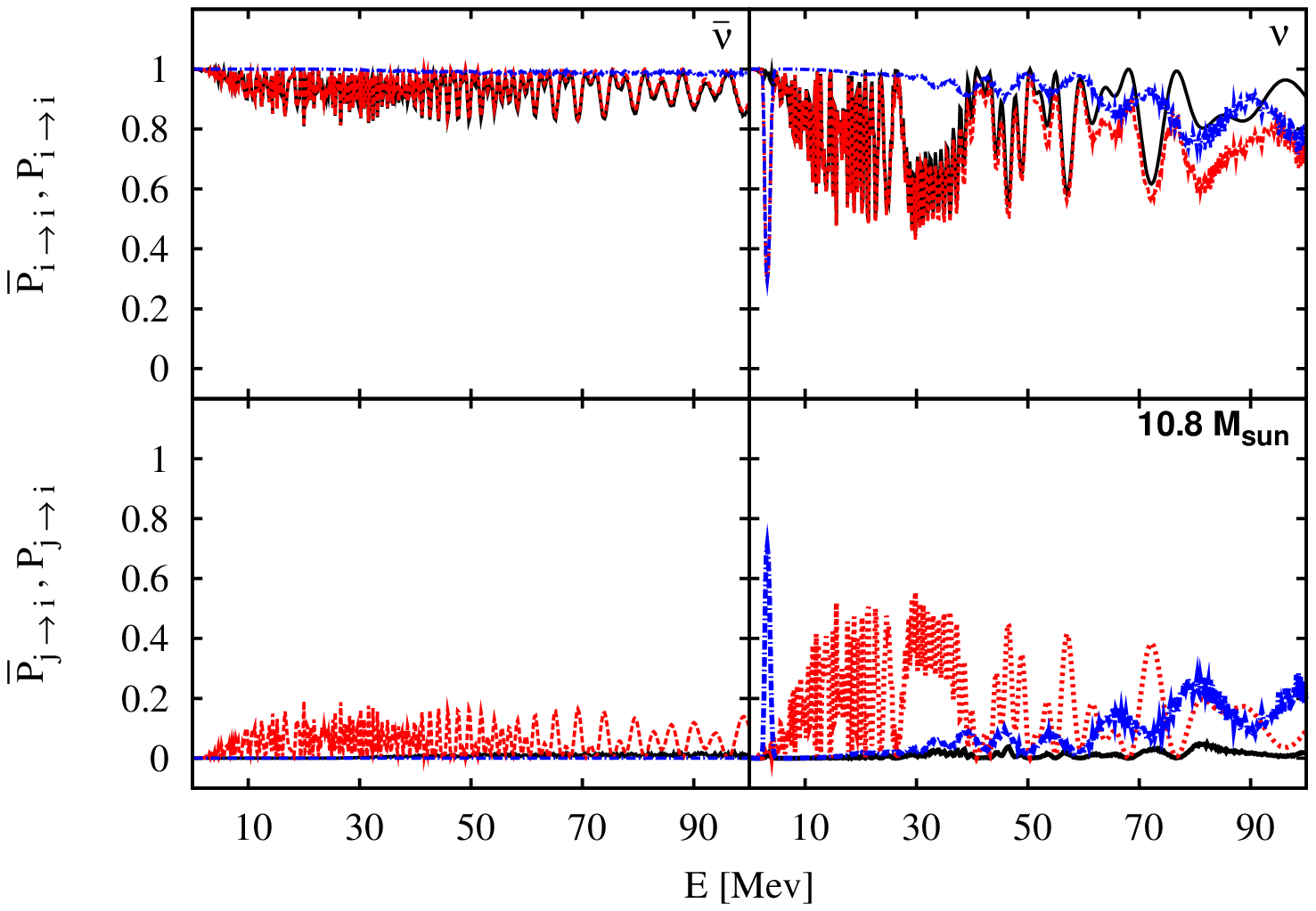}
\caption{\label{fig:m10_t5t8t105} (color online).
Matter survival and transition probabilities for the \Mten{}
model. All results are for a full profile calculation.  Top to
bottom quartets have probabilities from 4.8~s, 7.8~s and
10.5~s.  On the left side is the IH and on the right side is
the NH.  Each quartet has the same layout as the middle quartet
of Fig.~\ref{fig:nmIH_3sec} and line styles have the same
meaning.} 
\end{figure*}

\begin{figure*}
\includegraphics[width=0.49\linewidth]{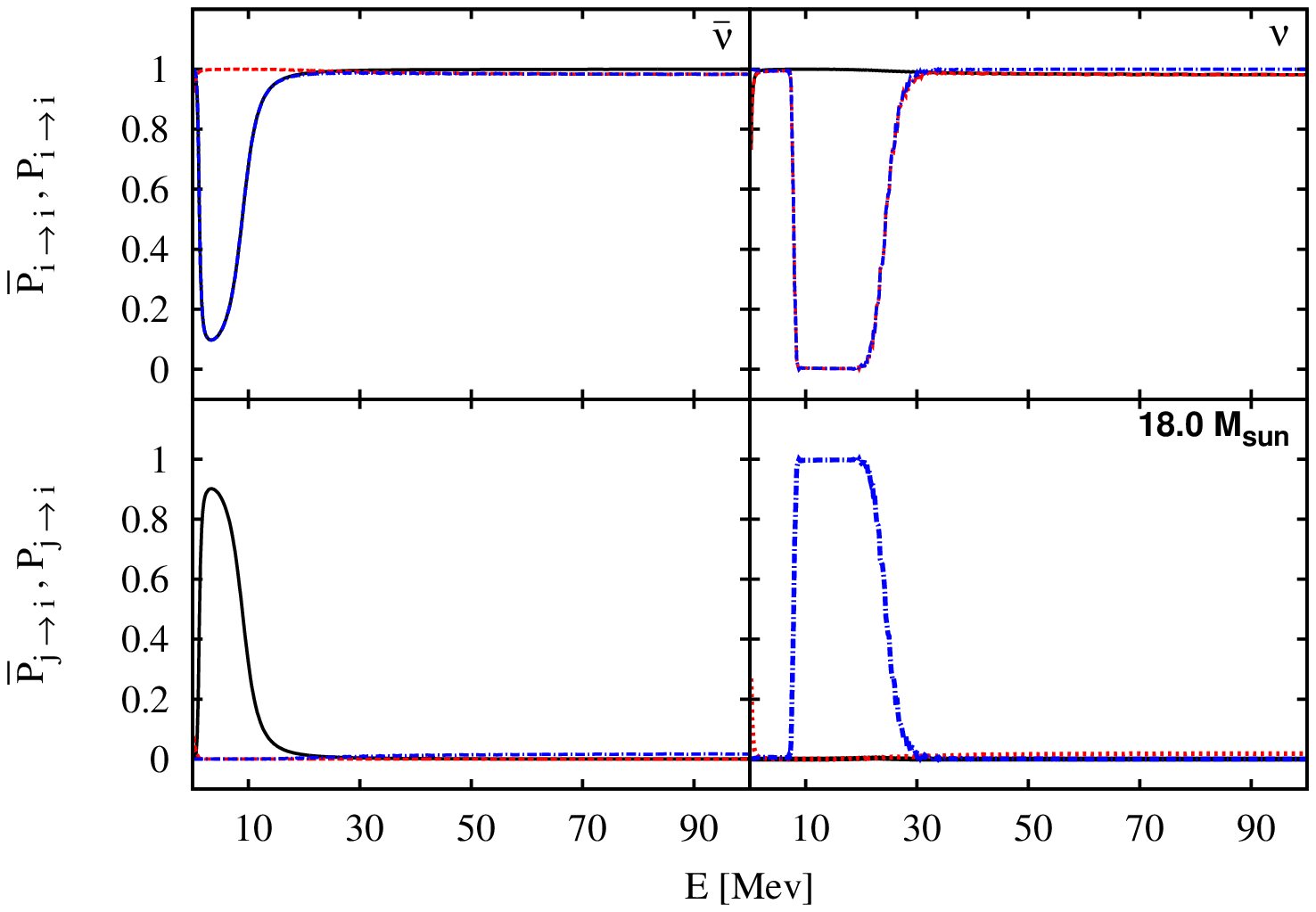}
\includegraphics[width=0.49\linewidth]{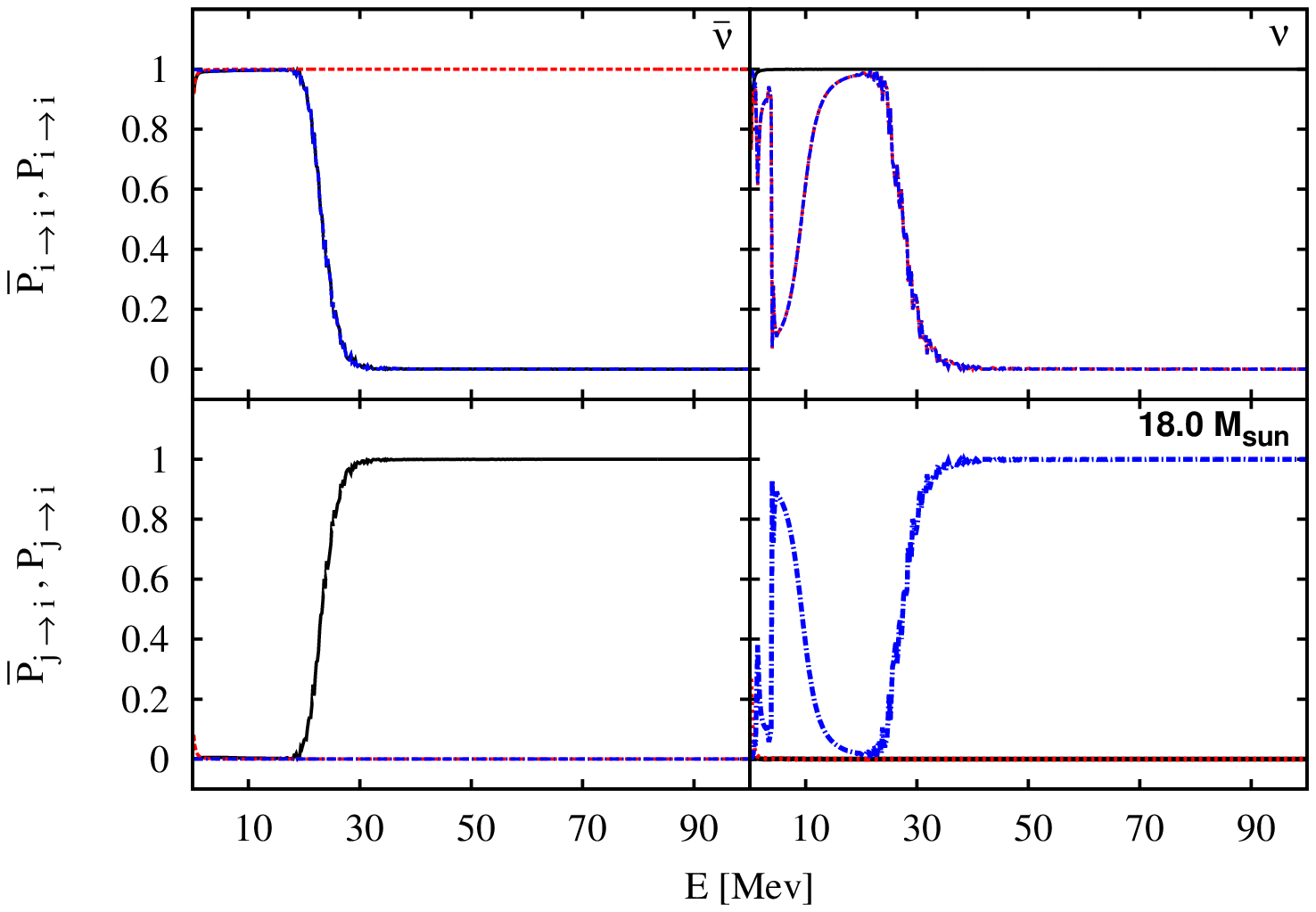}
\includegraphics[width=0.49\linewidth]{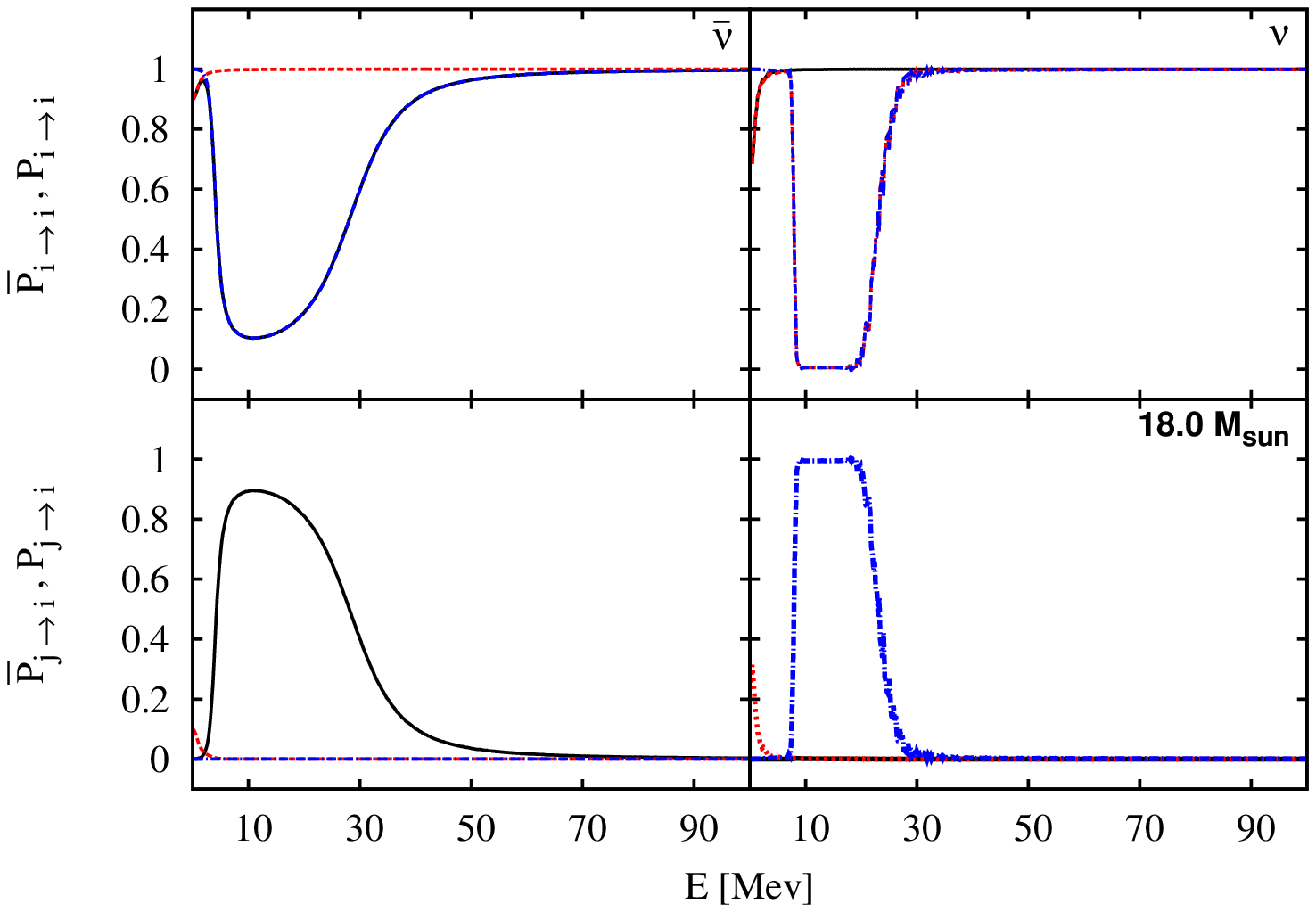}
\includegraphics[width=0.49\linewidth]{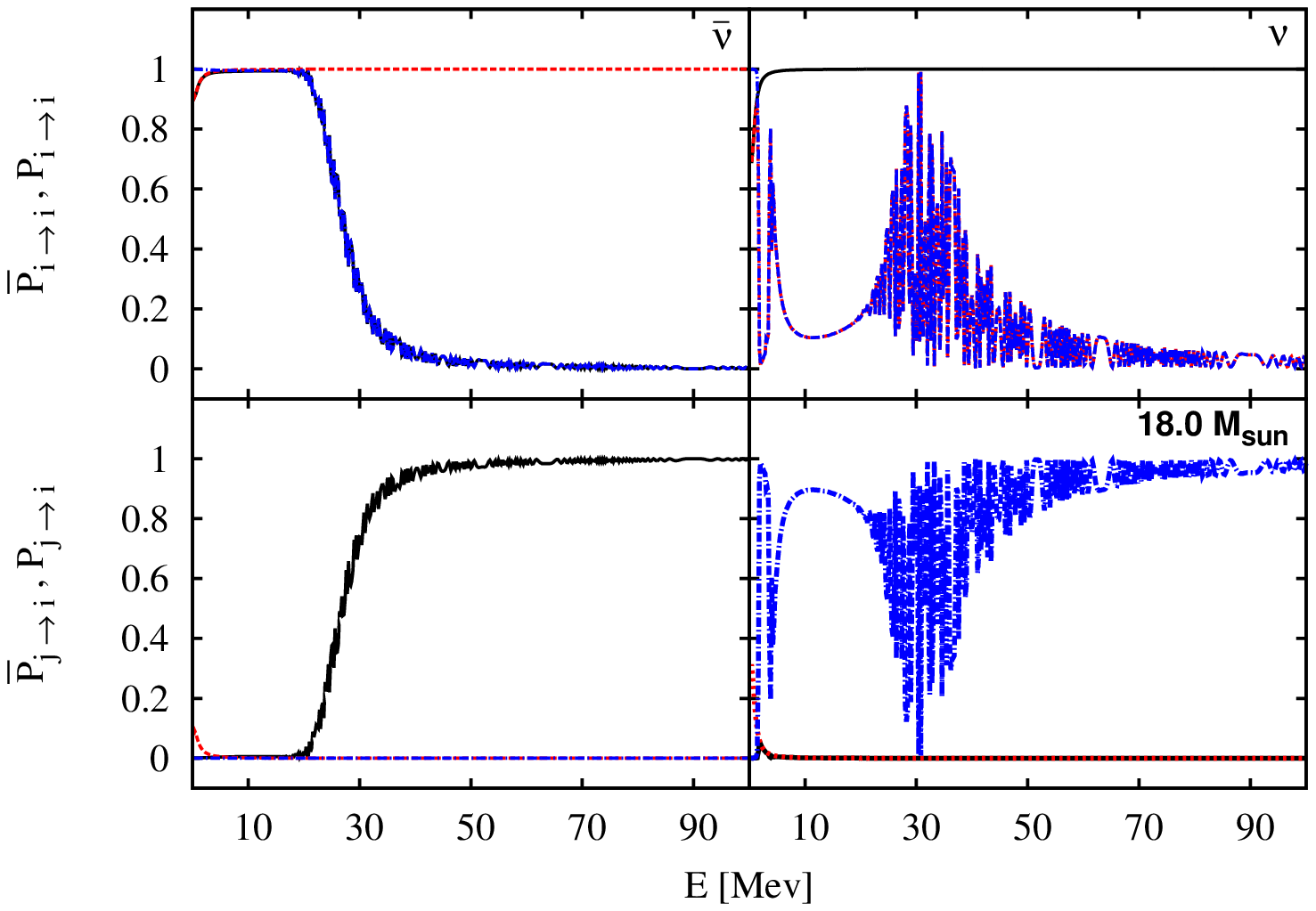}
\includegraphics[width=0.49\linewidth]{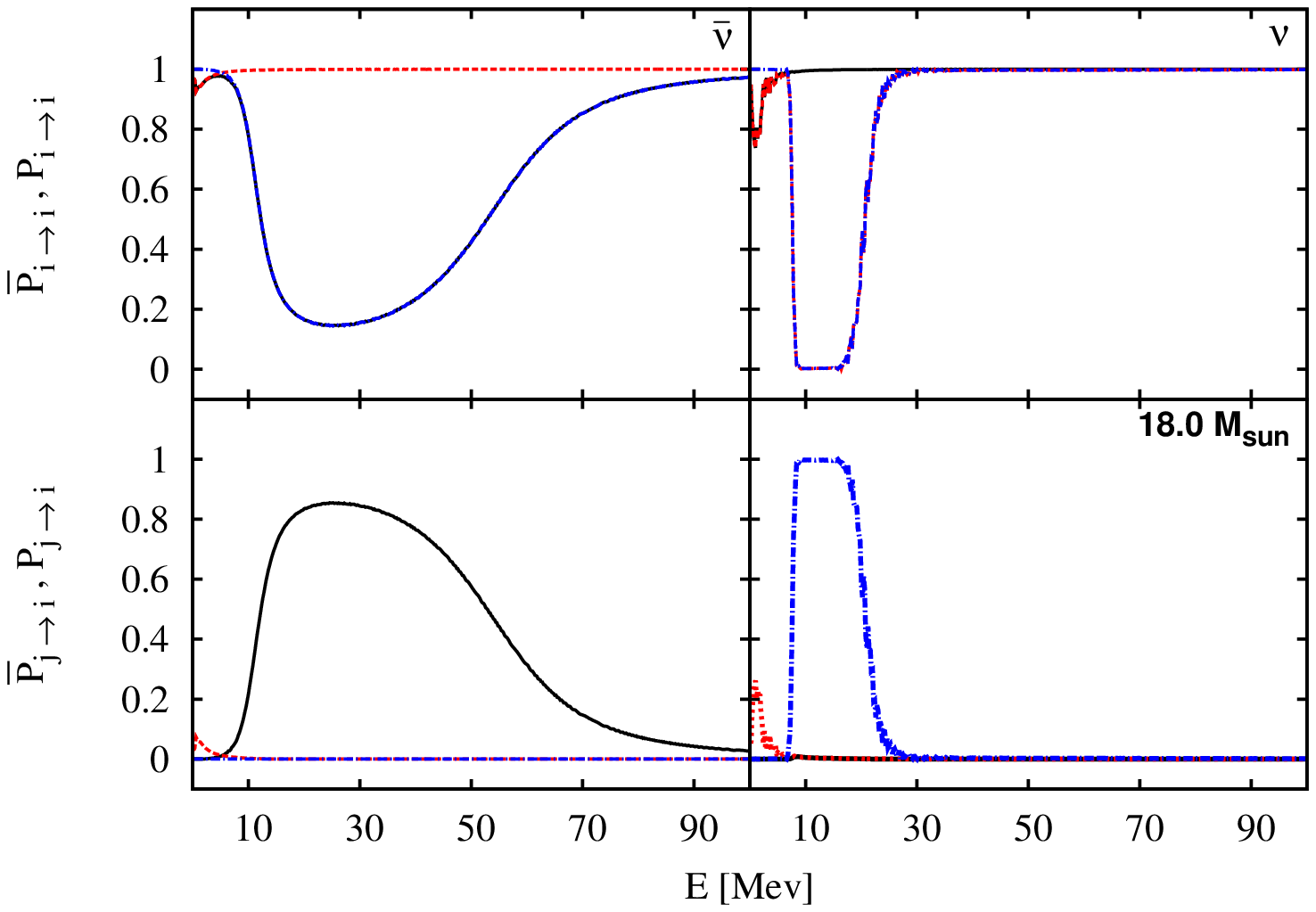}
\includegraphics[width=0.49\linewidth]{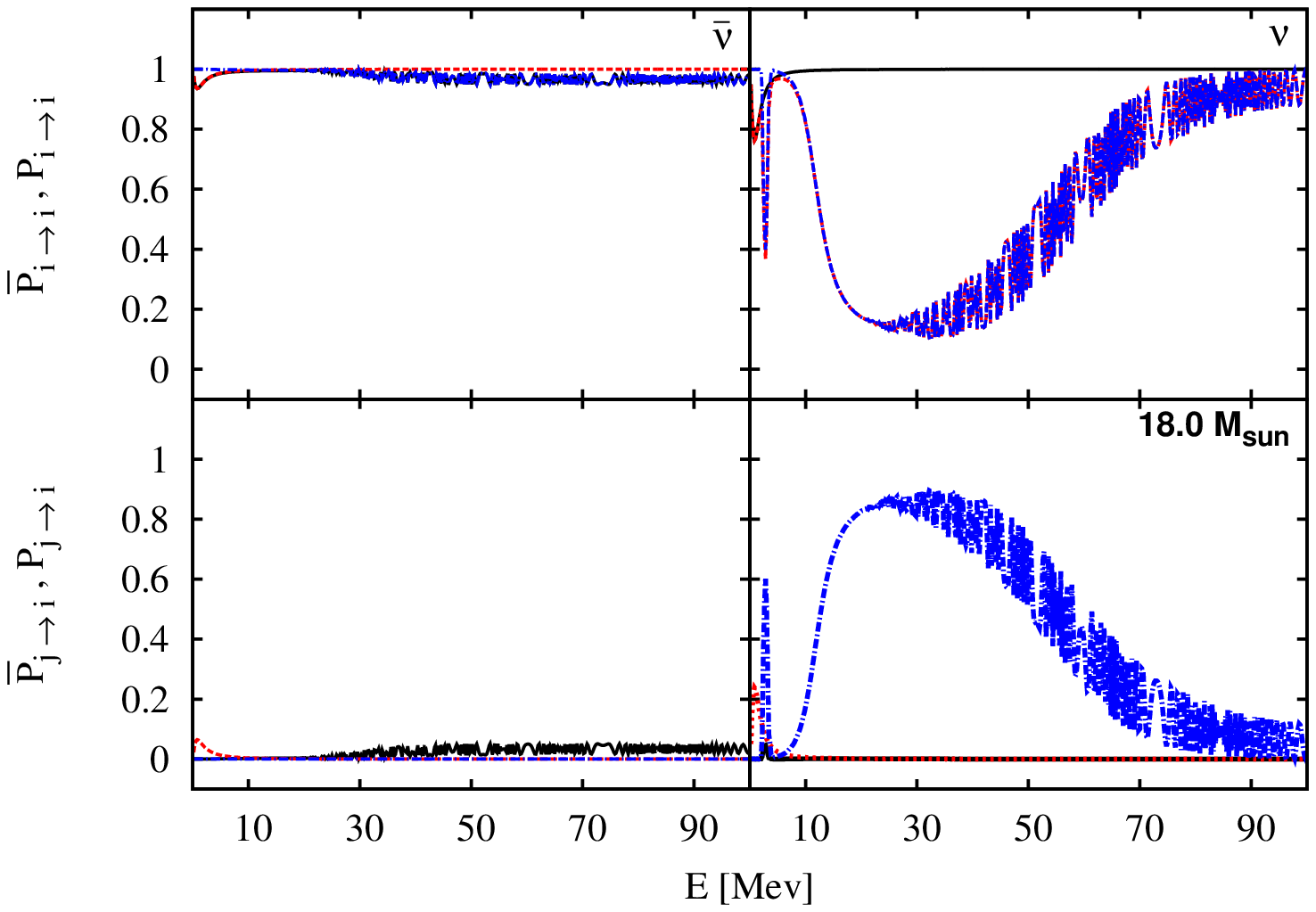}
\caption{\label{fig:m18_t5t8t11} (color online).  As in
Fig.~\ref{fig:m10_t5t8t105}, but for the \Met{} model.  From
top to bottom the quartets are at 5~s, 8~s and 11~s.  Left side
is the IH and the right side is the NH.} 
\end{figure*}

In the previous subsections we investigated the detailed time
evolution of important MSW and collective features separately.
In this section we will show the combined 
probabilities for a (anti-)neutrino traversing the full profile.
We show the survival probabilities in both hierarchies for three 
late time profiles of the \Mten{} and \Met{}
progenitors in Figures~\ref{fig:m10_t5t8t105} and 
\ref{fig:m18_t5t8t11} respectively.
For the \Mten{} progenitor we show results at 4.8~s (top
quartets), 7.8~s (middle quartets) and 10.5~s (bottom quartets).
The corresponding density profiles 
are shown in Fig.~\ref{fig:rhos_late} (red lines) along with 
the appropriate density profiles for the \Met{} model (black 
lines). Fig.~\ref{fig:m18_t5t8t11} show the results at 5~s (top
quartets), 8~s (middle quartets) and 11~s (bottom quartets) for
the \Met{} progenitor. The complete simulation of the \Met{} 
model runs to a post bounce time of
21.4 seconds, and going into a meticulous discussion would
require almost a paper on its own. Fortunately the evolution of
the density profile over the last 10 s is rather limited, thus
we show only a couple of examples of the late time results in
Fig.~\ref{fig:m18_t5t8t11}.

The two progenitors have several features in common, and a few
differences that can easily be related back to the behavior of
their respective density profiles.
The most obvious feature displayed by the two progenitor's
probabilities is the double split in the IH for neutrino states 2
and 3 (right panels of the left most quartets in 
Figures~\ref{fig:m10_t5t8t105} and \ref{fig:m18_t5t8t11}), that 
was discussed at length above. Equally clear is the
spectral split in the anti-neutrino states \bone{} and \bthree{}
in the NH (left panels of the right most quartets in 
Figures~\ref{fig:m10_t5t8t105} and \ref{fig:m18_t5t8t11}). 
This split is present in both progenitors at the two 
earlier snapshots, and then it has disappeared by the last 
snapshot at $\sim$11~s.
The time evolution of this split was discussed previously for
the \Mten{}. These two very prominent collective features are in 
sharp contrast to the absence of collective features in 
anti-neutrinos in the IH, occurring in both progenitors. 
A slightly curious result of this investigation is the
time duration over which the collective features remain visible.
One would expect the collective features to disappear from 
the probabilities as the interaction strength diminishes over 
time with the decreasing density.
We mentioned in Sec.~\ref{sec:coll_time_evolv} that there is no
analytical way of predicting the exact evolution of collective
features from the current understanding of neutrino
self-interactions. 
However, based on Fig.~\ref{fig:rhos_late} it would appear 
plausible that the largest effect of self-interaction takes 
place within the first $\sim$150--200~km where the densities 
remain on the same order as time progresses. Thereby resulting in 
similar collective features over time and across progenitors.

For both progenitors the behavior of the collective features in 
neutrino states 2 and 3 in the NH is significantly tangled with 
the evolution of the MSW features in the same energy region
(right panels of the right most quartets in 
Figures~\ref{fig:m10_t5t8t105} and \ref{fig:m18_t5t8t11}). 
We alluded to this fact towards the end of the previous subsection.
This entangled behavior of collective and MSW features leads us
to some of the features that differ between the two progenitors.
Unsurprisingly these features are dominantly MSW induced, and
easily related back to the differences in the density profiles.

The underlying collective features in neutrino states 2 and 3 in
the NH are two sharp spectral splits at 2 and 4~MeV, and a
slightly softer spectral split at $\sim$30~MeV.
The MSW then adds a ``shock'' feature, very similar to the one 
seen in the IH anti-neutrino states \bone{} and \bthree{} for 
the \Met, to the probabilities.
Tackling one progenitor at the time, we'll start with
the \Mten{} model:
The two low energy collective splits are visible throughout the 
times we are examining although they are clearly in the
process of vanishing at 10.5~s.
The MSW effect of the contact discontinuity at 4.8~s is to 
generate a split for neutrinos with energies above $\sim$30~MeV 
(cf. Fig.~\ref{fig:shockMove_dens}). This split is 
overlapping in energy with the collective swap also around
30~MeV. Therefore we see an initial decline in the survival
probabilities $P_{22}$ and $P_{33}$ prior to 30~MeV, followed by
a swap back to nearly unit survival probability just after
30~MeV.  
At 7.8~s the contact discontinuity has moved out of the H resonance
region, and the corresponding split just below $\sim$30~MeV in 
neutrino states 2 and 3 have gone, leaving only the collective 
split at 30~MeV before that too is gone by 10.5~s.

Moving to the \Met{} progenitor we see a similar story, only
without any obscuring phase effects as the \Met{} develops only
a forward shock.
As with the \Mten{} model the underlying collective effects in 
neutrino states 2 and 3 in the NH (left panels of the right most 
quartets of Fig.~\ref{fig:m18_t5t8t11}) are two spectral splits 
at 2 and 4~MeV, with an additional split at 30~MeV.
In this case however the MSW effect is caused by the forward
shock. At 5~s the forward shock is affecting neutrinos of
energies from roughly 1 to 10 MeV, whereas the affected neutrinos
have energies of 10 to roughly 60 MeV at 11~s (cf. 
Fig.~\ref{fig:shockMove_dens18}).
By 11~sec the collective interaction have subsided so much that
the split at 30~MeV has vanished. A remnant of the low energy
double split remains, but the main flavor conversion is induced
by the MSW effect due to the forward shock.
At 5~s initially the collective effect causes the swap of
neutrinos with energies between 2 and 4~MeV. Then the shock
feature arises and forces neutrinos with energies already from 
1~MeV to swap, but when the shock impacts the already swapped 
neutrinos in the energy region 2--4~MeV it swaps them back, 
making their survival probability approach unity. The forward 
shock continues to affect
neutrinos with energies up to $\sim$10~MeV causing the drop in
the survival probabilities $P_{22}$ and $P_{33}$ below 10~MeV.
At 8~s the impact of the forward shock is on neutrinos with 
energies between 4 and $\sim$29~MeV. We see how this causes the
survival probabilities to drop in between the
self-interaction induced splits at 4 and 30~MeV.

The remaining MSW features are quickly summarized and explained:
A swap in the anti-neutrino states \bone{} and \bthree{} in the
IH of the intermediate and higher energy neutrinos in the 
\Mten{} progenitor is caused by the diabatic contact discontinuity. 
By 7.8~s all features in the density profile of the
\Mten{} model have moved out of the H resonance region and into
the L resonance region. This is reflected in the survival
probabilities for neutrino states 1 and 2 in both hierarchies,
as they drop from unity at low and intermediate
energies. The multiple diabatic resonances caused by the contact
discontinuity, reverse and forward shocks (visible in
Fig.~\ref{fig:rhos_late}) lead to dominating phase effects.

In the \Met{} progenitor the forward shock, although moving to
higher energy resonant densities, remains in the H
resonance region (black lines in Fig.~\ref{fig:rhos_late}). 
The impact of this was already mentioned above for neutrino 
states 2 and 3 in 
the NH, and is plainly visible in anti-neutrino states \bone{}
and \bthree{} in the IH too. Although the shock itself does not
reach into the L resonant densities the width of the L resonance
causes a small mixing of neutrino states 1 and 2 at very low
energies, an effect visible in both hierarchies of the 8~s and 
the 11~s results.

\section{Conclusions}       \label{sec:conclusions}
In this paper we have aimed at uncovering the rich phenomenology
that arises in the neutrino spectra from the combination of
self-interactions, MSW interactions and turbulence in
core-collapse supernovae. We have calculated the neutrino flavor
evolution through realistic density profiles from 1D numerical
simulations and found that the signal can, potentially, provide great
insight into what takes place inside a core-collapse supernova,
but that insight comes at a price.  The various flavor
transformation processes leave complex features that depend
sensitively on a wealth of quantities: the detailed ratios of
energies and luminosities between neutrino flavors, the density
profile of the progenitor star and its time development
especially with respect to shocks and turbulence, and the
magnitude of the turbulence.  Many of these aspects have been
investigated separately hitherto, but from the investigation
presented here, it is clear that for a comprehensive picture one
needs to tackle all of these effects together.

We have calculated the flavor conversion caused by collective,
MSW and turbulence effects separately and in combination as
neutrinos leave a core-collapse supernova. We find that the
self-interaction induced spectral splits are similar at equal
times for all three progenitor masses investigated here. The
similarities can be ascribed to two factors: The ratio between
the number fluxes of the different neutrino flavors are almost
identical for the three progenitors, and the differences between
the matter density profiles inside 1000~km are only moderate,
indicating that the neutrino densities are somewhat similar. We
have pointed out several prominent features in the survival
probabilities that persist and evolve over time. These include
spectral splits in both hierarchies for both neutrinos and
anti-neutrinos, and are consistent with previous findings,
albeit different in detail.

Moving further out into the supernova mantle we have found that
the impact of MSW effects are significant only at later times,
and that of our three progenitors the \Mten{} model shows the
largest effects in the neutrino survival probabilities.  We have
shown how one can in principle follow the progression of the
shock front in the star by following how the spectral split induced
by the MSW H resonance moves in energy. We have shown how this
shock feature moves rapidly over almost two orders of magnitude
in energy for the \Mten{} model, but only over an order in
energy for the \Met{} model.  Should it be possible to identify
this shock feature in a future observation of neutrinos from a
galactic supernova, it could potentially give us a hint about
the progenitor mass. For the \Mten{} progenitor we were
furthermore able to follow the shock front moving into the MSW L
resonant density layers. As a consequence we can present, for the
first time, results of calculations on how this resonance impacts
the neutrino flavor probabilities. We found the anticipated
conversions between neutrino matter states 1 and 2, and found
them to have approximately the same magnitude in both
hierarchies as expected due to the largeness of $\theta_{12}$.

Through calculations combining both collective interactions and
MSW interactions we have shown that the net effect can lead to a
wash out of features that would be expected from treating the
two effects separately. The subsequent influence of the MSW
interaction will (partially) undo the impact of the
self-interaction. Consequently future investigations should
include both effects when predicting observable features.

We have shown that including moderate amounts of turbulence has
a limited effect on the flavor evolution for the two heaviest
progenitors. On the lightest progenitor the impact of turbulence
hinges sensitively on the instantaneous density profile.  Over a
few seconds the impact of turbulence goes from moderately
significant at 1~s to non-existent at 3~s.  Increasing the
amplitude of turbulence we found that large amounts of
turbulence can obscure features induced by the neutrino
self-interactions or the MSW effect. This general conclusion
holds for all of our three progenitors although in this paper we
have only presented results from the \Mten{} model.
We find that the impact of large amounts of turbulence can be
summarized as both obstructive and constructive.  On the
constructive side we find that large amounts of turbulence will
lead to novel mixing phenomena in the anti-neutrino channel in
the NH as discussed in Sec.~\ref{sec:turb_larger}.  Furthermore,
large turbulence leads to increased mixing between neutrino
states 1 and 2 at low energies in both hierarchies. 
On the destructive side large amounts of turbulence will obscure
some signal features.\\

We expect some of the spectral splits and large scale features
in the neutrino probabilities will be observable in a future
neutrino signal. Especially the NH neutrino split at
$\sim$30~MeV between states 2 and 3 or the anti-neutrino split
between states \bone{} and \bthree{} around 25~MeV might be
visible. In an upcoming paper \cite{Lund:2013obs} we will
examine in more detail the prospect of observing some of the
features described  in this paper. However, presently it is
clear that with the energy resolution of current detector
technology the high frequency oscillations imposed on the
survival probabilities by the phase effect will not be
observable. For diagnostic purposes in a future observed
neutrino signal we find that the NH generally always has a
spectral split around 30~MeV in the neutrino sector and, in most
snapshots, we also find a similar split around 25~MeV in the
anti-neutrino sector. Both of these splits are induced by
collective effects and appear to survive MSW effects and the
addition of moderate amounts of turbulence, although the
neutrino split does not survive large turbulence.
The spectra in the IH are much more dependent upon time but a
persistent feature is the double spectral split in neutrinos at
$\sim$8~MeV and 23--35~MeV. This feature is again induced by
collective effects and survives the impact of the MSW
interaction in the outer regions. Unlike the neutrino feature
discussed above for the NH, this double split survives both the
addition of moderate and large amounts of turbulence. 
The upper split is in an energy region where it is could
actually be observed. However, this requires that future
detectors focused on neutrinos are built, and with a large future
neutrino detector to compliment current (and future)
anti-neutrino detectors we may gain evidence for the hierarchy
from an observation of spectral splits in a neutrino signal.

In order to decode a real neutrino burst signal from a Galactic
supernova we cannot continue to investigate collective and MSW
effects separately, but have to treat them in combination. With
our upcoming paper predicting the observed signals, we aim to
show how these features will appear in an observation and how we
might use such observations to learn about the neutrino mass
hierarchy, neutrino interactions in dense matter, the mechanism
of core-collapse supernovae and potentially the progenitor mass.

\section*{Acknowledgments}
We acknowledge support by the Department of Energy under grant
number DE-SC0006417. We gratefully acknowledge access to data 
from Tobias Fischer.


\end{document}